\documentclass[finalnew]{agujournal2019}
\usepackage{url} %this package should fix any errors with URLs in refs.
\usepackage{lineno}
\usepackage[inline]{trackchanges} %for better track changes. finalnew option will compile document with changes incorporated.
\usepackage{soul}
\soulregister\cite7
\soulregister\citeA7
\soulregister\ref7
\usepackage{amsmath}
\usepackage{upgreek}
\usepackage{tabularx}

\draftfalse
\journalname{JGR: Space Physics}

\begin{document}

%% ------------------------------------------------------------------------ %%
%  Title
%
% (A title should be specific, informative, and brief. Use
% abbreviations only if they are defined in the abstract. Titles that
% start with general keywords then specific terms are optimized in
% searches)
%
%% ------------------------------------------------------------------------ %%

\title{Not So Fast: A New Catalog of Meteor Persistent Trains}

%% ------------------------------------------------------------------------ %%
%
%  AUTHORS AND AFFILIATIONS
%
%% ------------------------------------------------------------------------ %%

\authors{L. E. Cordonnier\affil{1,2}, K. S. Obenberger\affil{1}, J. M. Holmes\affil{1}, G. B. Taylor\affil{2}, and D. Vida\affil{3}}

\affiliation{1}{Space Vehicles Directorate, Air Force Research Laboratory, Kirtland AFB, NM, USA}
\affiliation{2}{Department of Physics and Astronomy, University of New Mexico, Albuquerque, NM, USA}
\affiliation{3}{Department of Physics and Astronomy, University of Western Ontario, London, ON, Canada}

%% Corresponding Author:
% Corresponding author mailing address and e-mail address:
\correspondingauthor{Logan Cordonnier}{lcordonnier@unm.edu}

%% Keypoints, final entry on title page.

%  List up to three key points (at least one is required)
%  Key Points summarize the main points and conclusions of the article
%  Each must be 140 characters or fewer with no special characters or punctuation and must be complete sentences

\begin{keypoints}
\item New observations of meteor persistent trains are not consistent with many of the previous assumptions
\item Most persistent trains were left by relatively slow and dim meteors which are traditionally not expected to produce them
\item We found that meteor properties such as terminal altitude and dynamical origin affect the likelihood that persistent trains develop
\end{keypoints}

%% ------------------------------------------------------------------------ %%
%
%  ABSTRACT
%
% A good Abstract will begin with a short description of the problem
% being addressed, briefly describe the new data or analyses, then
% briefly states the main conclusion(s) and how they are supported and
% uncertainties.
%% ------------------------------------------------------------------------ %%

\begin{abstract}
This paper presents the results of a nearly two year long campaign to detect and analyze meteor persistent trains (PTs)---self-emitting phenomena which can linger up to an hour after their parent meteor. The modern understanding of PTs has been primarily developed from the Leonid storms at the turn of the century; our goal was to assess the validity of these conclusions using a diverse sample of meteors with a wide range of velocities and magnitudes. To this end, year-round observations were recorded by the Widefield Persistent Train camera, 2nd edition (WiPT2) and were passed through a pipeline to filter out airplanes and flag potential meteors. These were classified by visual inspection based on the presence and duration of trains. Observed meteors were cross-referenced with the Global Meteor Network (GMN) database, which independently detects and calculates meteor parameters, enabling statistical analysis of PT-leaving meteors. There were 4726 meteors codetected by the GMN, with 636 of these leaving trains. Among these were a large population of slow, dim meteors that left PTs; these slower meteors had a greater train production rate relative to their faster counterparts. Unlike prior research, we did not find a clear magnitude cutoff or a strong association with fast meteor showers. Additionally, we note several interesting trends not previously reported, which include PT eligibility being primarily determined by a meteor's terminal height and an apparent dynamical origin dependence that likely reflects physical meteoroid properties.
\end{abstract}

%% ------------------------------------------------------------------------ %%
%
%  TEXT
%
%% ------------------------------------------------------------------------ %%

\section{Introduction}
\begin{sidewaysfigure}
    \centering
    \noindent\includegraphics{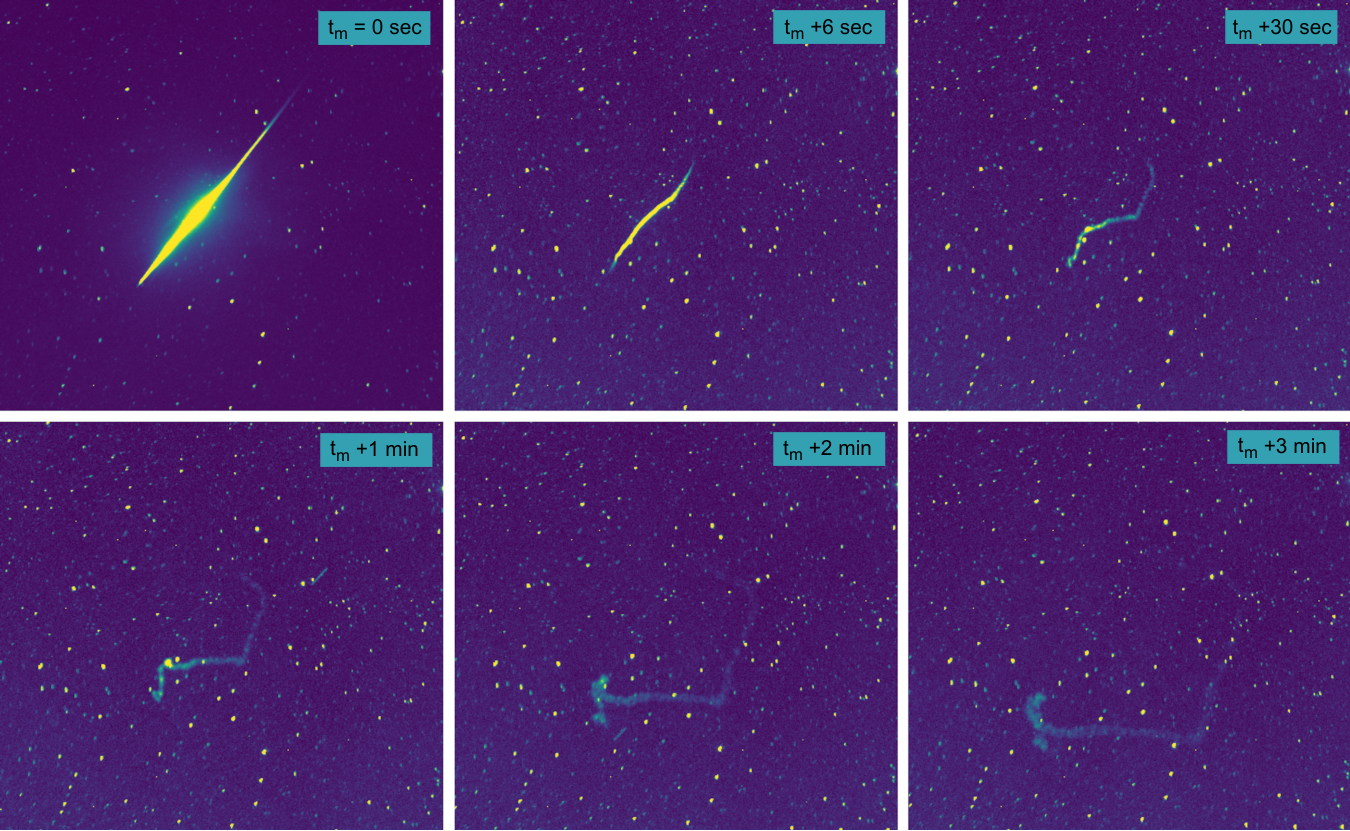}
    \caption{The evolution of a PT. The t$_m$=0 sec panel shows the parent meteor, the other timestamps reflect the time elapsed since the meteor. The colormaps of the PT panels are identically scaled; the colormap maximum of the meteor panel is higher to better show its structure. This PT lasted at least 22 minutes before drifting out of our field of view. Each panel shows a patch of sky approximately 26$^\circ$ by 27$^\circ$. The corresponding movie file (Movie S3) can be found in the supporting information.}
    \label{fig:PT_image}
\end{sidewaysfigure}

Persistent trains (PTs) are fascinating luminous phenomena which can occasionally be observed lingering after the passage of a meteor; their enduring luminosity arises from self-emitting chemiluminescent mechanisms, i.e. they are not the same as meteoric smoke trails which are instead illuminated by scattering sunlight. PTs also often have intricate structures that undergo distortion governed by winds in the upper mesosphere/lower thermosphere (MLT) regions of the atmosphere---wind shears are capable of producing large-scale morphologies such as loops, spirals, and bright knots in the trains. As evidenced by their name, PTs have been observed to last on the order of minutes up to more than an hour \cite{Beech1987}. Figure \ref{fig:PT_image} illustrates how a PT evolves over a span of several minutes; movie files for this event and a curated selection of other PTs are included in the supporting information. 
Following the terminology put forth by the International Astronomical Union (IAU) and by \citeA{Borovicka2006}, the generic term \textit{train} refers to the ``light or ionization left along the trajectory of the meteor after the meteor has passed" \cite{Koschny2017}. The qualifier \textit{persistent} as used in this work applies to trains which have existed sufficiently long to exhibit the continuum phase of emission. The continuum phase is the third and final stage of PT evolution identified by \citeA{Borovicka2006}, the earlier two being the afterglow and recombination phases. 

\subsection{PT Formation and Morphology}
In brief, the afterglow phase lasts a few seconds and is a result of radiative deexcitation from species which were collisionally excited in the warm air plasma that remains after the passage of the meteor, where temperatures are initially several thousand Kelvin \cite{JenniskensBook}. Spectra from this phase indicate that the main emission comes from low excitation lines associated with metals ablated from the meteor, including neutral Na, Mg, K, Ca, Cr, Mn, and Fe and ionized Ca; also seen is the green oxygen line (557.7 nm) arising from the metastable $^1$S to $^1$D transition in atmospheric oxygen atoms \cite{Borovicka2000}. As the air plasma cools, the recombination phase begins to dominate. This phase generally lasts several tens of seconds and differs from the afterglow mainly by the presence of lines with higher excitation energies. Recombination processes between cations and free electrons (or anions) are primarily responsible for this emission, since they are capable of populating higher energy levels than is possible via thermal collisions at these temperatures \cite{Borovicka2006}. \citeA{Abe2004} examined the recombination spectra of several Leonid trains between 300 and 930 nm and found emission lines in the near infrared (NIR), visible, and ultraviolet (UV) regimes. They pointed out several prominent features, including the Mg I 518 nm line and the Na D-line doublet (589 nm) which are often the two brightest lines in the visible portion. The latter has long been thought to result from the Chapman mechanism \cite{Chapman1939, Chapman1955} where sodium catalyzes the conversion of ozone and atomic oxygen to diatomic oxygen according to
\begin{linenomath*}
\begin{subequations}
    \label{eqn:NaRxn}
    \begin{align}
        Na+O_{3} &\rightarrow NaO+O_{2}\\
        \label{eqn:second_step}
        NaO+O &\rightarrow Na\left(^2\textnormal{S},^2\textnormal{P}\right)+O_{2}\\
        Na\left(^2\textnormal{P}\right) &\rightarrow Na\left(^2\textnormal{S}\right)+h\nu(\textnormal{D}_{1}, \textnormal{D}_{2})
    \end{align}
\end{subequations}
\end{linenomath*}
with the ozone coming from ambient atmosphere. At UV wavelengths, the emission was dominated by Mg I 383 nm for early times, and also showed strong Fe I and hydroxyl OH(A-X) lines. Lastly, in NIR they found an assortment of neutral meteoric metal lines as well as O$_2$(0,1) ($\sim$865 nm) emission coming from the $\left(b^1\Sigma_{g}^{+}\right) \rightarrow \left(X^3\Sigma_{g}^{-}\right)$ transition which lasted around 14 seconds. This excited state of O$_2$ can be produced in step \ref{eqn:second_step} above \cite{Hapgood1980}, can also result from similar oxidation processes with Fe, Me, or Ca \cite{Kruschwitz2001}, or can result from oxygen atoms (originating as atmospheric O$_2$ that underwent UV-photodissociation) exciting other atmospheric O$_2$ molecules \cite{Zinn2005}. 

The fading of the recombination lines signals the start of the continuum phase, which lasts for the remainder of the PT's existence. This stage is characterized by broad continuum emission predominantly associated with molecules, the leading candidate being iron oxide \cite<FeO;>[]{Jenniskens1998}. According to \citeA{Jenniskens1998}, the reaction giving rise to this emission could go as
\begin{linenomath*}
\begin{subequations}
    \label{eqn:FeO}
    \begin{align}
        Fe+O_3 &\rightarrow FeO\left(^5\Delta\textnormal{ etc.}\right)+O_2\\
        FeO+O &\rightarrow Fe+O_2
    \end{align}
\end{subequations}
\end{linenomath*}
with the FeO being produced in an excited state to form the ``orange arc" (570-630 nm) band emission \cite{West1975, Popov2020}. Though there may still be atomic emission lines present, these are generally a small fraction of the total luminosity; in one instance, the integrated intensity of the FeO emission was about 40 times greater than that of the Na D-lines, though still not sufficient to account for the total luminosity \cite{Kruschwitz2001}. Other metal oxides engaged in processes similar to reaction \ref{eqn:FeO} such as MgO, AlO, and CaO could also be involved \cite{Baggaley1976}. Additionally, molecular species like NO$_2$ and SO$_2$ have been suggested contributors as well \cite{Borovicka2000, Murad2001}. Emission in IR wavelengths has been detected during the continuum phase---\citeA{Hapgood1980} observed a PT exclusively in the 700-900 nm NIR regime for more than half an hour and attributed the luminosity to the same O$_2$(0,1) emission mentioned previously; \citeA{Kruschwitz2001} also supports O$_2$ as the source of the NIR component. Alternatively, \citeA{Clemesha2001} and \citeA{Vasilyev2021} assert that OH, not O$_2$, is instead responsible. Lastly, heated atmospheric molecules, e.g. CO, CO$_2$, CH$_4$, and H$_2$O, were identified in the mid-IR (3-13 $\upmu$m) regime with temperatures of $\sim$300 K \cite{Russell1998}. Due to the absence of clear spectral structures during the continuum phase, it is difficult to pin down which molecules listed above contribute to the overall luminosity, and in what proportions they do so. 

In toto, the current understanding of PT evolution and long-enduring luminosity is governed by these three stages, though some challenges remain. The individual species present in the spectra during all stages of PT evolution can differ on a train-by-train basis, further complicating the matter. A substantial caveat with this PT evolution scheme is that it was primarily developed using observations of the Leonids, especially those from the meteor storms around the turn of the century, and therefore may not be entirely applicable to our population of meteors presently under study. This is due to the Leonids being one of the fastest meteor showers, with most of its meteors having initial velocities in excess of 68 km/s, whereas the majority of our observed meteors had initial velocities less than 42 km/s.

The overall shape and structural evolution of persistent trains is largely governed by winds in the upper atmosphere; tracking their movements can provide insight into atmospheric wind patterns. Horizontal winds in the region of interest for PTs (i.e. around 90 km) are typically on the order of a few tens of m/s, though their magnitude and direction exhibit seasonal trends \cite{Yuan2008}. A case study of a bright Perseid meteor and its associated PT was performed by \citeA{Spurny2014} which revealed velocities consistent with those of atmospheric winds, showing horizontal motions ranging from 27 m/s to 81 m/s depending on the time and location of the measurement in the train. They also found an overall upward vertical motion, with typical speeds of 2 m/s and a maximum uplift of 13.5 m/s. A complete description of PT emission intensity, duration, and morphological evolution also requires consideration of diffusion processes, which affect movement of meteoric metals out of the train and of fresh atmospheric O$_3$ into the train; sufficient quantities of both of these reactants are necessary for continued PT luminosity. Diffusivity is inversely related to pressure, which therefore introduces an inherent altitude dependence to the diffusion rate. This dependence, taken in conjunction with mesopause O$_3$ content (which is quite concentrated in a relatively narrow range of altitudes), results in complex interactions that affect the duration and time evolution of PTs. Further investigation into the importance of diffusion in PT dynamics should be pursued.  

A curious morphology has also been observed in a large number of PT events, the so-called ``double train" phenomenon, where two distinct, parallel lanes of luminous trails are observed. There have been several proposed explanations for how these double trains could form. One of the earliest theories, put forth by \citeA{Trowbridge1907}, was that the meteor train was a single, hollow cylinder which would appear brighter on either side due to the increased look-through thickness in these regions (i.e. ``limb brightening"). This hollow structure would arise due to the chemiluminescent reactions being concentrated at the outer boundary of the meteor's train; there would be a depletion of atmospheric constituents at the center resulting in less emission from that region \cite{Clemesha2001}. However, this idea has lost popularity since recent, detailed photographs of PTs show that there are indeed two distinct trains. Another early explanation for double trains relied on fragmentation of the parent meteoroid into two main pieces which then individually ablate and create two separate, parallel PTs (e.g. as discussed in \citeA{Jenniskens2003}). This theory has also become disfavored since fragmentation is expected to often produce more than two pieces, and PTs with triple trains or more have rarely been observed \cite{Kelley2013}. Modern research has used numerical computations to suggest that the cylinder of the PT first radially expands and then buoyancy effects cause the formation of a pair of counterrotating linear vortices which become the double train \cite{Zinn2005}. \citeA{Kelley2013} instead suggest that convective instabilities rather than buoyancy are responsible for the initial formation of these vortices. It should be noted that none of the PTs in our observations exhibited clear signs of this double train phenomenon, likely due to our level of resolution.

Lastly, there is one more emission regime that has only recently been associated with PTs. Dubbed ``meteor radio afterglows" (MRAs), these self-emitting phenomena could potentially result from transition radiation arising from suprathermal electrons produced after interactions with ablated meteoric species, though other plausible explanations exist \cite{Obenberger2020}. MRAs have been observed in the high and very high radio frequencies (3-30 and 30-300 MHz, respectively) and have durations lasting tens of seconds up to several minutes \cite{Obenberger2014, Dijkema2021}. As of yet, no MRAs have been observed with lifetimes comparable to the longest lasting optical PTs (e.g. half an hour or longer). Their radio emission is nearly isotropic \cite{Varghese2019}, broadband \cite{Obenberger2015, Varghese2021}, and interestingly shows a strong altitudinal dependence; MRAs are not typically observed below $\sim$90 km \cite{Obenberger2016}. This altitude is approximately the mean height where PTs are found. \citeA{Obenberger2020} reported four meteors that each exhibited both a PT and MRA, indicating that they potentially share a common energy source. Additionally, the four MRAs coincided with the location of PT activity rather than features of the parent meteor (e.g. flares), strengthening claims of a shared mechanism. It is worth noting that not every observed PT in their study had a corresponding MRA---the relationship between these two phenomena will be explored in more detail in a forthcoming paper.

\subsection{Prior Catalogs and Assumptions}
To contextualize the departure of our results from previous catalogs and observations, it is necessary to understand the nature and limitations of these foundational works. The first recognized attempt at describing the physics behind meteor trains came from C. C. Trowbridge in the early 20th century \cite{Beech1987}. Trowbridge compiled observations and reported on the altitudes of 13 trains; he noted that they form in a narrower range relative to the general meteor population, which he connected to favorable atmospheric conditions within that range \cite{Trowbridge1907}. The altitudes he found are still in good agreement with those presented in section \ref{alt}. A few years later, he stated that PTs are more frequently seen during meteor showers, particularly during the Perseids and Leonids, and posits that the high velocity of these showers could contribute to PT formation \cite{Trowbridge1911}. The work started by Trowbridge was later continued by Charles Olivier who collated PT observations from historical records, scientific publications, and observer accounts. The extensive results of this undertaking were published across three papers \cite{Olivier1942,Olivier1947,Olivier1957} and in total provided data for 2073 train events (though some of these were sunlit rather than self-emitting). He went on to calculate that about 1 in 750 (0.13\%) observed meteors ended up producing a PT, indicating that these events are relatively rare. Olivier's contemporaries made use of this catalog to demonstrate, for example, a positive correlation between the relative occurrence of PTs and solar sunspot number, with an above average number of PTs coinciding with solar maximum \cite{Kresak1949}. Interestingly, the maximum of solar cycle 23 aligned quite well with the Leonid storms at the turn of the century, possibly contributing to the considerable number of PTs observed. \citeA{Baggaley1978} used Olivier's catalog to determine that for PTs lasting at least a minute, 96\% are associated with meteors brighter than -2 mag and that only 1 visual meteor out of 5000 creates a PT longer than 5 minutes. Despite the undoubted value of these early catalogs, they do have limitations that hamper their usefulness for PT population statistics, mainly arising from the inhomogeneity of observations. The compilation of meteor and PT data from different observers around the world who are using a variety of instruments (e.g. naked eye, binoculars, telescopes, cameras, etc.) and who have a predilection for observing during prominent meteor showers is all but certain to introduce biases in the resulting statistics. 

The most recent large-scale effort to observe and catalog PTs was centered around the Leonid meteor storms of 1999-2002, as the Leonids have long been associated with PTs. A variety of observing campaigns were planned and conducted, including the Leonid Multi-Instrument Aircraft Campaign \cite<MAC; e.g.>[]{Jenniskens1999}, meteor train observation (METRO) campaign in Japan \cite{Higa2005}, sodium lidar measurements at the Air Force Research Laboratory's Starfire Optical Range \cite<SOR;>[]{Kelley2000}, as well as those of assorted meteor radars and observatories. Results of this directed effort greatly propelled the understanding of PTs; it enabled the development of the detailed PT evolution scheme outlined above \cite{Borovicka2000}, revealed the presence of MIR emission \cite{Russell1998}, and resulted in the first measurements of a PT via lidar \cite{Kelley2000}, among other things. Many of the papers published about PTs in the early-to-mid 2000s were connected to the Leonids in one way or another. The METRO campaign is particularly germane as it resulted in a catalog of PTs, reminiscent of Olivier's. The 176 trains listed therein were detected from 1998-2002 and were primarily photographed by amateur observers, producing a wealth of stunning images \cite{Higa2005}. The distribution of a campaign manual to these amateur observers helped to standardize data collection and mitigated some of the homogeneity issues faced by previous catalogs. \citeA{Higa2005} went on to determine that for Leonid meteors brighter than -2 mag, there was about a 20\% chance of PT formation. Analysis of the heights of 18 multi-station Leonid PTs revealed a nearly constant beginning altitude as a function of local time, whereas the ending altitude showed slight dependence on the local time, which they suggest could imply a dependence on the penetrating trail length of the parent meteor \cite{Yamamoto2004}. Despite the bounty of information obtained during the METRO campaign, as well as the others listed above, the results are highly Leonid-centric and therefore cannot be assumed a priori to be representative of all PTs.

After the excitement surrounding the Leonid storms waned, so too did the research pertaining to PTs. Aside from the association with MRAs, not much has fundamentally changed in our understanding of PTs since then. \citeA{Borovicka2006} summarizes a couple of general assumptions regarding PTs, which are likely heavily informed by the Leonid storms. He states that only bright, high-velocity fireballs will produce PTs, and that meteors which are brighter and faster tend to approximately result in brighter, longer lasting trains. \citeA{Borovicka2006} also states that PTs are always associated with the region of meteor maximum brightness (e.g. a bright flare). These assumptions, among others, are difficult to reconcile with the meteors and trains presented in this paper.

\section{Data Collection and Processing}
\subsection{Equipment}
\begin{figure}
    \centering
    \includegraphics{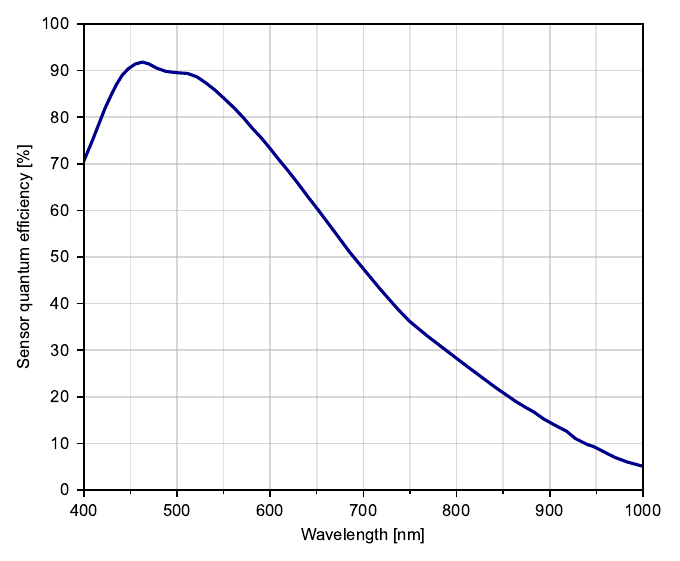}
    \caption{The spectral response of the sensor used in the WiPT2 camera; it exhibits moderate sensitivity in the NIR regime. This figure has been adapted from data available on the manufacturer's website (\url{https://www.zwoastro.com/product/asi6200/}).}
    \label{fig:QE}
\end{figure}

\begin{figure}
    \centering
    \includegraphics{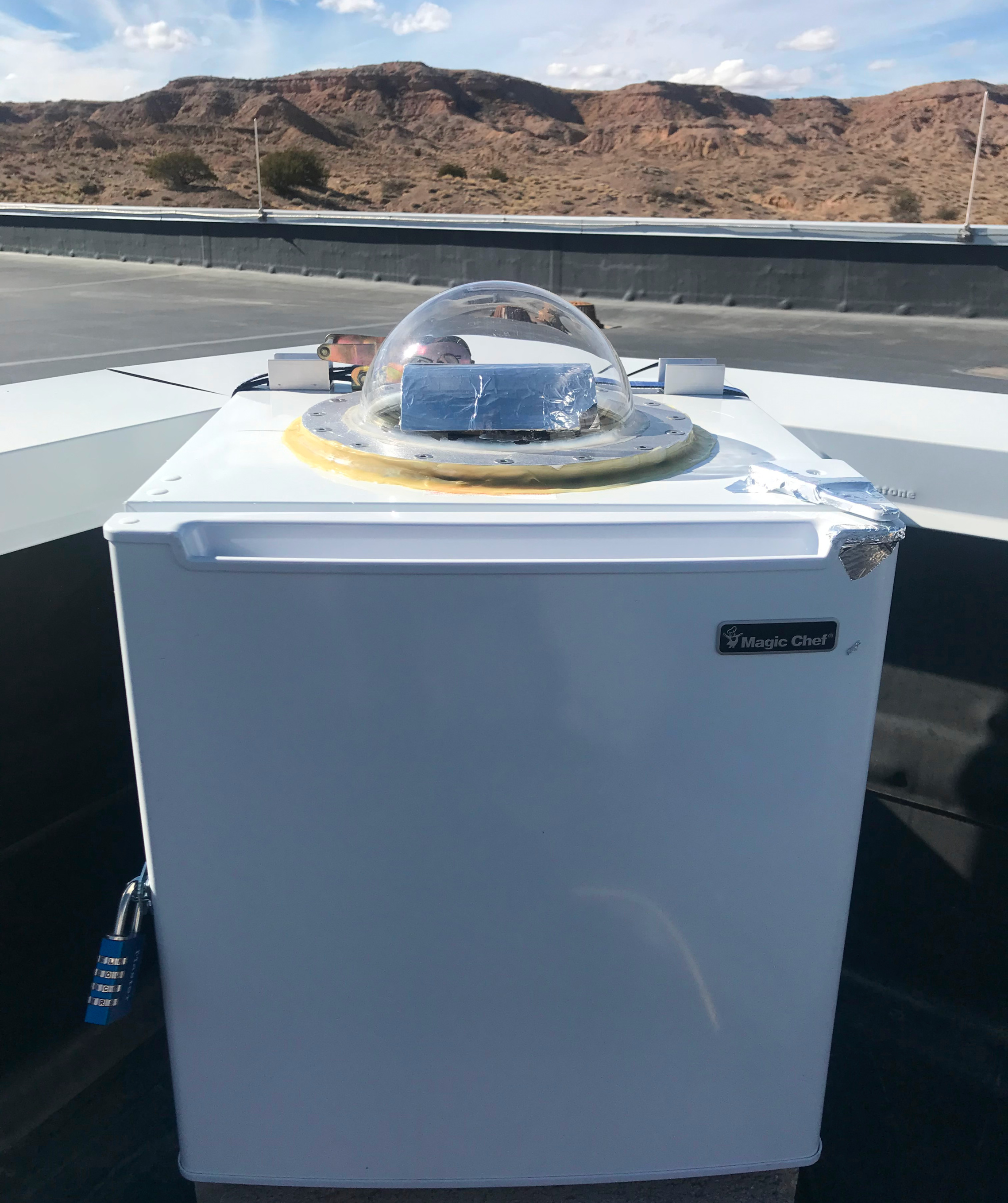}
    \caption{The deployed WiPT2 setup. All electronics are housed inside the minifridge; also seen are the clear plastic dome protecting the camera as well as the reflective clamshell sun shield within it.}
    \label{fig:setup}
\end{figure}

The Widefield Persistent Train Camera 2nd Edition (WiPT2 Cam) was deployed at the Sevilleta National Wildlife Refuge in New Mexico on 5 October 2021. The WiPT2 uses a ZWO ASI6200MM camera with an IMX455 CMOS sensor capable of being cooled 35$^{\circ}$C below the ambient temperature. The sensor consists of a 9576$\times$6388 array of 3.76$\times$3.76 $\upmu$m pixels which are first binned into blocks of 4$\times$4 pixels before being saved; this gives an effective pixel size of 15.04 $\upmu$m. It is also fairly sensitive to NIR emission, having a quantum efficiency (QE) of about 45\% at 700 nm and 15\% at 900 nm. The complete spectral response of this sensor is provided in Figure \ref{fig:QE}.
A Canon EF 15mm f/2.8 fisheye lens with a 180$^{\circ}$ diagonal angle of view is mounted atop the camera. This configuration provides a plate scale of about 3.45 arcmin/pixel near the image center. The images from the camera are saved to, and later processed on, a System 76 Meerkat computer with 4 TBs of storage. The camera, computer, and other electronics are housed inside a repurposed Magic Chef mini refrigerator which has a cutout on the top for a plastic dome. An automated clamshell sun shield is also installed inside the mini refrigerator that blocks much of the plastic dome when shut; this is kept closed when the camera is not operating to protect it from the sun, as well as to keep the interior electronics cooler. The exterior view of this assembly can be seen in Figure \ref{fig:setup}.
The camera is configured to take long exposure (5 second) images and has an inherent $\sim$1.5 second readout time between images. A full hour of data collection therefore yields approximately 552 images. The camera only collects data while the moon is not in the sky (to prevent the images from becoming overexposed) and requires the sun to be at least 15$^{\circ}$ below the horizon to ensure that any trains observed are due to intrinsic emission rather than reflected sunlight. The full resolution images saved by the camera are further binned into blocks of 4$\times$4 pixels (giving plate scales near image center of 13.8 arcmin/pixel) for archival purposes and for use in the meteor detection pipeline as their smaller size lends itself to faster processing. The computer has enough disk space to store nearly five months of full resolution data; once it reaches capacity, the oldest data is automatically deleted in order to maintain storage space on the computer.  

\subsection{Meteor and PT Identification}
\label{pipeline}
An automated Python pipeline was developed to detect meteor-like objects. These objects were manually examined to determine if they were truly meteors; if so, they were further scrutinized to search for the presence of trains. Instead of a traditional two-frame image subtraction scheme, the pipeline uses a sequence of three consecutive images (referred to here as the `before', `current', and `after' frames) and looks for bright objects that only appear in the current frame, i.e. they flash for a single frame and then (mostly) disappear. This is implemented by creating a binary image which traces where all three of the following conditions are true, on a pixel-by-pixel basis: (1) the current frame is greater than the before frame, (2) the current frame is greater than the after frame, and (3) the current frame is greater than the background (average of before and after frames) by more than 20\%. The 20\% value was empirically chosen as it provides a good balance between meteor sensitivity and cloud rejection, the latter being a pervasive and difficult problem to solve. Objects with less than ten connected pixels are filtered out as these are likely due to star scintillation or noise spikes. Remaining objects are assigned a unique label; labeled objects are passed to scikit-image's \texttt{measure.regionprops\_table} which is configured to return their bounding box, centroid location, major axis length of a fitted ellipse, and orientation of the fitted ellipse. 

Flying objects such as airplanes and satellites abound in the widefield images, and also easily pass through the above pipeline. Because of this, it is necessary to implement a filtering process which makes use of the properties returned by \texttt{measure.regionprops\_table} to automatically connect objects sharing similar properties. In order to associate two or more flying objects together as the same source (e.g. airplane or satellite), they must satisfy four conditions. First, they have to be separated by no more than two frames---though one frame separation is generally sufficient, planes near the image's edge are not as reliably visible hence the extra buffer. Next, the objects' centroids need to be within three major axis lengths of the (temporally) first object; this imposes a travel distance limit on how far the objects can be separated. Third, the major axis lengths of both objects have to be within a factor of two which helps ensure that the speed of the two objects is comparable. Last, the orientations of both objects have to be within 45$^{\circ}$ of one another to match the travel direction. If all four requirements are met, the objects can reasonably be assumed to arise from the same flying source traveling across multiple frames, and are therefore excluded from further analysis.

This is the extent of the automatic processing; the remaining objects of interest are each made into a short (few second) video based on $\sim$3 minutes of recorded data. Identification of the object in the video is done by visual inspection, and is assigned one of three labels: meteor, meteor with train, or other. Here, `other' includes objects such as airplanes, satellites, clouds, lightning, and spiders/insects. As the image files are natively saved in hour-long increments, if a particular hour of observation looks predominantly cloudy, the entire hour file is skipped. Videos are preferred to still images for classification as quite a few of the trains are very faint and can only be detected via their motion. Typical conditions in a clear, dark patch of sky provide a limiting apparent stellar magnitude around +8 mag which we take to also represent the dimmest train visible under the best of conditions.

\subsection{Astrometric Calibration}
\label{astrometry}
Astrometric calibration enables any given pixel coordinate $(x,y)$ in an image to be converted into a celestial coordinate, such as altitude and azimuth (alt/az). This is a necessary step for determining the trains' beginning and ending heights (detailed in section \ref{alt}). A semi-automated Python script was written to perform the calibrations, though it still requires the user to manually click on prompted stars. One of the first considerations is how the sky is being mapped onto the sensor; the Canon fisheye lens in our setup is well modeled by an equisolid angle projection function. This means that the distance, $r$, along the sensor measured from the optical axis is related to the entry angle, $\theta$, of incoming light relative to the optical axis by
\begin{linenomath*}
\begin{equation}
    \label{eqn:distortion}
    r = k_1f\sin{\left(\theta/k_2\right)}
\end{equation}
\end{linenomath*}
where $f$ is the focal length and ideally $k_1=k_2=2$ \cite{Bettonvil2005}. In practice though, both $k$ values are found to be about 2.6 for our system. In Equation \ref{eqn:distortion}, $r$ is measured in physical distance along the sensor, with the same units as $f$, however dividing $r$ by the effective pixel size gives the distance in number of pixels instead---something much easier to calculate directly from the images.

\begin{figure}
    \centering\makebox[\textwidth]{\includegraphics[width=1.2\textwidth]
    {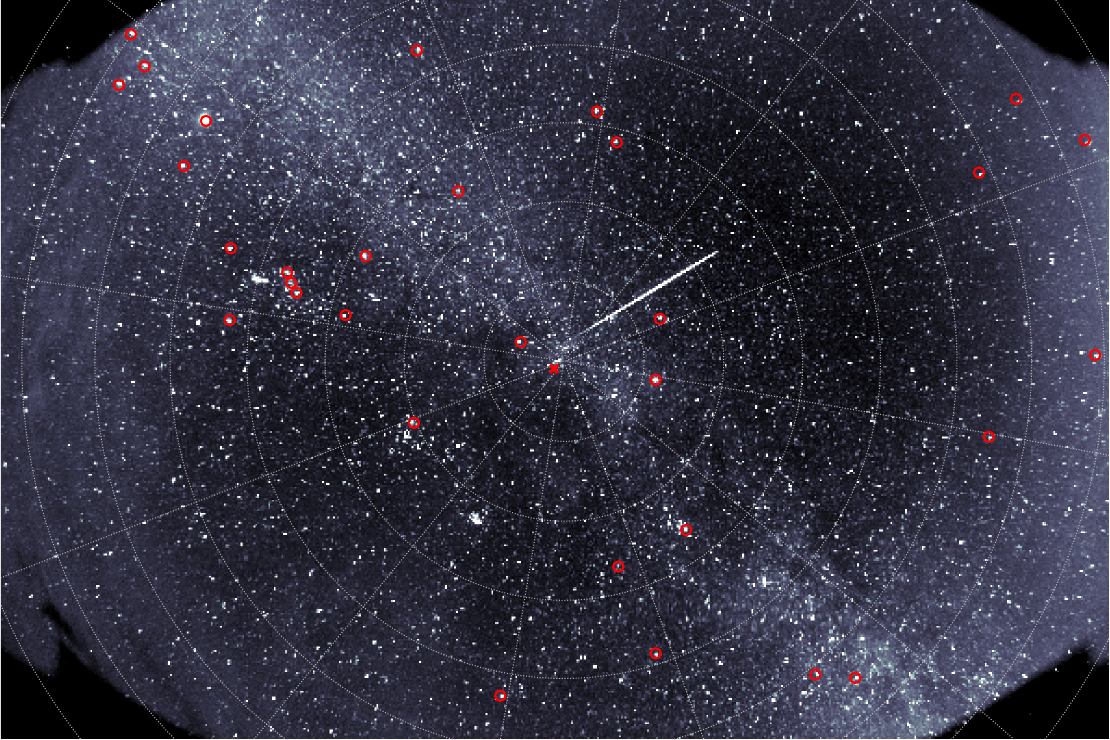}}
    \caption{An image taken by the WiPT2 camera with astrometric overlay. The image is in archival resolution (the same used by the meteor detection pipeline). The red circles show the calculated positions for the calibrator stars. The red ``x" indicates the center of the image, which is slightly offset from the true zenith (convergence of dotted lines). Dotted azimuth lines are separated by 30$^\circ$ and dotted altitude lines are separated by 10$^\circ$. The meteor near the center gives an idea of how the astrometric errors compare to a meteor's size.}
    \label{fig:astrometric_fit}
\end{figure}

Spherical trigonometry can be employed using five parameters to determine the alt/az of any $(x,y)$ point in the image: $k_1$, $k_2$, zenith $x$ and $y$ coordinates ($x_z$ and $y_z$), and azimuth of image center ($az_c$). The determination of these five parameters is detailed in \ref{app:Astrometry}. Ideally these parameters would remain nearly constant, such that only a few calibrations spread throughout the campaign would be necessary to verify their integrity. Unfortunately, the clamshell sun shield used in our setup forcefully snapped shut at the end of each night's observation, causing the camera to slightly unscrew from its mount each time. This led to a gradual change in $x_z$, $y_z$, and $az_c$ with time ($k_1$ and $k_2$ are inherent to the lens and were not affected). Interestingly, the change in these values could be modeled quite well by a quadratic function, allowing most of the affected days' parameters to be interpolated rather than manually recalculated. This issue was fixed with a set screw on MJD 59906 and has since shown good parametric stability. A typical calibration produces a maximum azimuth residual of about 2$^{\circ}$ with average azimuth residuals around 0.4$^{\circ}$, and a maximum altitude residual of about 0.4$^{\circ}$ with average altitude residuals around 0.2$^{\circ}$. Combined, the average separation between calculated and true positions across fifty different calibrations is found to be about 0.21$^{\circ}$ (12.8$'$); this is about the same size as a pixel. Granted, the maximum error of 2$^{\circ}$ would correspond to about ten pixels. We note that there are regions of sensor distortion present in our data that are ignored in the fitting process which could potentially account for some of the large errors. Figure \ref{fig:astrometric_fit} shows the results of a typical astrometric calibration. The red circles indicate the calculated position of a star; distortion from the wide-angle lens causes stars far from the image's center to have larger apparent errors, however the errors between the true and calculated positions remain fairly consistent across the image. A meteor is included near the center of the image in order to contextualize how the astrometric errors compare to the size of a meteor. 

\citeA{Bannister2013} were able to obtain comparable residual errors on images from New Mexico State University's SkySentinel video network using a similar approach, except they used eight different (yet closely related) parameters optimized with MATLAB's \texttt{fsolve} routine. Though additional work could be done to characterize and reduce the errors in our astrometric calibrations, these errors are still typically smaller than those due to other factors involved in finding the train heights. As this is currently the main use case for performing the calibrations in the first place, we find the calibrations to be satisfactory for our purposes. 

\subsection{Global Meteor Network}
An indispensable resource for this work is the Global Meteor Network \cite<GMN;>[]{Vida2019, Vida2021}. The GMN consists of low-cost, high sensitivity CMOS video cameras deployed around the world. As of mid-2023, there were at least 1000 cameras spread across 40 countries; most of these stations are funded, maintained, and deployed through the efforts of amateur astronomers, though some are also operated by professional institutions. Each camera runs open-source meteor detection software which automatically detects, calibrates both astrometry and photometry, and records meteor events. These data are sent to a central GMN server which correlates the multi-station observations of individual meteors and computes their corresponding trajectories. Fortuitously, the network has a high density of stations in New Mexico. There are 20 GMN stations located within a 100 km radius of our WiPT2 camera, with 51 unique stations contributing to the cross-correlations of our meteor observations. One meteor in our study was simultaneously observed by as many as 14 stations. The three most prolific stations during our campaign (US0005, US000H, US000K) were involved in 67\% of the meteor detections. The most applicable GMN parameters for the PT analyses discussed in section \ref{result} are: the meteors' geodetic coordinates, associated meteor shower, average velocity, photometric mass, peak absolute magnitude, height of peak absolute magnitude, and Tisserand's parameter with respect to Jupiter. Altitudinal data is measured relative to the WGS84 ellipsoid.

There are, however, caveats with a couple of these properties. Individual GMN cameras saturate around -1 mag on average \cite{Vida2021}; the actual saturation limit scales with the apparent angular velocity, i.e. faster meteors tend to have higher apparent angular velocities and therefore spend less time illuminating any given pixel in the camera's sensor. This pushes the saturation limit for faster meteors to brighter magnitudes. The GMN does not employ a saturation correction---this is partially mitigated if a more distant station records the same meteor and does not saturate, though at present the GMN does not report which events fall into this fortuitous category. For this reason, peak absolute magnitude values brighter than about -1 mag cannot be assumed entirely accurate. Absolute magnitude, hereafter shortened to ``magnitude", refers to the apparent magnitude of a meteor normalized to a distance of 100 km. The GMN determines the photometric mass via the discretized form of 
\begin{linenomath*}
\begin{equation}
    \label{mass}
    M_{ph}=\frac{2}{\tau}\int\frac{I}{v^2}dt
\end{equation}
\end{linenomath*}
where the integral is over the meteor's period of luminous flight, $\tau$ is the luminous efficiency, $v$ is the meteor's velocity, and $I$ is the radiated power \cite{Vida2020}. The radiated power is related to a meteor's magnitude. Regrettably, whenever the GMN cameras saturate, the photometric mass will consequently also be underestimated. The reported magnitude and mass values, though potentially saturated, still impose a hard lower limit on their true values. We primarily use magnitude and mass in this work as a tracer of qualitative trends rather than as a source of strong quantitative conclusions, partially mitigating saturation concerns. Other estimates of meteoroid mass exist, such as the dynamic mass \cite<e.g.>[]{Gritsevich2008}, though photometric mass still provides the best estimate short of running a full ablation and fragmentation model. It is worth noting that both the GMN and our WiPT2 system are brightness-limited which introduces inherent observational biases in the types of meteors that are detectable, namely that the lower mass threshold seen in the GMN is 2-3 orders of magnitude lower for the fastest meteors relative to the slowest \cite{Vida2021}. 

\section{Results}
\label{result}
The data presented here were collected between 5 October 2021 and 1 July 2023 and represent over 2810 hours of observations. Application of the pipeline described in section \ref{pipeline} resulted in the detection of 872,992 total objects. Of these, 476,863 objects were automatically filtered out by the airplane detection process and another 357,916 were discarded because they were associated with cloudy skies. This left 38,213 objects to be manually reviewed, which took nearly two months to completely process. Most of these were false positives---only 7465 were identified as meteors, of which 849 exhibited trains. It is almost certain that some airplanes have been erroneously included among the 7465 meteors, though using the GMN as a cross-check effectively weeds these out. The eventual inclusion of additional WiPT cameras will allow for internal validation of meteors, as it enables the same sort of triangulation and anti-coincidence techniques employed by the GMN. Regardless, we are quite confident that each of the 849 train-producing meteors has been classified correctly. Of the meteors we observed, 4726 of them were co-detected by the GMN, with 636 of these producing detectable trains. It is important to clarify that the ensuing results and conclusions are based on PTs which were observable by the WiPT2 camera. Since PTs essentially trace the number and strength of exothermic chemical reactions, a camera with greater sensitivity than the WiPT2 would be able to detect fainter emissions from these reactions. This increased sensitivity would enable it to detect PTs over a wider range of altitudes and for longer durations relative to the WiPT2; it would also be able to observe PTs that are too dim for the WiPT2 to register. Because of this, the rigid dichotomy between meteors which do or do not leave PTs is not universal, but is instead based on whether a PT is bright enough to be distinguished by the particular camera setup (or a naked eye observer, etc.). Other factors besides the observation equipment can also affect the relative detectability of meteors and their trains, including the presence of thin, high-altitude clouds and the meteor's location on the sky (e.g. whether it is on/off the galactic plane, its nearness to the horizon and light pollution sources, etc.). We did not explicitly account for these various factors in the analyses that follow. The remainder of this section is devoted to analyzing the detected meteors to determine what conditions are most favorable for the production of an observable train.

\subsection{Duration}
\label{duration}
Exact duration values were difficult to establish as many of the trains became nearly imperceptible when viewed as still frames or low frame rate movies, so we instead opted to sort the PTs into custom-width duration bins. The bin cutoff values were chosen to keep the number of trains in each bin relatively uniform. These categorical duration values were obtained by inserting a flash into the PT videos at the appropriate frame and determining whether the PT disappeared before or after the flash, with borderline cases being promoted to the next duration bin. There is no single duration value that can unambiguously distinguish whether a train is in the continuum phase, though based on prior observations this phase usually begins 30--40 seconds after meteor passage for Leonids \cite{Borovicka2003, Abe2004}. As we do not have spectral information nor the fine temporal resolution needed to see the continuum's characteristic brightening, we have to assume that this timeline extrapolates to all varieties of meteors and therefore chose 1 minute after meteor passage as the minimum threshold for continuum emission (and by extension, for qualification as a PT). A more conservative estimate might put this threshold at 2 minutes, though the resulting conclusions remain essentially the same. Trains lasting less than the 1 minute cutoff will be referred to nominally as short duration trains (SDTs), solely to differentiate them from PTs and the generic term `train'. 

There are many extrinsic factors that can bias the observed duration of meteor trains, including background brightness (e.g. whether the meteor is on or off the galactic plane), disappearance behind a cloud, or moving out of the camera's field of view. Additionally, our camera is mildly sensitive to NIR emission. As PTs are known to emit in NIR \cite<e.g.>[]{Hapgood1980}, this could enhance PT detectability and duration relative to naked eye observations. With these considerations in mind, the train count per duration bin is provided in Table \ref{tab:duration}.
\begin{table}[htbp]
\centering
\caption{\textit{PT Breakdown by Duration Bin$^a$}}
\begin{tabular}{c c c c c c}
\hline
Duration bin & Train & Obscured train & M$_\textnormal{peak}$ & Mass & $v_\textnormal{avg}$ \\
{[}minutes] & count & count & & [grams] & [km/s] \\
\hline
$<$1 & 58 & 0 & -0.32 & 0.30 & 27.72 \\
1-2 & 119 & 11 & -0.35 & 0.37 & 26.69 \\
2-5 & 185 & 12 & -0.62 & 0.40 & 29.54 \\
5-10 & 139 & 17 & -0.97 & 0.54 & 28.65 \\
10-15 & 54 & 6 & -1.13 & 0.72 & 28.72 \\
$>$15 & 35 & 0 & -1.07 & 1.53 & 26.45 \\
\hline
\multicolumn{6}{p{0.85\textwidth}}{$^a$The M$_\textnormal{peak}$ (peak absolute magnitude of parent meteor), Mass (meteor mass), and $v_\textnormal{avg}$ (average meteor velocity) columns provide the median value within in each respective duration bin.}
\end{tabular}
\label{tab:duration}
\end{table}
The obscured train column indicates the number of trains which were obscured in one way or another before they naturally faded out; they are included with the last bin in which they could clearly be seen, though their actual durations are likely longer. These obscured durations are therefore excluded from ensuing statistics when comparing different bins. Considering only the 578 trains lasting longer than a minute, we find that at least 12.2\% of meteors produced observable trains---nearly two orders of magnitude larger than \citeA{Olivier1957}'s estimate. This substantial increase reflects the improvement in sensitivity afforded by an automated camera setup versus naked eye observations. It should be noted before continuing that the chosen duration bins are merely for convenience (excepting the one minute demarcation) and are not meant to reflect different stages in the continuum emission responsible for driving PTs.

\begin{figure}
    \centering\makebox[\textwidth]{\includegraphics[width=1.2\textwidth]
    {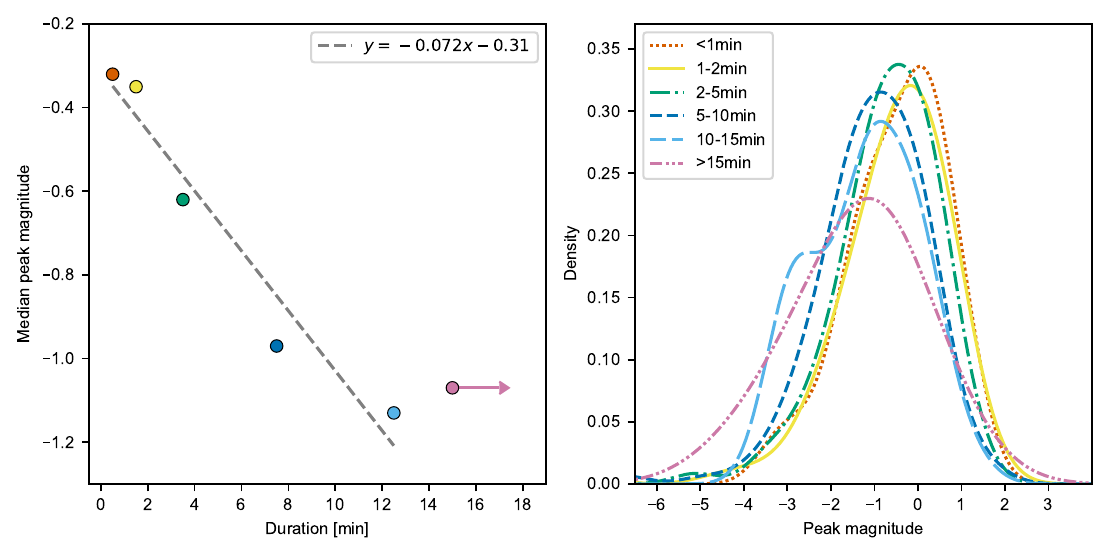}}
    \caption{Relationship between PT duration and meteor peak magnitude. (Left) The median peak magnitude value of each duration bin plotted against the midpoint of the bin, the `longer than 15 minutes' point is placed at its lower bound and is excluded from the linear regression fit. (Right) Gaussian KDE distributions for each duration bin; longer lasting PTs are shifted towards brighter peak magnitudes.}
    \label{fig:trend_mag}
\end{figure}

\begin{figure}
    \centering\makebox[\textwidth]{\includegraphics[width=1.2\textwidth]
    {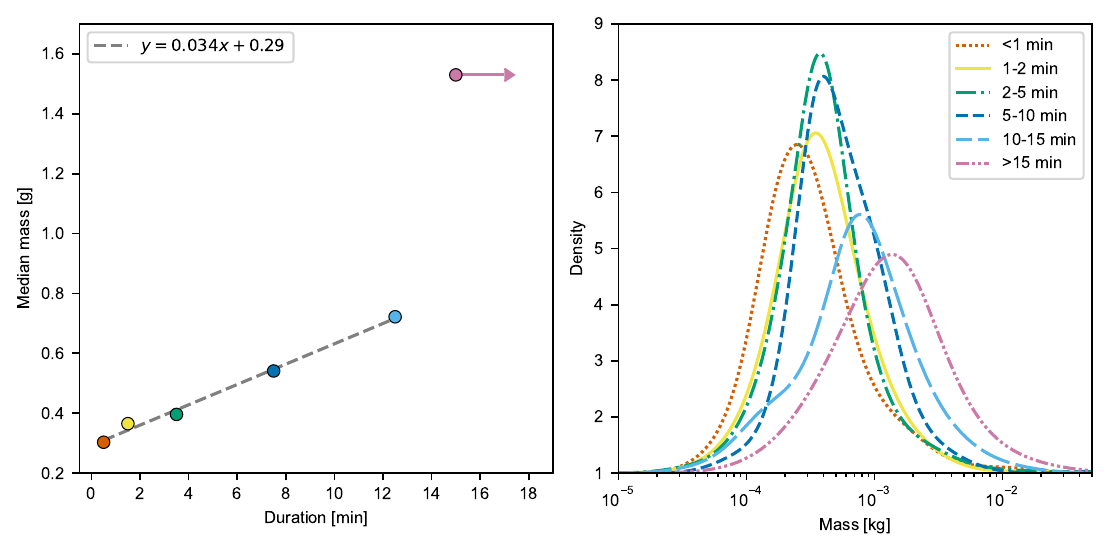}}
    \caption{Relationship between PT duration and meteor mass. (Left) The median meteor mass value of each duration bin plotted against the midpoint of the bin, the `longer than 15 minutes' point is placed at its lower bound and is excluded from the linear regression fit. (Right) Gaussian KDE distributions for each duration bin; longer lasting PTs are shifted towards larger meteor masses.}
    \label{fig:trend_mass}
\end{figure}

\begin{figure}
    \centering\makebox[\textwidth]{\includegraphics[width=1.2\textwidth]
    {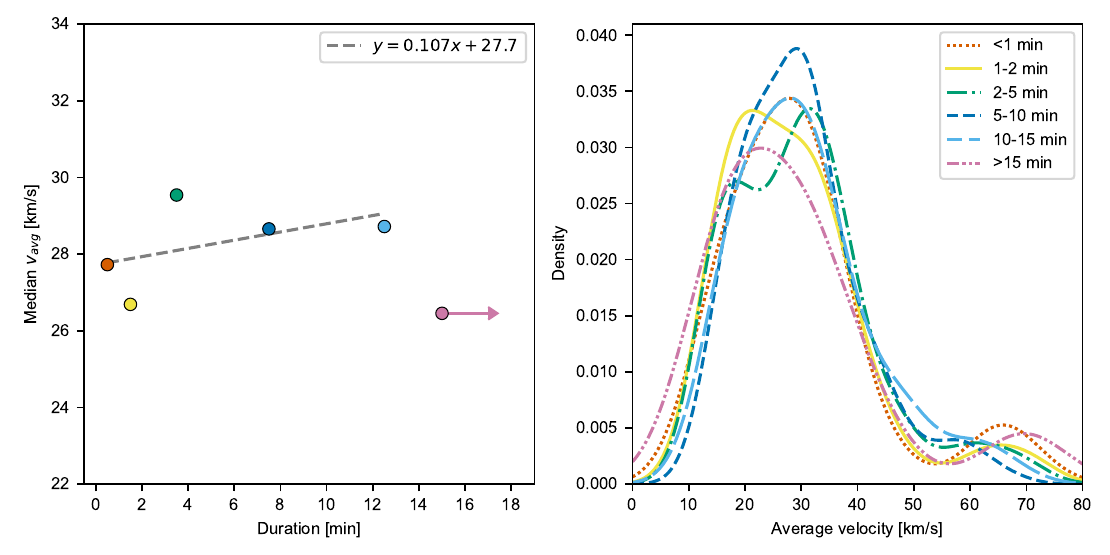}}
    \caption{Relationship between PT duration and meteor average velocity. (Left) The median meteor average velocity for each duration bin is plotted against the midpoint of the bin, the `longer than 15 minutes' point is placed at its lower bound and is excluded from the linear regression fit. The slight positive slope is tenuous at best; inclusion of the standard error and/or the 15 minute point could invert the slope. (Right) Gaussian KDE distributions for each duration bin, though a clear trend is not evident.}
    \label{fig:trend_velocity}
\end{figure}

The dimmest meteor in our sample had a peak absolute magnitude of 2.6 mag as determined by the GMN, meaning that judiciously placed arbitrary observers could have visually detected all the meteors. Accordingly, we find that about 5.3\% of all our visual meteors produced PTs longer than 5 minutes; this value is significantly higher than reported by \citeA{Baggaley1978} where only 1 visual meteor in 5000 (0.02\%) produced the same. In partial support of the claims made by \citeA{Borovicka2006}, a meteor's peak magnitude does appear to correlate with PT duration. This relationship is shown in Figure \ref{fig:trend_mag}. The left panel displays the median peak magnitude value for each duration bin plotted against the bins' midpoints; the right panel shows the kernel density estimate (KDE) using Gaussian kernels with bandwidths determined by Scott's Rule \cite{Scott1992} for each duration subset. The regression line in the left panel uses the method of least squares and does not include the `longer than 15 minutes' bin as it does not have a well-defined midpoint.
The median meteor peak absolute magnitude (M$_\textnormal{peak}$) for each of the duration bins is also listed in Table \ref{tab:duration}. Though the actual values are likely biased by camera saturation, they nonetheless indicate that longer lasting trains tend to be produced by brighter meteors. As perhaps expected, the derived photometric mass values exhibit a similar trend, with longer duration PTs associated with more massive meteoroids. Figure \ref{fig:trend_mass} shows this trend; it was prepared in the same manner as Figure \ref{fig:trend_mag}. The regression lines in Figures \ref{fig:trend_mag} and \ref{fig:trend_mass} should be treated cautiously as both are based on potentially saturated values, though they do at least impose lower limits. The KDE plots in these figures do exhibit a progressive shift as the duration increases, corroborating the trend.
Borovi{\v{c}}ka's claim regarding a velocity dependence cannot be substantiated however, as we do not see a correlation between the average velocity of a meteor and the duration of its PT. This is exemplified in Figure \ref{fig:trend_velocity}---the slope of the regression line is 0.107 but its standard error is 0.112, suggesting that a slope of zero is entirely possible. Additionally, the inclusion of the `more than 15 minutes' data point would easily shift the regression to have a slightly negative slope instead.

Amongst the 35 meteors in the `longer than 15 minutes' category, almost half (17) of them are classified as sporadic, with the remainder being associated with a variety of meteor showers. One shower in particular, the $\tau$-Herculids, produced 5 PTs with this duration; it will be discussed separately in section \ref{sec:tau}. All but 3 of these 35 meteors had average velocities below 45 km/s, and 14 had peak magnitudes dimmer than -1 mag. The potential for long-lived PTs to form under these hitherto implausible circumstances, and with such relative frequency, reveals a deficiency in the current understanding of PTs. One such PT-producing meteor (recorded on 2022-04-23 04:09:58 UTC), highlighted here due to its unconventionality, was observed traveling with an average velocity of only 15.0$\pm$0.2 km/s at a grazing entry angle of 8.6$\pm$1.1$^{\circ}$. It had a peak magnitude of 1.05 mag which occurred at an altitude of 85.7 km; the estimated photometric mass for this sporadic meteor was about 0.3 g.

\subsection{Altitude}
\label{alt}
As we have only a single station, it is impossible to determine train altitudes using triangulation. Instead, we rely on the GMN which reports a meteor's beginning and ending locations in geodetic coordinates (latitude, longitude, and height). These are converted to Earth-centered, Earth-fixed (ECEF) coordinates and are connected by a straight line which is discretized into 5000 points. Each ECEF point along the track is converted to an alt/az value as seen by WiPT2. Separately, the train's initial and final pixels are manually identified in the image and converted to an alt/az value per the astrometric conversion described in section \ref{astrometry}. From here, the minimum separations between the alt/az values of the train's endpoints and the alt/az values of the meteor's linear track are determined. The best-matched point along the linear track is then converted from ECEF back to geodetic coordinates, providing the altitude of that point and consequently of the beginning/end of the train. The error in determining this altitude, strictly from an astrometric perspective, is found similarly. A conservative astrometric error of 0.5$^{\circ}$ was used. The points along the meteor track which were separated by 0.5$^{\circ}$ from the best-matched point were used to determine the possible upper and lower bounds in altitude. The larger of the two altitudinal differences from the best-matched value was taken as the astrometry error. This method of determining astrometry errors inherently causes forshortened meteors to have greater errors relative to those that graze by; these forshortened meteors have small apparent angular extents due to the camera's perspective. The choice of a 0.5$^{\circ}$ astrometric error typically produced altitudinal errors of around 1 km, though for highly forshortened meteors this rose as high as 7 km. The astrometry error was combined in quadrature with the GMN's reported 3$\upsigma$ error in meteor height to give an estimated total altitudinal error.

There are a few caveats in determining the altitude values. The technique of using our single station with the GMN means that we can only accurately match points when the train is relatively close to the original meteor. As a result of this, we are tracing the location of the initial train which may not necessarily correspond to the region that persists the longest. Distortion by winds in the upper atmosphere also makes it difficult to ensure that train features remain near their initial altitudes. \citeA{Hawkins1959} empirically found that an SDT (lasting 18 seconds) exhibited a rate of decay that followed a well-defined function of height, with rapid decay outside the 85 to 100 km range and greater persistence within 88-98 km. Since we are able to determine only the upper and lower altitude boundaries of the trains, we do not have information on how this range evolves with time nor can we pinpoint the altitude region associated with the greatest persistence as they were able to do. Additionally, determination of the beginning and end of a train can be quite subjective, with subsequent reclassification of the same train resulting in differences up to a kilometer or two. With these considerations in mind, the true error of these measurements is hard to quantify though it is likely to be up to a couple kilometers.  

Several criteria are used to determine which trains are best suited for characterization: the train's beginning and end need to be unobstructed by clouds or bright background regions, the GMN reported meteor location has to coincide reasonably well with our image, and the meteor cannot be too forshortened (to avoid causing sizable altitudinal errors). A total of 160 PTs were deemed suitable and were selected for analysis. Table \ref{tab:altitude} provides the weighted average of PTs' beginning, center, and end altitudes along with the associated weighted standard deviations ($\sigma$). Also calculated is the uncertainty of the weighted average (following the $\pm$ sign), as each estimated PT altitude has some inherent level of uncertainty.
\begin{table}[htbp]
\centering
\caption{\textit{Average Altitudes for Meteors and PTs$^a$}}
\begin{tabular}{c c c c c c c}
\hline
  & Begin ht. & $\sigma$ & Center ht. & $\sigma$ & End ht. & $\sigma$ \\
\hline
PTs & 93.2$\pm$0.2 & 3.8 & 89.4$\pm$0.2 & 2.8 & 85.5$\pm$0.2 & 2.8 \\
PT meteors & 99.6 & 7.7 & 89.3 & 5.2 & 79.1 & 6.2 \\
All meteors & 101.8 & 10.1 & 93.7 & 9.5 & 85.5 & 9.9 \\
\hline
\multicolumn{7}{p{0.75\textwidth}}{$^a$All values are in kilometers. The beginning, center, and end height columns are averages; $\sigma$ (standard deviation) columns are associated with the preceding value. The $\pm$ values in the PT row are the uncertainties associated with the weighted average.}
\end{tabular}
\label{tab:altitude}
\end{table}
The weights represent the total altitudinal error arising from the fitting process plus a constant one kilometer error accounting for the subjectivity in choosing the start and end points. The larger standard deviation associated with the PT beginning height indicates slightly more variability compared to their ending altitudes, which may reflect differences in meteoroid strength. Table \ref{tab:altitude} also provides the average beginning, center, and ending altitudes for the meteors that produced these 160 PTs (``PT meteors") and for the total population of meteors observed (``All meteors"). PT meteors having a much lower average ending height is significant and will be treated more fully in the following section. 

\begin{figure}
    \centering
    \includegraphics{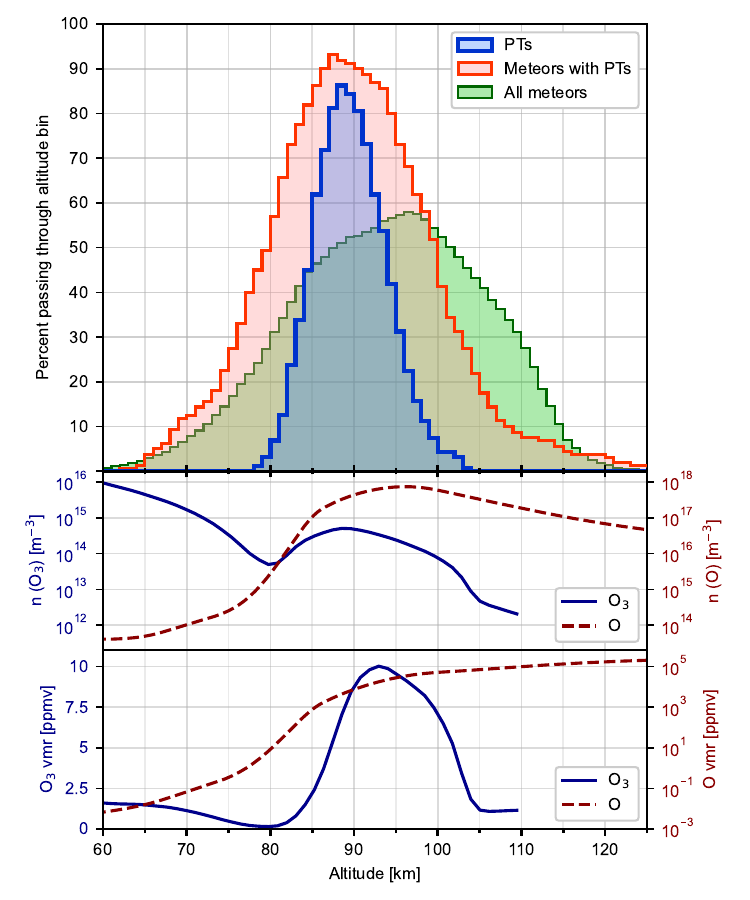}
    \caption{(Top) The percentage of meteors/PTs that pass through a given one kilometer wide altitude bin. Three populations are shown: the 160 PTs which had their altitudes determined, the parent meteors that produced these 160 trains, and the total population of 4726 meteors observed during the campaign. Each bin is incremented when a meteor/PT starts in, stops in, or passes through it. The relative narrowness of the PT zone can be clearly seen against the total meteor population. (Middle) The number densities of O and O$_3$ over the same altitude range. (Bottom) The volume mixing ratio (vmr) of O and O$_3$; the peak in O$_3$ aligns quite well with the narrow altitude range of PTs.}
    \label{fig:altitude_bins}
\end{figure}

The top panel of Figure \ref{fig:altitude_bins} shows the percentage of trains and meteors that pass through a given altitude bin for three different populations: the PTs themselves, the parent meteors of these PTs, and the entirety of meteors observed in this campaign. 
It is not hard to see that the region where PTs are found is noticeably narrower compared to the range of their parent meteors, which in turn is more concentrated relative to the total population of meteors we observed. The fact that PTs occur within a small range of altitudes has been thoroughly documented \cite<e.g.>[]{Trowbridge1907, Olivier1957, Yamamoto2004}; our PT altitude range (or ``PT zone") is quite consistent with their values, though Olivier and Yamamoto reported slightly higher beginning heights of around 100 km. Atmospheric influences are likely responsible for this restricted range. Since a steady supply of O and O$_3$ is needed to drive the proposed chemistry (viz. Reactions 1 and 2), their density profiles will dictate where PTs can form \cite{Yamamoto2004}. The nighttime ozone profile in the upper mesosphere (60-110 km) has a local maximum in ozone density at an altitude of about 90 km, which decreases to a local minimum around 80 km \cite{Smith2013}. The local extrema in the profiles of both ozone density and volume mixing ratio (vmr) align quite well with the PT distribution. These profiles can be seen in the middle and bottom panels of Figure \ref{fig:altitude_bins}, respectively. 

The atomic O number density and derived vmr were obtained from the MSIS 2.1 model \cite{Emmert2022}, which was used to generate atmospheric profiles directly above the WiPT2 camera. Ten such profiles, evenly spaced in time, were generated for each night of the observing campaign. All profiles from all relevant nights were subsequently averaged together, the result of which is displayed in Figure \ref{fig:altitude_bins}. Ozone data was retrieved from the Sounding of the Atmosphere using Broadband Emission Radiometry (SABER) instrument aboard the Thermosphere, Ionosphere, Mesosphere Energetics, and Dynamics (TIMED) satellite. The custom data tool provided by the SABER team was used to retrieve the 9.6 $\upmu$m O$_3$ data (version 2.0) up to the instrument's upper limit of 110 km; all nighttime events with tangent points occurring within a $\pm$5$^{\circ}$ latitude and $\pm$5$^{\circ}$ longitude region about the WiPT2 camera between October 2021 and June 2023 were gathered. All retrieved O$_3$ vmr profiles were aggregated and the altitude range was divided into 50 bins---the median O$_3$ value and altitude midpoint for each bin were determined and have been plotted in Figure \ref{fig:altitude_bins}. The O$_3$ number density graph was obtained similarly, however each vmr profile was first multiplied by its corresponding total density profile prior to the binning process. The secondary O$_3$ maximum at $\sim$90 km is readily apparent in both of the bottom panels of Figure \ref{fig:altitude_bins}.  

\begin{sidewaysfigure}
    \centering
    \noindent\includegraphics{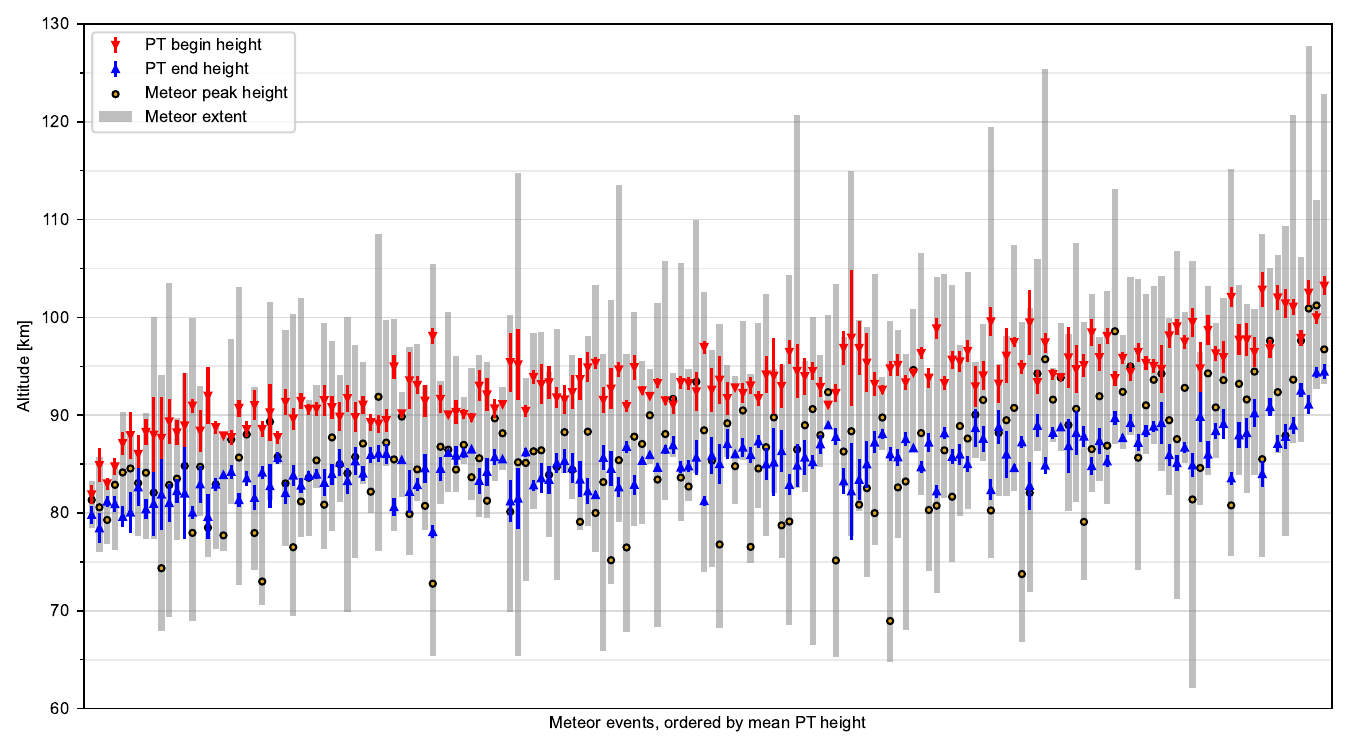}
    \caption{Altitude breakdown for all 160 individual PT events. For each, the meteor's altitude range is shown (gray bar) along with the height where the meteor's maximum brightness occurred (yellow dot). The PTs beginning/end heights show error bars that account for the altitudinal error due to astrometric uncertainty. The x-axis is sorted by increasing mean PT height in order to enhance readability, as no trend is evident when sorted temporally. Note the large number of events for which the PT range does not overlap with its parent meteor's location of maximum brightness.}
    \label{fig:altitude_individual}
\end{sidewaysfigure}

Looking at the train locations within individual meteors also provides additional insight into the altitude dependence. Figure \ref{fig:altitude_individual} shows the altitude range for both the parent meteor and its corresponding PT, with error bars indicating the associated altitudinal fitting error.
The meteors have been ordered along the x-axis by PT center height to increase clarity, as there does not appear to be any clear seasonal trend in their beginning or end heights. Perhaps most interesting is how the point of meteor maximum brightness aligns with the PT region: in 49 cases the location of the meteor's peak magnitude occurred below the PT, another 5 meteors peaked above their trains. In extremis, the greatest discrepancy between a meteor's brightest point and the bottom of its PT boundary was over 16 km. These instances contradict the assertion made in \citeA{Borovicka2006} that PTs are always produced cospatially with the region of meteor maximum brightness. Instead, it is evident that the train can be decoupled from the meteor's brightest region---provided that the meteor is still able to deposit sufficient reactant material at the correct altitude range.

\subsection{Magnitude and Mass}
\label{sec:magmass}
\begin{figure}
    \centering
    \includegraphics{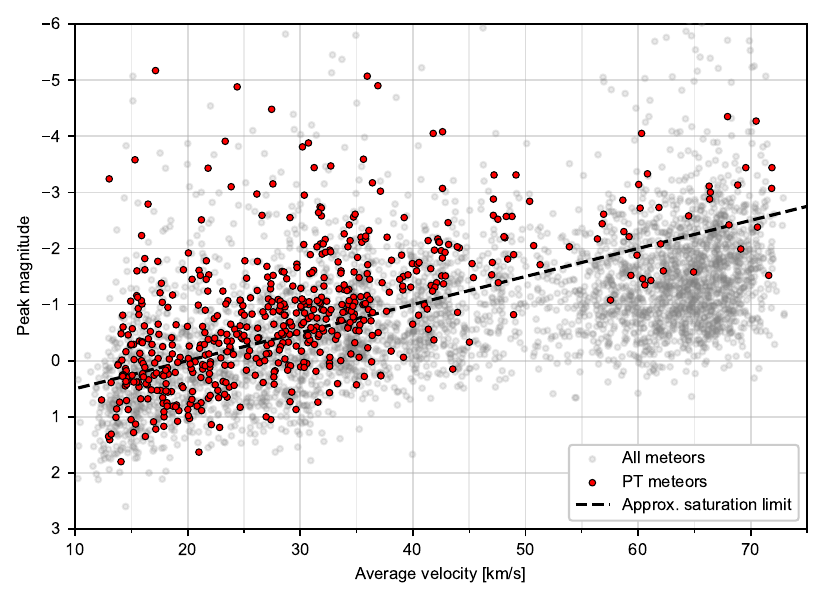}
    \caption{The effect of peak meteor magnitude and meteor average velocity on PT formation. Gray dots show the total population of meteors observed during the campaign, the red dots are those meteors that left a PT. The dashed line shows the approximate magnitude where GMN cameras saturate and takes into account the dependence of saturation on meteor velocity. Therefore, markers above that line are likely to have underestimated peak magnitudes. Notably, there is no clear magnitude cutoff that is applicable to all velocities of meteors.}
    \label{fig:mag_velocity}
\end{figure}

The peak magnitude and photometric mass will be discussed together, in part due to their implicit relationship, as well as the aforementioned problems regarding camera saturation. For comparative purposes, 475 of the 578 meteors which left observable PTs (82\%) were determined to be dimmer than -2 mag, whereas per Olivier's catalog only 4\% of PT meteors met this condition \cite{Baggaley1978}. Evidently the bright meteors do not have as strong of a monopoly on PTs as was thought previously. There is, however, still some truth in the claim that fast-moving meteors in particular need to be relatively bright, as is apparent in Figure \ref{fig:mag_velocity}.
The fast meteors (`fast' implying average velocities higher than 50 km/s) have a dearth of PTs, with only about 2\% of these meteors producing observable PTs, compared to the slow-moving ($\leq$ 50 km/s) meteors which have a PT occurrence frequency of 19\%. All the PT-producing meteors in the fast category were clustered among the brightest half of the fast meteors, leaving the dimmer group without any PTs. Because the slow meteors can be relatively dim and still produce PTs, the deficiency seen in fast meteors cannot simply be attributed to a magnitude cutoff. A potential explanation for this shortage is more easily understood through the lens of meteor mass.

\begin{figure}
    \centering
    \includegraphics{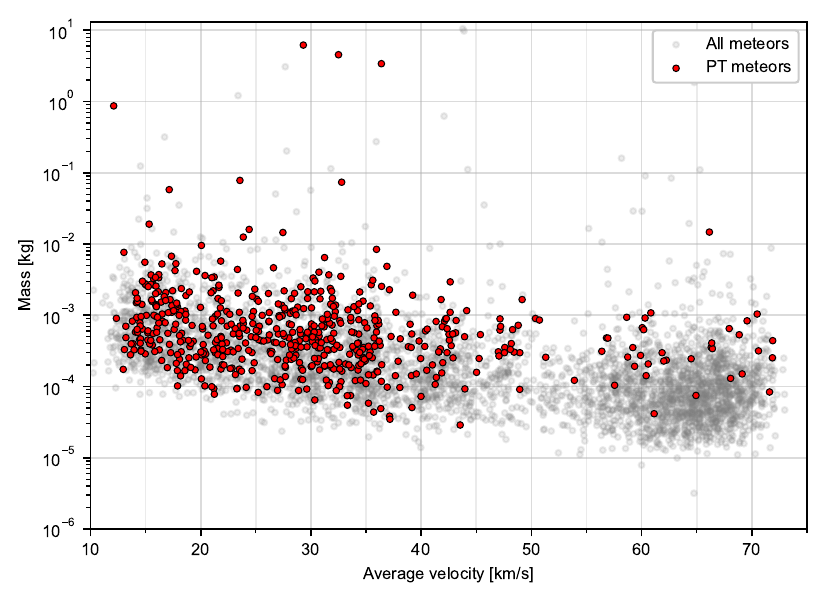}
    \caption{The significance of meteor photometric mass for the formation of PTs. Gray dots show the total population of meteors observed during the campaign, the red dots are those meteors that left a PT. As these mass values depend on the potentially saturated brightness values, only the lowest mass meteors can be assumed entirely accurate. Indications of a mass cutoff can be seen around $3\times10^{-5}$ kg.}
    \label{fig:mass_velocity}
\end{figure}

\begin{figure}
    \centering
    \noindent\makebox[\textwidth]{\includegraphics[width=1.2\textwidth]{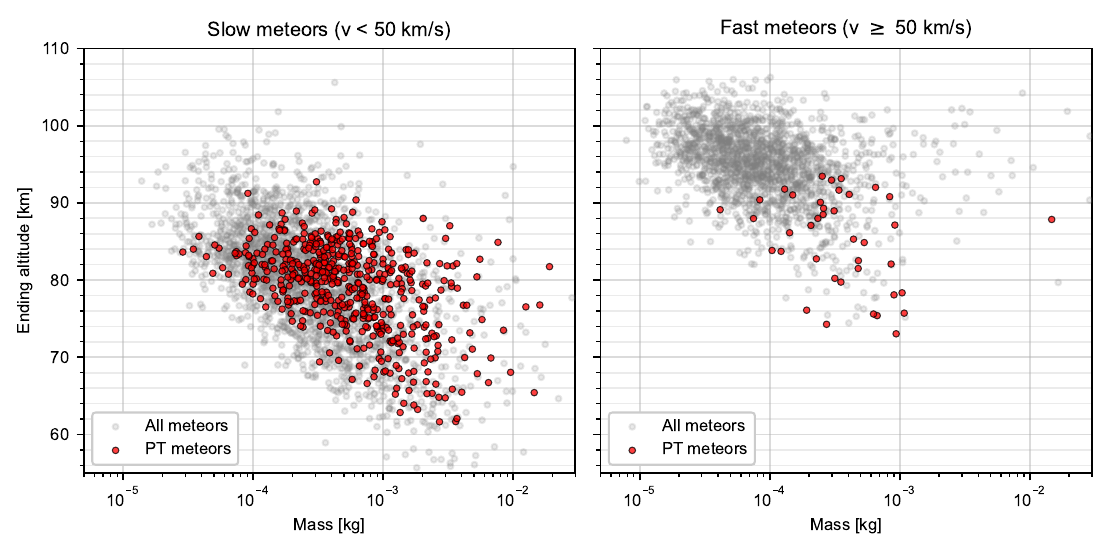}}
    \caption{Comparison between the terminal height of slow ($<$ 50 km/s) and fast ($\geq$ 50 km/s) meteors as a function of mass. Gray dots show the total population of meteors observed during the campaign, the red dots are those meteors that left a PT. Meteors that end above $\sim$93.5 km appear unlikely to leave PTs as this is outside the typical PT zone. Many of the fast, early-ending meteors have relatively low masses, contributing to their apparent mass and magnitude cutoff.}
    \label{fig:depth}
\end{figure}

The distribution of meteor photometric masses shown in Figure \ref{fig:mass_velocity} is again certainly biased by camera saturation, though we assume that the general trends seen within result from physical mechanisms.
The apparent PT mass cutoff primarily seen for the fast meteors is associated with the same dim, non-PT population mentioned above. While this could reflect an actual minimum mass necessary to drive observable PT reactions, which would by no means be an unreasonable assertion, it could also be caused by the dynamics of a meteor's atmospheric entry. This is illustrated in Figure \ref{fig:depth}, which shows how a meteor's mass relates to the depth it is able to penetrate into the atmosphere for both the fast and slow meteors.
The fast, lowest mass meteors which left observable PTs still had decent train durations of 2-5 minutes, slightly disfavoring the minimum material notion. As has been already been well established here and in numerous other studies, the region where trains form is relatively narrow compared to the full range of their parent meteors; this is nicely reflected in Figure \ref{fig:depth}, which shows an altitude cutoff for PT-producing meteors. Meteors which do not penetrate low enough (i.e. below $\sim$93.5 km) therefore might not be depositing sufficient material into the PT zone to fuel the chemiluminescent reactions. This $\sim$93.5 km cutoff altitude happens to align quite well with the average height where we saw PTs begin forming (93.2 km), thus supporting this idea. Other meteor properties, such as the rate of mass loss (as traced by the shape of the meteor's light curve), specific ablation characteristics (if and how the meteor fragments), and meteoroid composition, are expected to work in tandem with the meteor's terminal height to determine whether a PT can develop.

Bearing all this in mind, examination of the fast meteors' penetration profile reveals that as mass decreases, meteors have higher terminal heights; at sufficiently low masses, few meteors punch down into the PT zone which gives rise to an apparent mass cutoff for PT production. This is also concomitantly manifested in meteor luminosity as a magnitude limit, which explains the population of dim, fast meteors without PTs seen in Figure \ref{fig:mag_velocity}. The terminal height of a meteor is governed in part by meteoroid properties such as mass (magnitude), velocity, and entry angle---any combination of these properties culminating in a meteor which makes it past the critical altitude threshold seems to have the potential to leave a PT, regardless of its magnitude or velocity values. It is other innate factors (e.g. meteoroid strength and/or composition) which likely affect PT viability once the meteor has entered into the PT zone. 

\begin{figure}
    \centering
    \includegraphics{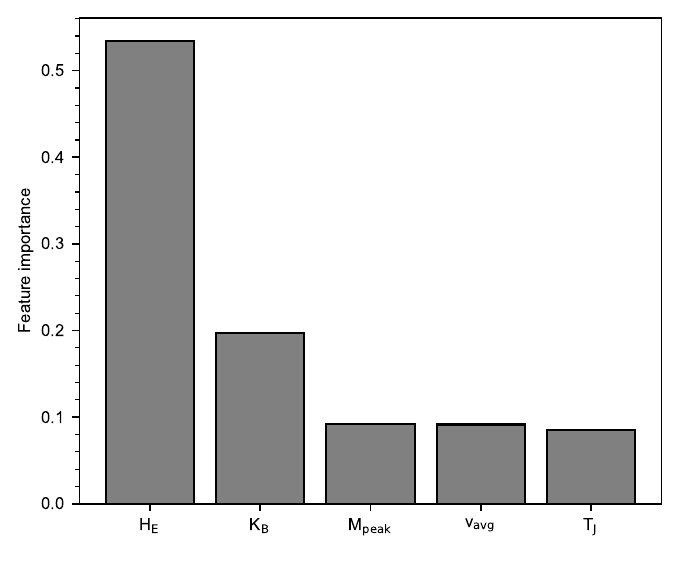}
    \caption{Relative feature importance from the gradient boosting classification algorithm. The terminal height ($H_E$) accounts for more than half of the total importance, with the $K_B$ parameter being the next most significant.}
    \label{fig:features}
\end{figure}

To quantify the importance of individual meteor parameters in producing PTs, we apply a gradient boosting classification algorithm \cite{Friedman2001} to the data set. The features included the meteor terminal height ($H_E$), the meteor's peak magnitude, the average velocity, as well as the Tisserand parameter with respect to Jupiter ($T_J$) and $K_B$ parameter, both of which will be defined and discussed further in the next section. We used a classic 80-20 train/test split to evaluate the quality of the classification and the Synthetic Minority Oversampling TEchnique (SMOTE) algorithm to balance the data set, producing a total of 4090 meteors each with and without a PT. We achieved an 81.7\% accuracy in predicting whether a meteor might produce a PT based only on these five features. Figure \ref{fig:features} shows a plot of relative feature importance in the classification---the meteor terminal height dominates the decision tree with 53\% of total importance, followed by the $K_B$ parameter at 19\%. The rest of the parameters had an importance of less than 10\%. 
This quantitative analysis strongly suggests that PT production predominantly relies on meteors reaching a certain threshold height and having a certain meteoroid strength (as traced by the $K_B$ parameter). Conversely, meteor peak magnitude and average meteor velocity appear to be relatively insignificant in determining whether a PT is created. 

\subsection{Dynamical Origin and Composition}
The Tisserand parameter with respect to Jupiter ($T_J$) is commonly used to roughly classify meteoroids according to the dynamical origin of their parent bodies \cite<see>[]{Kikwaya2011, Subasinghe2016}. This value is derived from a restricted three-body problem and therefore relies on assumptions that are only approximately true, such as Jupiter being the sole perturber of meteoroids and the absence of non-gravitational forces (e.g. Poynting-Robertson drag). Despite this, the Tisserand parameter remains relatively invariant following interactions with Jupiter and serves as a useful tool to broadly distinguish between asteroidal and cometary origins. It is calculated via
\begin{linenomath*}
\begin{equation}
    \label{Tisserand}
    T_J=\left(\frac{a_J}{a}\right)+2\sqrt{\left(\frac{a}{a_J}\right)\left(1-e^2\right)}\cos{i}
\end{equation}
\end{linenomath*}
where $a_J$ is Jupiter's semimajor axis and $a$, $e$, and $i$ are the meteoroid's semimajor axis, eccentricity, and inclination, respectively. The boundary limits are: $T_J\leq2$ for nearly isotropic comets (NICs) including Halley type comets, both of which originate in the Oort cloud, $2<T_J\leq3$ for Jupiter family comets (JFCs) coming from the Kuiper belt, and $T_J>3$ for asteroids within the main belt. There are known exceptions to this schema (Damocloids and comet 2P/Encke, for two) and sometimes the categorical boundaries are slightly adjusted in order to better reflect observed phenomena \cite<as in>[]{Jewitt2015}. Regardless, this Tisserand scheme allows for broad classification based on shared characteristics. 

\begin{figure}
    \centering
    \includegraphics{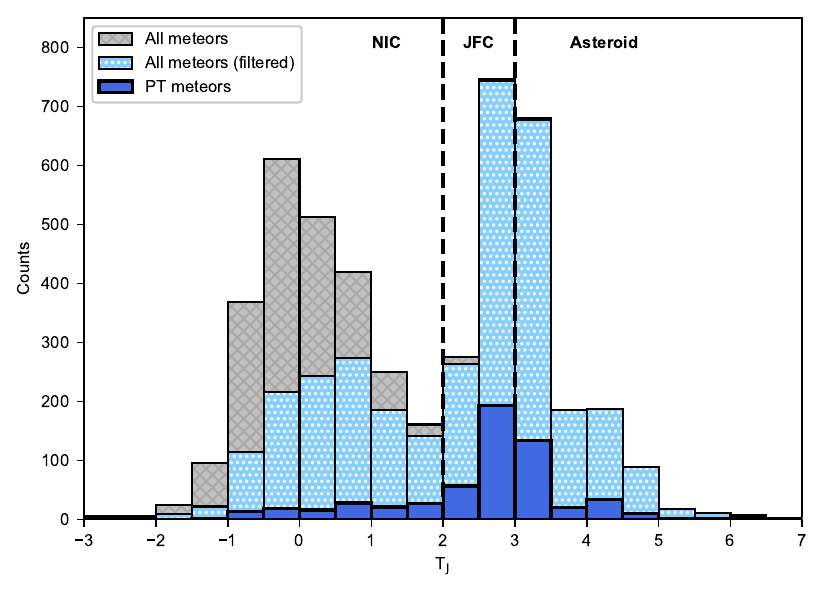}
    \caption{Histogram of Tisserand parameters. The gray hatched bars represent all meteors observed during this campaign, the light blue dotted bars are the subset of meteors that had terminal heights below 93.5 km, and the solid blue bars are the subset that left PTs. The dashed lines separate the different dynamical classes.}
    \label{fig:TJ_histogram}
\end{figure}

Using the Tisserand parameters reported by the GMN, we found the majority of meteoroids on orbits associated with NICs (52.2\%), with the remainder being fairly evenly split between the other two classes (22.1\% for JFC orbits and 25.7\% for asteroidal orbits). These proportions are inherently biased by the brightness-limited nature of the GMN; since NIC meteoroids have high velocities, they are detectable even when their masses are much lower than JFC or asteroidal meteoroids. As proposed in section \ref{sec:magmass}, we can filter out meteors unlikely to produce observable PTs and also mitigate the observational bias by excluding meteors which end above 93.5 km. Of the meteors that remain, 1222 came from NIC meteoroids, 1028 from JFC meteoroids, and 1209 from asteroidal meteoroids. The PT-producing meteors accounted for 127, 250, and 201 of these, respectively, which means that 10.4\% of candidate NIC meteoroids resulted in a visible PT, followed by 24.3\% for JFCs and 16.6\% for asteroids. The relative distribution of meteors and PTs across the different dynamical classes can be seen in Figure \ref{fig:TJ_histogram}.
Previous meteor train surveys have not explicitly investigated the dynamical origin of PT meteors, though based on the assumption of ``fast and bright" meteors and the focus on classic examples of PT-producing meteor showers (i.e. Leonids, Orionids, and Perseids), NIC meteoroids would likely dominate. Interestingly, we find this category to be the least fruitful at producing PTs. 

Different physical properties among the three Tisserand dynamical categories presumably contribute to their viability for making PTs. \citeA{Kikwaya2011} determined that bulk densities for small meteoroids on NIC orbits are usually between 360-1510 kg m$^{-3}$, for JFC orbits the average density is about 3100 kg m$^{-3}$, and asteroidal orbits have the highest associated densities, around 4200 kg m$^{-3}$. However, more recent analysis by \citeA{Vojacek2019} puts the average density of meteoroids on JFC orbits at a more reasonable 850 kg m$^{-3}$---much closer to the density obtained for their nearly isotropic cometary relatives. While \citeA{Kikwaya2011}'s JFC average density is consistent with chondritic-like meteorites or porous bodies containing inclusions of highly refractory material, \citeA{Vojacek2019}'s density is instead suggestive of cometary material, though they state that it is still possible to get individual high-density meteoroids from cometary sources. \citeA{Buccongello2024} examined 41 shower meteors with known parent bodies and found average bulk densities of 602 kg m$^{-3}$ and 345 kg m$^{-3}$ for JFC and NIC orbits, respectively. This corroborates the lower JFC bulk density value obtained by \citeA{Vojacek2019}. It is generally accepted that cometary material is fragile and easily fragmented during atmospheric entry, while asteroidal material is conversely stronger and harder to break apart. However, most observable meteors with diameters larger than $\sim$1 mm tend to exhibit indications of fragmenting regardless of their Tisserand classification \cite{Subasinghe2016}.

The $K_B$ parameter, introduced by \citeA{Ceplecha1958}, has often been used to characterize a meteoroid's strength and composition based on its beginning height. It is defined as 
\begin{linenomath*}
\begin{equation}
    \label{KB}
    K_B = \log{\rho_B}+2.5\log{v_\infty}-0.5\log{\cos{z_R}}
\end{equation}
\end{linenomath*}
where $\rho_B$ is the atmospheric density at the onset of ablation (in g cm$^{-3}$), $v_\infty$ is the meteor's pre-atmospheric velocity (in cm s$^{-1}$), and $z_R$ is the zenith angle of the radiant \cite{Ceplecha1958}. We calculated the atmospheric densities using the MSIS 2.1 model \cite{Emmert2022}. The $K_B$ parameter fundamentally assumes that all meteoroids begin the luminous portion of their flight at the same surface temperature. After removing the influence of velocity and entry angle, the remaining altitudinal dependence should then be closely related to the meteoroid's material properties. \citeA{Ceplecha1988} divided $K_B$ values into five classes: ``asteroidal" ($K_B \geq 8$), group A ($7.3 \leq K_B < 8$), group B ($7.1 \leq K_B < 7.3$), group C ($6.6 \leq K_B < 7.1$), and group D ($K_B < 6.6$). He also associated each group with a particular composition, where the ``asteroidal" group is comprised of ordinary chondrites, group A of carbonaceous chondrites, and groups B, C, and D of dense, regular, and soft cometary material, respectively \cite{Ceplecha1988}. \citeA{Ceplecha1988} derived the cutoff values between groups using Super-Schmidt cameras; a correction is needed to account for the differing sensitivity of GMN cameras. Comparing the peaks in \citeA{Ceplecha1967}'s bimodal $K_B$ distributions of sporadic meteors with our own, we find a correction constant of approximately $-0.10$ needs to be added to Equation \ref{KB} in order for the GMN-derived $K_B$ values to be relatively consistent with Ceplecha's cutoff values. Though densities have commonly been ascribed to each group, recent data \cite<as discussed in>[]{Moorhead2017} suggest that $K_B$ values do not strongly correlate with density; they instead recommend the Tisserand parameter as a more robust density proxy. 

\begin{figure}
    \centering
    \includegraphics{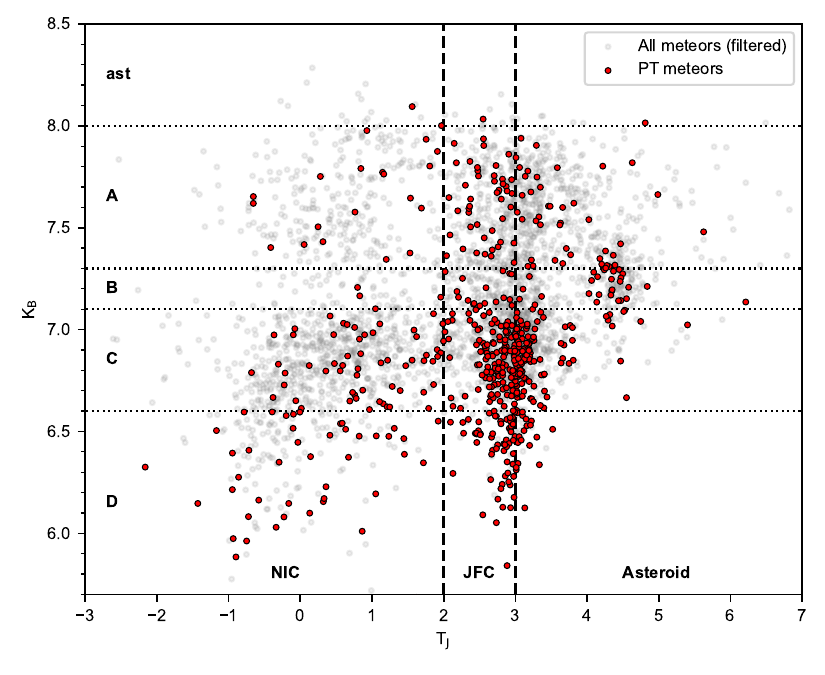}
    \caption{Relationship between $K_B$ and $T_J$. The population of meteors ending below 93.5 km and the subset that produced PTs is shown, with demarcations indicating the different Ceplecha types and dynamical classes. ``Ast" and group A are associated with asteroidal material and groups B, C, and D with dense, regular, and soft cometary material, respectively.}
    \label{fig:KB_TJ_scatter}
\end{figure}

\begin{figure}
    \centering\makebox[\textwidth]{\includegraphics[width=1.25\textwidth]{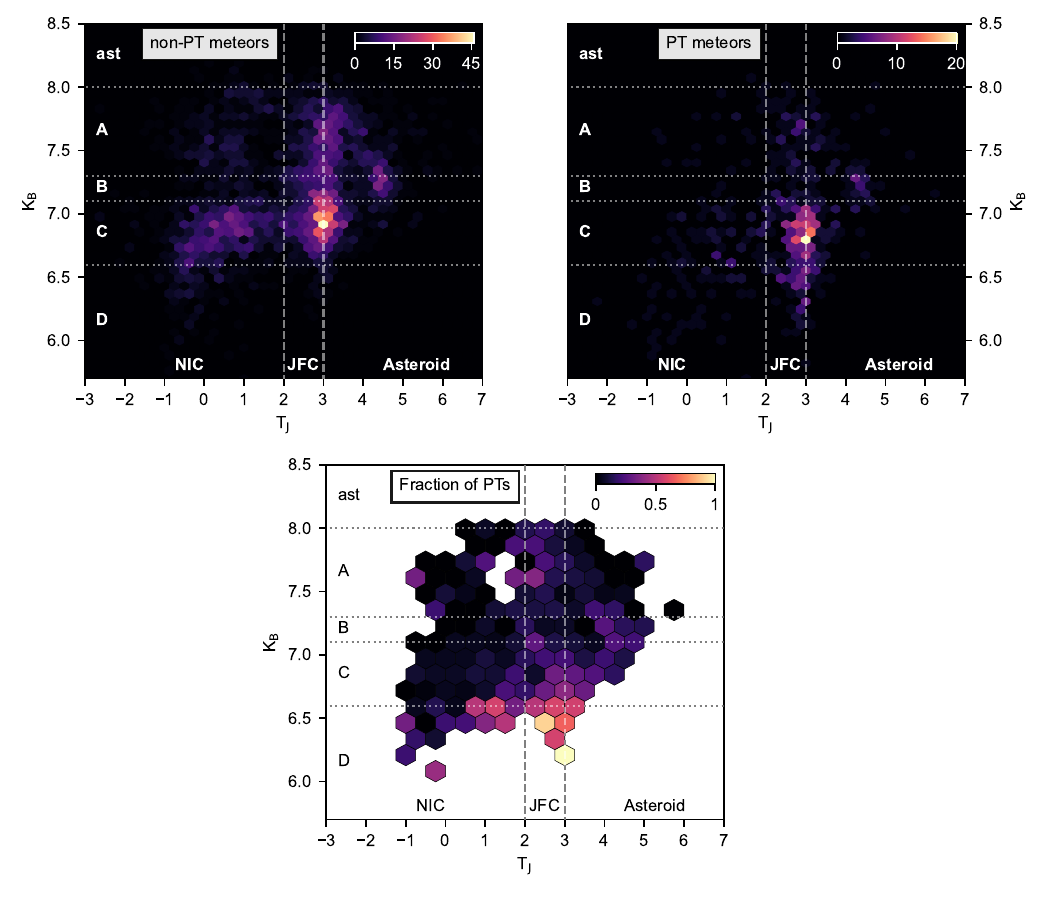}}
    \caption{Relationship between $K_B$ and $T_J$, with demarcations indicating the different Ceplecha types and dynamical classes. (Top Left) Density heat map for the sample of meteors without observed PTs, which have been filtered according to the 93.5 km cutoff altitude. (Top Right) Density heat map for the meteors with observed PTs. (Bottom Middle) Heat map for the fraction of meteors ending below 93.5 km that left PTs; fractions have only been computed for cells containing five or more total meteors.}
    \label{fig:KB_TJ_hexes}
\end{figure}

Figure \ref{fig:KB_TJ_scatter} plots the $K_B$ parameter against the Tisserand parameter with boundaries indicating the different dynamical classes and Ceplecha types.  
The majority of PT meteors are found clustered in the JFC and asteroidal dynamical classes, though many of the asteroidal meteors near the boundary likely evolved from JFC orbits to asteroidal-type orbits due to Poynting-Robertson drag. In regard to $K_B$, most PT meteors are associated with groups C and D: the regular and soft cometary material, respectively. The group D subset is especially interesting near the JFC/asteroidal boundary, as the majority of meteors in this population left PTs; 72\% of meteors with $T_J > 2$ and $K_B < 6.6$ produced a PT. This is illustrated in Figure \ref{fig:KB_TJ_hexes}, which shows density heat maps for the subsets of meteors that did and did not leave observable PTs, filtered by the 93.5 km altitude cutoff.
It also includes a heat map showing the fraction of meteors ending below 93.5 km that left PTs. To avoid noise due to small number statistics, fractional values have only been computed for the cells which contained at least five total meteors. An enhancement in the fraction of PTs can be clearly seen near and below the D group boundary, suggesting that meteoroid fragility influences PT production. The interplay between having weak cometary material and suitable chemical composition may ultimately underlie JFC meteors having PT production rates almost 50\% greater than asteroidal meteors and more than double that of NIC meteors, though further investigation is needed.

\subsection{Association with Meteor Showers}
\label{sec:showers}
\begin{figure}
    \centering
    \includegraphics{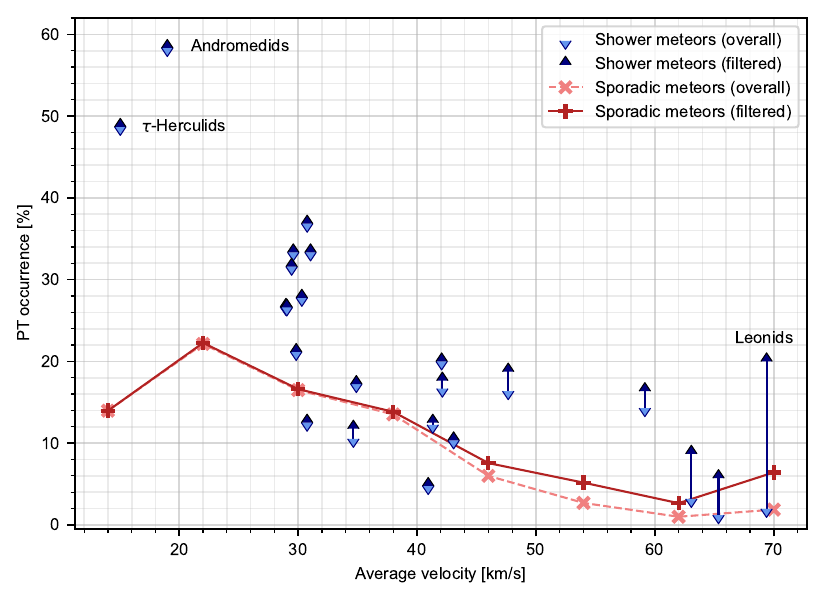}
    \caption{Occurrence rate of PTs for sporadic meteors and several meteor showers. The sporadic meteors are grouped into bins of 8 km/s, with the marker indicating the bin's midpoint; meteor shower velocities are found by averaging the constituent meteors' average velocities. Percentages are shown both with respect to the total observed sample (``overall") as well as just those that ended below 93.5 km (``filtered"). The Andromedids, $\tau$-Herculids, and Leonids are individually labeled as they are discussed in detail in the text.}
    \label{fig:showers}
\end{figure}

PTs have been primarily linked to well-known meteor showers like the Perseids and Leonids ever since they were first studied in earnest; our year-round observing campaign allowed us to probe both the popular and smaller meteor showers, as well as sporadic meteors, with improved statistical detail. Table \ref{tab:showers} provides abridged data for only the meteor showers that produced trains---complete shower statistics for all detected meteors and the breakdown of PTs by duration bin can be accessed from the Open Research section. 
\begin{table}[htbp]
\tiny
\centering
\caption{\textit{Breakdown of All Train-producing Meteor Showers$^a$}}
\begin{tabularx}{\textwidth}{l>{\centering\arraybackslash}X>{\centering\arraybackslash}X>{\centering\arraybackslash}X>{\centering\arraybackslash}X>{\centering\arraybackslash}X>{\centering\arraybackslash}X>{\centering\arraybackslash}X>{\centering\arraybackslash}X}
\hline
Shower name & IAU code & $v_\textnormal{avg}$ (km/s) & Total meteors & ``Deep" meteors & SDTs & PTs & \%PT (total) & \%PT (deep) \\ 
\hline
Sporadic & ... & 40.9 & 2730 & 2004 & 27 & 280 & 10.3 & 14.0 \\ 
Geminids & GEM & 34.9 & 238 & 238 & 6 & 41 & 17.2 & 17.2 \\ 
$\tau$-Herculids & TAH & 15.0 & 76 & 76 & 3 & 37 & 48.7 & 48.7 \\ 
Southern Taurids & STA & 29.0 & 94 & 94 & 2 & 25 & 26.6 & 26.6 \\ 
s-Taurids & STS & 31.0 & 60 & 60 & 1 & 20 & 33.3 & 33.3 \\ 
Andromedids & AND & 19.0 & 24 & 24 & 0 & 14 & 58.3 & 58.3 \\ 
Northern Taurids & NTA & 29.8 & 52 & 52 & 0 & 11 & 21.2 & 21.2 \\ 
$\sigma$-Hydrids & HYD & 59.2 & 63 & 55 & 0 & 9 & 14.3 & 16.4 \\ 
$o$-Eridanids & OER & 29.6 & 27 & 27 & 1 & 9 & 33.3 & 33.3 \\ 
April Lyrids & LYR & 47.7 & 55 & 48 & 0 & 9 & 16.4 & 18.8 \\ 
Southern $\chi$-Orionids & ORS & 30.8 & 19 & 19 & 0 & 7 & 36.8 & 36.8 \\ 
Quadrantids & QUA & 41.3 & 49 & 48 & 0 & 6 & 12.2 & 12.5 \\ 
f-Taurids & FTR & 29.5 & 19 & 19 & 1 & 6 & 31.6 & 31.6 \\ 
64-Draconids & SFD & 25.2 & 12 & 12 & 1 & 5 & 41.7 & 41.7 \\ 
$\tau$-Arietids & TAR & 30.3 & 18 & 18 & 0 & 5 & 27.8 & 27.8 \\ 
$\lambda$-Cetids & LCT & 29.0 & 15 & 15 & 1 & 4 & 26.7 & 26.7 \\ 
November $\eta$-Taurids & NET & 30.2 & 8 & 8 & 0 & 4 & 50.0 & 50.0 \\ 
December Monocerotids & MON & 42.1 & 20 & 20 & 0 & 4 & 20.0 & 20.0 \\ 
Northern December $\delta$-Arietids & DNA & 27.8 & 11 & 11 & 0 & 3 & 27.3 & 27.3 \\ 
A2-Taurids & ATS & 28.7 & 5 & 5 & 0 & 3 & 60.0 & 60.0 \\ 
Northern October $\delta$-Arietids & NOA & 30.2 & 9 & 9 & 0 & 3 & 33.3 & 33.3 \\ 
December $\alpha$-Draconids & DAD & 42.1 & 18 & 17 & 0 & 3 & 16.7 & 17.6 \\ 
$\tau$-Taurids & TAT & 28.7 & 14 & 14 & 3 & 3 & 21.4 & 21.4 \\ 
$\delta$-Arietids & DAT & 30.7 & 24 & 24 & 2 & 3 & 12.5 & 12.5 \\ 
November Orionids & NOO & 43.1 & 29 & 29 & 1 & 3 & 10.3 & 10.3 \\ 
Ursids & URS & 34.6 & 19 & 17 & 0 & 2 & 10.5 & 11.8 \\ 
90-Herculids & NHE & 40.8 & 8 & 6 & 0 & 2 & 25.0 & 33.3 \\ 
Southern October $\delta$-Arietids & SOA & 30.3 & 10 & 10 & 1 & 2 & 20.0 & 20.0 \\ 
Comae Berenicids & COM & 63.1 & 63 & 23 & 1 & 2 & 3.2 & 8.7 \\ 
Orionids & ORI & 65.4 & 168 & 35 & 1 & 2 & 1.2 & 5.7 \\ 
October Leporids & OLP & 32.0 & 4 & 4 & 0 & 2 & 50.0 & 50.0 \\ 
m-Taurids & MTA & 25.5 & 12 & 12 & 0 & 2 & 16.7 & 16.7 \\ 
A1-Taurids & ATU & 29.9 & 13 & 13 & 0 & 2 & 15.4 & 15.4 \\ 
$\kappa$-Serpentids & KSE & 47.6 & 2 & 2 & 0 & 2 & 100.0 & 100.0 \\ 
68-Virginids & OAV & 30.0 & 7 & 7 & 0 & 2 & 28.6 & 28.6 \\ 
$\xi$-Arietids & XAR & 30.3 & 7 & 7 & 0 & 2 & 28.6 & 28.6 \\ 
$\beta$-Bootids & BBO & 49.0 & 3 & 2 & 0 & 1 & 33.3 & 50.0 \\ 
R-Lyrids & RLY & 38.1 & 1 & 1 & 0 & 1 & 100.0 & 100.0 \\ 
January $\kappa$-Leonids & JKL & 40.5 & 5 & 5 & 0 & 1 & 20.0 & 20.0 \\ 
Southern December $\delta$-Arietids & DSA & 22.9 & 5 & 5 & 0 & 1 & 20.0 & 20.0 \\ 
$\epsilon$-Scorpiids & EPS & 38.8 & 1 & 1 & 0 & 1 & 100.0 & 100.0 \\ 
$\kappa$-Cepheids & KCE & 38.7 & 2 & 2 & 0 & 1 & 50.0 & 50.0 \\ 
February Leonids & FLO & 33.4 & 5 & 5 & 0 & 1 & 20.0 & 20.0 \\ 
$\xi$-Scorpiids & XSC & 30.7 & 1 & 1 & 0 & 1 & 100.0 & 100.0 \\ 
February $\sigma$-Leonids & FSL & 33.3 & 2 & 2 & 0 & 1 & 50.0 & 50.0 \\ 
p-Taurids & PTS & 26.9 & 5 & 5 & 0 & 1 & 20.0 & 20.0 \\ 
e-Velids & EVE & 42.7 & 3 & 2 & 0 & 1 & 33.3 & 50.0 \\ 
February $\mu$-Virginids & FMV & 66.2 & 3 & 2 & 0 & 1 & 33.3 & 50.0 \\ 
$\chi$-Taurids & CTA & 41.6 & 4 & 4 & 0 & 1 & 25.0 & 25.0 \\ 
November $\zeta$-Perseids & NZP & 29.7 & 3 & 3 & 0 & 1 & 33.3 & 33.3 \\ 
Northern $\delta$-Aquariids & NDA & 36.9 & 4 & 4 & 0 & 1 & 25.0 & 25.0 \\ 
$\eta$-Lyrids & ELY & 43.4 & 4 & 4 & 0 & 1 & 25.0 & 25.0 \\ 
August $o$-Aquariids & AOA & 38.4 & 2 & 2 & 0 & 1 & 50.0 & 50.0 \\ 
$\alpha$-Virginids & AVB & 21.2 & 5 & 5 & 1 & 1 & 20.0 & 20.0 \\ 
$\upsilon$-Andromedids & UAN & 59.2 & 3 & 1 & 0 & 1 & 33.3 & 100.0 \\ 
$\beta$-Comae Berenicids & BCO & 28.6 & 2 & 2 & 0 & 1 & 50.0 & 50.0 \\ 
$\eta$-Andromedids & EAN & 39.1 & 1 & 1 & 0 & 1 & 100.0 & 100.0 \\ 
January $\pi$-Puppids & PIP & 26.4 & 5 & 5 & 0 & 1 & 20.0 & 20.0 \\ 
$\omega$-Taurids & FTA & 25.2 & 2 & 2 & 0 & 1 & 50.0 & 50.0 \\ 
$\nu$-Eridanids & NUE & 65.8 & 30 & 1 & 0 & 1 & 3.3 & 100.0 \\ 
Camelopardalids & CAM & 17.7 & 1 & 1 & 0 & 1 & 100.0 & 100.0 \\ 
Southern $\delta$-Aquariids & SDA & 40.9 & 21 & 21 & 0 & 1 & 4.8 & 4.8 \\ 
Northern $\delta$-Cancrids & NCC & 29.8 & 10 & 10 & 0 & 1 & 10.0 & 10.0 \\ 
$\lambda$-Ursae Majorids & LUM & 61.2 & 2 & 1 & 0 & 1 & 50.0 & 100.0 \\ 
February $\pi$-Leonids & FPL & 32.8 & 2 & 2 & 0 & 1 & 50.0 & 50.0 \\ 
November $\theta$-Aurigids & THA & 39.2 & 2 & 2 & 0 & 1 & 50.0 & 50.0 \\ 
22-Aquilids & TAL & 65.6 & 9 & 2 & 0 & 1 & 11.1 & 50.0 \\ 
December $\sigma$-Virginids & DSV & 67.1 & 8 & 2 & 0 & 1 & 12.5 & 50.0 \\ 
November $\gamma$-Bootids & NGB & 47.0 & 1 & 1 & 0 & 1 & 100.0 & 100.0 \\ 
December Canis Minorids & CMI & 37.4 & 7 & 7 & 0 & 1 & 14.3 & 14.3 \\ 
110-Herculids & OTH & 50.7 & 1 & 1 & 0 & 1 & 100.0 & 100.0 \\ 
Southern March $\gamma$-Virginids & SMV & 40.6 & 7 & 7 & 0 & 1 & 14.3 & 14.3 \\ 
December $\phi$-Cassiopeiids & DPC & 18.7 & 1 & 1 & 0 & 1 & 100.0 & 100.0 \\ 
Leonids & LEO & 69.4 & 51 & 5 & 0 & 1 & 2.0 & 20.0 \\ 
$\zeta$-Cygnids & ZCY & 44.6 & 12 & 10 & 0 & 1 & 8.3 & 10.0 \\ 
August Draconids & AUD & 23.4 & 3 & 3 & 1 & 0 & 0.0 & 0.0 \\ 
$\rho$-Piscids & RPI & 38.1 & 2 & 2 & 1 & 0 & 0.0 & 0.0 \\ 
Northern $\delta$-Piscids & NPI & 30.7 & 3 & 3 & 1 & 0 & 0.0 & 0.0 \\ 
15-Leonids & FLE & 64.7 & 3 & 2 & 1 & 0 & 0.0 & 0.0 \\ 
Southern $\mu$-Sagittariids & SSG & 25.0 & 1 & 1 & 1 & 0 & 0.0 & 0.0 \\ 
\hline
\multicolumn{9}{p{\textwidth}}{$^a$ Columns represent the average velocity of the shower ($v_\textnormal{avg}$), the total count of meteors, the number of meteors that terminated below 93.5 km (``deep" meteors), the number of SDTs (trains lasting less than minute), PTs, and the percentage of meteors that left a PT, relative to the total meteors and just the ``deep" meteors.}
\end{tabularx}
\label{tab:showers}
\end{table}
A key trend seen in Table \ref{tab:showers} is nicely visualized by Figure \ref{fig:showers}, namely that a connection exists between a meteor's average velocity and the likelihood it produces an observable PT.
To reduce the number of artificially low/high occurrence rates due to very small sample sizes, only meteors showers that produced at least 15 total meteors with 5 or more ending below 93.5 km are included in Figure \ref{fig:showers}. Sporadic meteors are also shown, though they had to first be aggregated into bins (each 8 km/s wide) before calculating their PT occurrence rates. The midpoints of each bin are indicated by the triangle markers. For both sporadic and shower meteors, the ``overall" and ``filtered" percentages are determined: overall PT percentages are calculated using all meteors, representative of what an observer could expect to see under similar circumstances, and filtered PT percentages are calculated using the meteors that terminate below 93.5 km, representing a normalized value better suited for direct comparisons. Unsurprisingly, occurrence rates increased after application of the altitude cut, with the greatest discrepancy seen among the fast meteors.  

\begin{figure}
    \centering
    \includegraphics{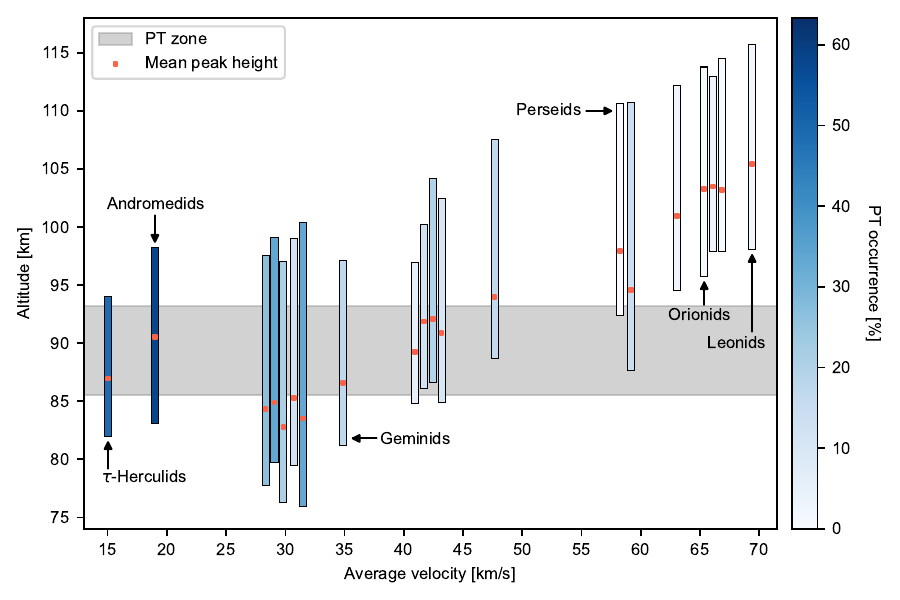}
    \caption{Relationship between altitude, velocity, and PT formation. Each bar is the average altitude range of a meteor shower, with the average location of its peak magnitude shown as a red marker. The `PT zone' shows the average altitude range where PTs form; the color bar indicates the percent of meteors that had an observed PT. Prominent showers are annotated on the graph. The IAU three-letter codes for all showers depicted are (left to right): TAH, AND, STA, OER, NTA, DAT, STS, GEM, SDA, QUA, MON, NOO, LYR, PER, HYD, COM, ORI, NUE, ETA, and LEO.}
    \label{fig:height_vs_vavg}
\end{figure}

However, it is unlikely that a meteor's velocity alone is responsible for driving this trend, as the consequences of the previous section also contribute. High velocity meteors are essentially singularly associated with NIC dynamical origin; only 0.8\% of meteors faster than 45 km/s come from the other two categories. Similarly, the slow regime is dominated by JFC/asteroidal origins, with only 3.2\% of meteors slower than 30 km/s coming from NIC orbits. Since JFC meteors, and to a lesser extent asteroidal meteors, produce many more detectable PTs relative to NIC meteors, the observed trend of slower meteors having greater PT production rates could instead arise from dynamical origin differences. If so, velocity may be merely acting as a proxy for physical meteoroid properties (as traced by dynamical origin) rather than a bona fide influence in determining the likelihood of PT formation. Alternatively, meteor velocity is related to the altitude range over which a meteor is visible, with faster meteors tending to start and end ablation higher than slower meteors. Previous sections have demonstrated the importance of altitude on PT formation, which would therefore result in a velocity dependence. Figure \ref{fig:height_vs_vavg} shows the connection between altitude, velocity, and PT production rates for all meteor showers having at least 20 observed meteors. 
Each bar represents the average altitude range of a different meteor shower, plotted against the shower's average velocity (which was computed using each member meteor's average velocity). The color scale indicates the rate of PT formation, and the gray `PT zone' band is the average altitude range where PTs are found (from Table \ref{tab:altitude}). The average altitude where each meteor shower generates its peak magnitude is also indicated to provide an idea of where the greatest mass deposition occurs. A clear trend is visible: slower meteors more reliably reach the PT zone, which in turn correlates to higher rates of PT production. It is also interesting to note that the two slowest showers (the $\tau$-Herculids and Andromedids) have mean peak heights which sit within the PT zone, in spite of the trend suggested by the remaining showers. This is likely related to their high fragility. Regardless of whether the velocity dependence is due to dynamical reasons or altitudinal ones, the implication for observational studies remains---slower meteors (from either sporadics or showers) have a higher PT-to-meteor ratio than their faster counterparts.

Before highlighting a few interesting meteor showers, it is worth pointing out that our observing strategy leaves us at the mercy of interloping clouds and the moon. As such, the peak date or time for a given meteor shower may not be fully captured. This can potentially bias our sample to different regions of the meteor stream, which may have different ages and/or properties than those associated with peak activity. Repeated yearly observations will help to ameliorate these concerns and provide better overall statistics. Additionally, the use of long exposure images and the camera's sensitivity to NIR emission means that our observed PT occurrence rate for any given shower may not be reflective of the rate observed by an analog naked eye observer. 

\subsubsection{Andromedids}
Remarkably, the Andromedids meteor shower had the highest proportion of detected PTs among the non-trivial showers (i.e. those with only a couple meteors), with 14 of the 24 meteors (58\%) forming PTs. The Andromedids are essentially the complete opposite of the archetypal PT meteor shower, being fairly faint (average peak magnitude: $\sim$0.3 mag) and slow (average velocity: 19 km/s). Historically, it has produced some excellent outbursts in the late 1800s after the parent comet 3D/Biela broke up, with some observers noting a large number of meteor trains lasting several seconds \cite{Smieton1885}. Biela was a JFC, which perhaps contributes to its abundance of trains, both then and now. These days the Andromedids are a quite weak shower, having a typical rate of only a few meteors per hour or less \cite{Wiegert2013}. There was, however, an unexpected Andromedid outburst detected on November 28, 2021; the Cameras for Allsky Meteor Surveillance (CAMS) surveys triangulated more than 120 Andromedid meteors in less than 10.5 hours \cite{Jenniskens2022}. We have observations spanning two years for this shower: in 2021 there were 10 PTs out of 18 meteors and 2022 saw 4 PTs out of 6 meteors. For the 2021 shower, 6 meteors occurred during the outburst with 4 of those leaving PTs. The PT rate during the outburst (67\%) is higher than the PT rate during the rest of the 2021 shower (50\%), though the small sample sizes make it difficult to definitively claim that the outburst (and/or the younger material it is likely composed of) was responsible. While both years showed activity during the last week of November, the 2021 shower also had a decent number of meteors/PTs during the first couple weeks of the month. Early November is generally associated with newer material, which then gets progressively older as December approaches \cite{Kronk2014}. \citeA{Wiegert2013} provides no information on the 2021 shower, but links the 2022 shower to the 1772 orbit of Biela, prior to its breakup. They predict moderate to strong activity in 2036; this event could provide a good opportunity to collect additional statistics from an otherwise weak shower. The particular characteristics of the Andromedids that make them especially suited to form trains will require additional study.

\subsubsection{Leonids}
We cannot overlook the Leonids in a discussion of PTs, though much has already been said on the subject. The Leonids produce some of the fastest meteors, with average velocities around 70 km/s. Their parent comet, 55P/Tempel–Tuttle, is a classic example of a Halley-type comet; the meteoroids correspondingly have very low bulk densities (400 kg m$^{-3}$) and a high porosity (83\%), making them quite fragile \cite{Badadzhanov2009}. The reoccurrence of spectacular meteor storms every 33 years or so is perhaps their biggest claim to fame, the last of which took place in 2002. Like the Andromedids, we observed the Leonids twice during the campaign. In 2021, the shower's peak activity coincided with a nearly full moon, limiting observations during this period. The following year, the moon was in its third quarter during peak activity, though this did not lead to more meteors being detected. We observed 34 meteors in 2021, with only one of these leaving an observable PT; none of the 17 meteors seen during the 2022 shower produced trains. Based on the most recent Leonid storms, \citeA{Higa2005} calculated a PT incidence rate of about 20\% for Leonid meteors brighter than -2 mag. Granted, these storms contained many large meteoroids capable of penetrating deeper in the atmosphere relative to the non-storm meteor population. When we apply the same magnitude criterion, we find only 3\% of meteors leaving PTs (1 in 33), though this value incorporates potentially inaccurate GMN magnitudes. The majority of our observed Leonid meteors ended above the 93.5 km altitude cutoff though---only 5 of the 51 made it into the PT zone. With just these five meteors, our PT occurrence rate jumps to 20\% which is in better agreement with \citeA{Higa2005}; this rate should be taken with caution, however, as the sample is quite small. Regardless, the non-outbursting Leonid showers we observed were not fruitful for PT viewing. 

\subsubsection{$\tau$-Herculids}
\label{sec:tau}
The $\tau$-Herculids were the second most prolific shower for PTs, both in terms of counts and occurrence rate (among the non-trivial showers). Out of 76 meteors, 37 left behind noticeable PTs (48.7\%) and all 76 penetrated below the altitude threshold. Though the $\tau$-Herculids are usually considered an inactive shower, their parent comet 73P/Schwassmann–Wachmann 3 underwent a large-scale fragmentation event in 1995 and divided into at least 4 main fragments; this breakup and subsequent crumbling was predicted to cause considerably increased meteor rates during its 2022 shower \cite{Egal2023}. As a testament to its low activity, the GMN observed only one multi-station event for each 2021 and 2023, but saw more than 1240 events in 2022. Two distinct activity peaks were noted during the 2022 outburst. According to simulations done by \citeA{Egal2023}, the first, smaller peak arose from older, pre-breakup trails, while the main activity peak was traced to the 1995 breakup. All but three of the $\tau$-Herculid meteors we observed had solar longitudes greater than 69.39, consistent with material from the splitting event. The remaining three were likely due to older trails, however one of these older meteoroids still produced a PT. Though Schwassmann–Wachmann 3 is a JFC, the estimated bulk density of its meteoroids is very low---around 250 kg m$^{-3}$ \cite{Egal2023}, which is even lower than the Leonids'. Consequently, this material is quite fragile and prone to fragmentation. Due to their extremely low speeds, the average beginning and end height for the $\tau$-Herculid meteors were 94 and 82 km respectively; this overlaps almost exactly with the PT zone, potentially contributing to the abundance of trains. \citeA{Koten2023} obtained spectra for a few $\tau$-Herculid meteors during the 2022 shower, all of which were dominated by Na D-lines, though Fe and Mg lines were also present. Though they did not see any molecular contribution (e.g. FeO), it is unclear whether their spectral data extended beyond the meteors' afterglow and into the continuum phase. However, they suggest the enhanced Na content could be due to the recency of the parent comet's disruption; meteoroids originating from the comet's interior were only exposed to space for 27 years prior to the 2022 outburst. Whether the freshness of this material played a role in their elevated PT formation rate remains to be seen. The $\tau$-Herculids are predicted to return with stronger showers in 2033 and 2049, again primarily due to material from the 1995 fragmentation \cite{Egal2023}, providing a good opportunity for additional study.

\section{Conclusions}
We have presented here the results from our meteor observing campaign and the PT catalog it produced. This catalog is unique among other PT catalogs in that the year-round, shower-agnostic observations were taken using the same automated camera setup in the same location, resulting in great data collection uniformity. Additionally, the detection and identification of our meteors/trains was accomplished using the same pipeline, with the derived meteor parameters being reduced and calculated in a consistent manner by the GMN. Taken altogether, this affords our catalog an excellent degree of homogeneity. Though the long exposure images and mild sensitivity to NIR mean we likely probe a slightly different population than used in previous surveys and studies, our results nonetheless challenge many long-held assertions and assumptions regarding PTs, as well as offer new avenues of exploration.

We now briefly summarize the main takeaways from this work, starting with previous information about PTs that has been corroborated by our data:
\begin{enumerate}
    \item PTs occur in a narrow altitude range relative to the full range spanned by meteors. We found the average beginning, center, and terminal heights for PTs to be 93.2, 89.4, and 85.5 km, respectively; these values are consistent with those previously reported. 
    \item The secondary O$_3$ maximum aligns remarkable well with the PT altitude range, strengthening the assertion that O$_3$ availability dictates where trains form.
    \item Meteors with brighter peak absolute magnitudes tend to produce PTs that endure longer. By extension, meteors with larger masses also produce longer lasting PTs.
\end{enumerate}
However, there is also a fair amount of previous research and assumptions that are inconsistent with the data from our catalog. The following points are worth reconsidering in light of our new observations:
\begin{enumerate}
    \item PTs are more common than previously claimed. Our camera saw $\sim$1 in 8 meteors leave a PT, compared to the 1 in 750 from \citeA{Olivier1957}'s estimate. Long duration PTs are also more common, with 1 in 19 meteors producing a PT lasting longer than five minutes compared to the previously estimated 1 in 5000.
    \item There appears to be no correlation between a meteor's velocity and the duration of its train; slow moving meteors routinely left PTs lasting longer than 15 minutes.
    \item A meteor's peak magnitude is not a consistent discriminator for whether a PT can form. The vast majority of our observed PTs were associated with meteors dimmer than the assumed -2 mag visual cutoff derived from the very fast Leonids.
    \item PTs are not always cospatial with their parent meteors' region of maximum brightness; a third of the meteors used in the altitude study had peak magnitude locations completely exterior to their PT range.
    \item High-velocity meteor showers typically had lower rates of PT production relative to their slower counterparts, even after applying normalization. This is because meteoroids of fast showers usually broke up before reaching the 93.5 km cutoff needed for PT generation.
\end{enumerate}
Based on our observations, we also suggest a few novel concepts that warrant additional investigation:
\begin{enumerate}
    \item The terminal altitude of a meteor appears to be the best discriminator of whether a PT is capable of forming, with meteor magnitude and velocity being less significant.
    \item A meteor's dynamical origin plays a role in the likelihood it leaves a detectable PT, with JFC meteoroids being most prolific and NIC meteoroids being the least. The exact mechanical or compositional reason behind this remains to be seen, though the $K_B$ parameter suggests weaker cometary material may contribute to this. 
    \item Meteoroid material produced in ``recent" large-scale fragmentation events (or during outburst/storm events) may have an improved chance at making PTs, as suggested by the $\tau$-Herculids and Leonid storms, though this probably depends on the nature of the parent body. 
\end{enumerate}

In addition to the open questions mentioned above, planned work for the future includes continuing the current campaign to obtain more robust year-round data, incorporating spectroscopic information into the observations to determine the species and timescales involved in non-Leonid PTs, and examining in greater detail the connection between PTs and meteor radio afterglows. These efforts will hopefully continue to shed light on this intriguing noctilucent phenomenon.

%  Numbered lines in equations:
%  To add line numbers to lines in equations,
%  \begin{linenomath*}
%  \begin{equation}
%  \end{equation}
%  \end{linenomath*}

%% Enter Figures and Tables near as possible to where they are first mentioned:
%
% DO NOT USE \psfrag or \subfigure commands.
%
% Figure captions go below the figure.
% Table titles go above tables;  other caption information
%  should be placed in last line of the table, using
% \multicolumn2l{$^a$ This is a table note.}
%
%----------------
% EXAMPLE FIGURES
%
% \begin{figure}
% \includegraphics{example.png}
% \caption{caption}
% \end{figure}
%
% Giving latex a width will help it to scale the figure properly. A simple trick is to use \textwidth. Try this if large figures run off the side of the page.
% \begin{figure}
% \noindent\includegraphics[width=\textwidth]{anothersample.png}
%\caption{caption}
%\label{pngfiguresample}
%\end{figure}
%
% If you get an error about an unknown bounding box, try specifying the width and height of the figure with the natwidth and natheight options. This is common when trying to add a PDF figure without pdflatex.
% \begin{figure}
% \noindent\includegraphics[natwidth=800px,natheight=600px]{samplefigure.pdf}
%\caption{caption}
%\label{pdffiguresample}
%\end{figure}
%
% PDFLatex does not seem to be able to process EPS figures. You may want to try the epstopdf package.

%%% End of body of article

%%%%%%%%%%%%%%%%%%%%%%%%%%%%%%%%
%% Optional Appendix goes here
%
% The \appendix command resets counters and redefines section heads

\appendix
\section{Astrometic Calibration Process}
\label{app:Astrometry}
In order to facilitate the star fitting, a set of reference stars with known celestial coordinates is needed. A catalog of the right ascension and declination (RA/Dec) for 85 bright stars was therefore compiled, though typically only 15-20 stars are in the camera's field of view for any given image. With this in hand, the first step is to create a naive astrometric fit that requires only the pixel coordinates of Polaris. Since the camera is located in the northern hemisphere, Polaris is always above the horizon and relatively simple to locate; this is done by summing together several minutes worth of consecutive images. Polaris remains essentially stationary (and with greatly enhanced brightness) in this summed image while the other stars begin to leave trails. The naive model assumes that the optical axis of the camera coincides with the astronomical zenith, both occurring at the center of the image. Both the naive and refined models assume that the center of the image (i.e. the center pixel) corresponds to the center of the projection (i.e. the origin point of Equation \ref{eqn:distortion}). Since Polaris has a known alt/az at our observation site, the line connecting its pixel coordinate to the image center then becomes the reference axis from which all other azimuths can be measured. The pixel locations of the reference stars can be determined using this reference axis along with Equation \ref{eqn:distortion} to convert their altitudes to pixel distances. The calculated pixel coordinates of the reference stars can then be overlaid on top of the image, which has been appropriately thresholded to only show relatively bright stars. The difference between the calculated position and actual position of the stars is fairly small---it is therefore straightforward to go through each reference star plotted in the image and select its corresponding, correct pixel coordinate position. This results in a list of true pixel coordinates and their associated alt/az values which are used to optimize the refined astrometric fit.

Two of the parameters to be optimized, $k_1$ and $k_2$, were already seen in Equation \ref{eqn:distortion}. In addition, the refined model no longer assumes the optical axis aligns with zenith, therefore the zenith pixel coordinates, $x_z$ and $y_z$, need to be determined. The offset zenith also means that a simple reference axis for measuring azimuths can no longer be used and instead requires spherical trigonometry. Determination of azimuths now relies on knowing the azimuth of the image center, $az_c$, based on the new zenith. Once optimized, these five parameters alone are sufficient to provide suitable astrometry for the image. 

Three separate optimization steps are performed, in part to reduce the sensitivity of the parameters to their initialization values. This is handled by the SciPy \texttt{optimize} package. First, the values of $k_1$ and $k_2$ are determined along with fitting RA/Dec values to the image's center (though these are not used beyond this step). Two different distances are computed for each star: the angular distance between the star's known RA/Dec and the ``floating" image center RA/Dec, as well as the angular distance derived from inverting Equation \ref{eqn:distortion} and plugging in the known pixel distance between the image center and the star. Minimizing the root mean square (RMS) of the differences between these two angular distances provides the best fit values for $k_1$ and $k_2$. The \texttt{optimize.brute} approach is used to do the minimization, with a \texttt{fmin} finishing function. The brute force approach is useful for finding the true, global minimum of the function, at the expense of increased computation time. 

Next, the zenith's $x_z$ and $y_z$ are found. This step relies of the spherical law of cosines in the form
\begin{linenomath*}
\begin{equation}
    \label{eqn:sloc}
    c=\arccos{\left(\cos{a}\cos{b}+\sin{a}\sin{b}\cos{\gamma}\right)}
\end{equation}
\end{linenomath*}
where $c$ represents the zenith angle, $a$ is angular distance between the star and the image's center, $b$ is the angular distance between the ``floating" zenith and the image's center, and $\gamma$ is the angle between the vectors that point toward the star and toward zenith from the image's center. Both $a$ and $b$ rely on Equation \ref{eqn:distortion} with the optimized values of $k$ found in the last step. Taking the complement angle of $c$ gives the star's altitude, which can be compared to the known altitude. Again, minimizing the RMS of the differences between the true and calculated altitudes provides the best fit values for $x_z$ and $y_z$. Here, \texttt{optimize.minimize} using the default L-BFGS-B solver \cite{LBFGSB} reliably produces values for the zenith's pixel coordinates. It should be cautioned that Equation \ref{eqn:sloc} is more sensitive to the input variables (and their errors) when $c$ is small due to the inverse cosine; however this does not pose a substantial problem for our fitting process. 

Lastly, the azimuth of the image's center is optimized, providing a fixed reference value from which other azimuths can be derived. To do this, a rearranged version of Equation \ref{eqn:sloc} is used
\begin{linenomath*}
\begin{equation}
    \label{eqn:sloc2}
    \alpha=\arccos{\left(\frac{\cos{a}-\cos{b}\cos{c}}{\sin{b}\sin{c}}\right)}
\end{equation}
\end{linenomath*}
where $\alpha$ is the angle between the vectors pointing to the image's center and to the star (origin at the zenith point); the other variables retain their meanings from above. This $\alpha$ represents the azimuthal offset of the star from the image center's reference azimuth. Finding the value of this reference azimuth is done using the \texttt{optimize.brute} method as above, where the best fit reference azimuth minimizes the RMS of the differences between the calculated and true star azimuths. After this, all five parameters necessary to perform the astrometric calibration have been determined.

%%%%%%%%%%%%%%%%%%%%%%%%%%%%%%%%%%%%%%%%%%%%%%%

\section{Open Research}
The raw image data for all PTs observed during this campaign can be found at \url{https://lda10g.alliance.unm.edu/~pasi/PTs/PT_images/} \cite{CordonnierImages}. Our derived data for these PTs (i.e. PT duration and altitude range) as well as their associated GMN data is located at \url{https://lda10g.alliance.unm.edu/~pasi/PTs/PT_tables/} \cite{CordonnierTables}. GMN meteor data are released under the CC BY 4.0 license; the complete GMN database that our research drew from can be accessed via \url{https://globalmeteornetwork.org/data/} \cite{Vida2019, Vida2021}. SABER O$_3$ data was retrieved from \url{https://saber.gats-inc.com/}.

All figures were produced using matplotlib v3.6.2 \cite<\url{https://matplotlib.org/};>[]{Hunter2007}. Data were organized, manipulated, and saved using h5py v2.10.0 (\url{https://www.h5py.org/}) and pandas v2.0.3 \cite<\url{https://pandas.pydata.org/};>[]{pandas2023, pandasBook}. Astrometric calibration made use of Astropy v5.1 \cite<\url{https://www.astropy.org/};>[]{Astropy2022} and PyMap3d v2.9.1 \cite<\url{https://pypi.org/project/pymap3d/};>[]{Hirsch2018}. The meteor detection pipeline relied on scikit-image v0.20.0 \cite<\url{https://scikit-image.org/};>[]{skimage2014}. The gradient boosting classification algorithm was applied using scikit-learn v1.3.1 \cite<\url{https://scikit-learn.org/};>[]{Pedregosa2011} with the SMOTE algorithm being performed by imbalanced-learn v0.11.0 \cite<\url{https://imbalanced-learn.org/};>[]{Guillaume2017}. The MSIS 2.1 atmospheric model was implemented via pymsis v0.8.0, a Python wrapper for the original Fortran source code \cite<\url{https://pypi.org/project/pymsis/;}>[]{pymsis, Emmert2022}.

% AGU requires an Availability Statement for the underlying data needed to understand, evaluate, and build upon the reported research at the time of peer review and publication.

% Authors should include an Availability Statement for the software that has a significant impact on the research. Details and templates are in the Availability Statement section of the Data and Software for Authors Guidance: \url{https://www.agu.org/Publish-with-AGU/Publish/Author-Resources/Data-and-Software-for-Authors#availability}

% It is important to cite individual datasets in this section and, and they must be included in your bibliography. Please use the type field in your bibtex file to specify the type of data cited. Some options include Dataset, Software, Collection, ComputationalNotebook. Ex: 
% \\
% \begin{verbatim}
% @misc{https://doi.org/10.7283/633e-1497,
%   doi = {10.7283/633E-1497},
%   url = {https://www.unavco.org/data/doi/10.7283/633E-1497},
%   author = {de Zeeuw-van Dalfsen, Elske and Sleeman, Reinoud},
%   title = {KNMI Dutch Antilles GPS Network - SAB1-St_Johns_Saba_NA P.S.},
%   publisher = {UNAVCO, Inc.},
%   year = {2019},
%   type = {dataset}
% }
% \end{verbatim}

%%%%%%%%%%%%%%%%%%%%%%%%%%%%%%%%%%%%%%%%%%%%%%%

\acknowledgments
The authors are thankful to Ji{\v{r}}{\'\i} Borovi{\v{c}}ka and an anonymous reviewer for their helpful comments which improved this work.
The authors thank Ralph Kelly and Jack Hines of Space Dynamics Laboratory for the design and construction of the WiPT2 system, including the automated clamshell sun shield and temperature control system. The authors recognize and thank the GMN station operators whose time and effort enabled the meteor observations used in this work.
The authors also appreciate the work that the SABER team put into processing and preparing their data, and for making it easily accessible. 
L.E. Cordonnier and G.B. Taylor acknowledge support for this research from the National Science Foundation under grant AST-1835400 and from the Air Force Research Laboratory (AFRL). L.E. Cordonnier acknowledges support from an appointment to the AFRL Scholars Program at Kirtland Air Force Base, administered by Universities Space Research Association (USRA) through a contract with AFRL. This research was sponsored in part by the Air Force Office of Scientific Research (AFOSR) Lab Task 23RVCOR002. D. Vida was supported in part by the NASA Meteoroid Environment Office under cooperative agreement 80NSSC21M0073. The views expressed are those of the authors and do not reflect the official guidance or position of the United States Government, the Department of Defense or of the United States Air Force. The appearance of external hyperlinks does not constitute endorsement by the United States Department of Defense (DoD) of the linked websites, or the information, products, or services contained therein. The DoD does not exercise any editorial, security, or other control over the information you may find at these locations.

% The acknowledgments should list:\\
% All funding sources related to this work from all authors\\
% Any real or perceived financial conflicts of interests for any author\\
% Other affiliations for any author that may be perceived as having a conflict of interest with respect to the results of this paper.\\
% It is also the appropriate place to thank colleagues and other contributors. AGU does not normally allow dedications.

%% ------------------------------------------------------------------------ %%
%% References and Citations

%%%%%%%%%%%%%%%%%%%%%%%%%%%%%%%%%%%%%%%%%%%%%%%
%
% \bibliography{<name of your .bib file>} don't specify the file extension
%
% don't specify bibliographystyle

% In the References section, cite the data/software described in the Availability Statement (this includes primary and processed data used for your research). For details on data/software citation as well as examples, see the Data & Software Citation section of the Data & Software for Authors guidance
% https://www.agu.org/Publish-with-AGU/Publish/Author-Resources/Data-and-Software-for-Authors#citation

%%%%%%%%%%%%%%%%%%%%%%%%%%%%%%%%%%%%%%%%%%%%%%%

\bibliography{sources}

\begin{thebibliography}{}

\bibitem [\protect \citeauthoryear {%
{Abe}%
\ \protect \BOthers {.}}{%
{Abe}%
\ \protect \BOthers {.}}{%
{\protect \APACyear {2004}}%
}]{%
Abe2004}
\APACinsertmetastar {%
Abe2004}%
\begin{APACrefauthors}%
{Abe}, S.%
, {Ebizuka}, N.%
, {Murayama}, H.%
, {Ohtsuka}, K.%
, {Sugimoto}, S.%
, {Yamamoto}, M\BHBI Y.%
\BDBL {}{Borovi{\v{c}}ka}, J.%
\end{APACrefauthors}%
\unskip\
\newblock
\APACrefYearMonthDay{2004}{}{}.
\newblock
{\BBOQ}\APACrefatitle {{Video and photographic spectroscopy of 1998 and 2001 Leonid persistent trains from 300 to 930 nm}} {{Video and photographic spectroscopy of 1998 and 2001 Leonid persistent trains from 300 to 930 nm}}.{\BBCQ}
\newblock
\APACjournalVolNumPages{Earth, Moon, and Planets}{95}{1-4}{265-277}.
\newblock
\begin{APACrefDOI} \doi{10.1007/s11038-005-9031-0} \end{APACrefDOI}
\PrintBackRefs{\CurrentBib}

\bibitem [\protect \citeauthoryear {%
{Astropy Collaboration}%
\ \protect \BOthers {.}}{%
{Astropy Collaboration}%
\ \protect \BOthers {.}}{%
{\protect \APACyear {2022}}%
}]{%
Astropy2022}
\APACinsertmetastar {%
Astropy2022}%
\begin{APACrefauthors}%
{Astropy Collaboration}%
, {Price-Whelan}, A\BPBI M.%
, {Lim}, P\BPBI L.%
, {Earl}, N.%
, {Starkman}, N.%
, {Bradley}, L.%
\BDBL {}{Astropy Project Contributors}%
\end{APACrefauthors}%
\unskip\
\newblock
\APACrefYearMonthDay{2022}{}{}.
\newblock
{\BBOQ}\APACrefatitle {{The Astropy Project: Sustaining and growing a community-oriented open-source project and the latest major release (v5.0) of the core package}} {{The Astropy Project: Sustaining and growing a community-oriented open-source project and the latest major release (v5.0) of the core package}}.{\BBCQ}
\newblock
\APACjournalVolNumPages{The Astrophysical Journal}{935}{2}{167}.
\newblock
\begin{APACrefDOI} \doi{10.3847/1538-4357/ac7c74} \end{APACrefDOI}
\PrintBackRefs{\CurrentBib}

\bibitem [\protect \citeauthoryear {%
{Babadzhanov}%
\ \BBA {} {Kokhirova}%
}{%
{Babadzhanov}%
\ \BBA {} {Kokhirova}%
}{%
{\protect \APACyear {2009}}%
}]{%
Badadzhanov2009}
\APACinsertmetastar {%
Badadzhanov2009}%
\begin{APACrefauthors}%
{Babadzhanov}, P\BPBI B.%
\BCBT {}\ \BBA {} {Kokhirova}, G\BPBI I.%
\end{APACrefauthors}%
\unskip\
\newblock
\APACrefYearMonthDay{2009}{}{}.
\newblock
{\BBOQ}\APACrefatitle {{Densities and porosities of meteoroids}} {{Densities and porosities of meteoroids}}.{\BBCQ}
\newblock
\APACjournalVolNumPages{Astronomy \& Astrophysics}{495}{1}{353-358}.
\newblock
\begin{APACrefDOI} \doi{10.1051/0004-6361:200810460} \end{APACrefDOI}
\PrintBackRefs{\CurrentBib}

\bibitem [\protect \citeauthoryear {%
{Baggaley}%
}{%
{Baggaley}%
}{%
{\protect \APACyear {1976}}%
}]{%
Baggaley1976}
\APACinsertmetastar {%
Baggaley1976}%
\begin{APACrefauthors}%
{Baggaley}, W\BPBI J.%
\end{APACrefauthors}%
\unskip\
\newblock
\APACrefYearMonthDay{1976}{}{}.
\newblock
{\BBOQ}\APACrefatitle {{The role of the oxides in meteoric species as a source of meteor train luminosity}} {{The role of the oxides in meteoric species as a source of meteor train luminosity}}.{\BBCQ}
\newblock
\APACjournalVolNumPages{Monthly Notices of the Royal Astronomical Society}{174}{}{617-620}.
\newblock
\begin{APACrefDOI} \doi{10.1093/mnras/174.3.617} \end{APACrefDOI}
\PrintBackRefs{\CurrentBib}

\bibitem [\protect \citeauthoryear {%
{Baggaley}%
}{%
{Baggaley}%
}{%
{\protect \APACyear {1978}}%
}]{%
Baggaley1978}
\APACinsertmetastar {%
Baggaley1978}%
\begin{APACrefauthors}%
{Baggaley}, W\BPBI J.%
\end{APACrefauthors}%
\unskip\
\newblock
\APACrefYearMonthDay{1978}{}{}.
\newblock
{\BBOQ}\APACrefatitle {{Meteor magnitudes and enduring trains}} {{Meteor magnitudes and enduring trains}}.{\BBCQ}
\newblock
\APACjournalVolNumPages{Observatory}{98}{}{8-11}.
\PrintBackRefs{\CurrentBib}

\bibitem [\protect \citeauthoryear {%
Bannister%
, Boucheron%
\BCBL {}\ \BBA {} Voelz%
}{%
Bannister%
\ \protect \BOthers {.}}{%
{\protect \APACyear {2013}}%
}]{%
Bannister2013}
\APACinsertmetastar {%
Bannister2013}%
\begin{APACrefauthors}%
Bannister, S\BPBI M.%
, Boucheron, L\BPBI E.%
\BCBL {}\ \BBA {} Voelz, D\BPBI G.%
\end{APACrefauthors}%
\unskip\
\newblock
\APACrefYearMonthDay{2013}{}{}.
\newblock
{\BBOQ}\APACrefatitle {A Numerical Analysis of a Frame Calibration Method for Video-based All-Sky Camera Systems} {A numerical analysis of a frame calibration method for video-based all-sky camera systems}.{\BBCQ}
\newblock
\APACjournalVolNumPages{Publications of the Astronomical Society of the Pacific}{125}{931}{1108--1118}.
\newblock
\begin{APACrefDOI} \doi{10.1086/673167} \end{APACrefDOI}
\PrintBackRefs{\CurrentBib}

\bibitem [\protect \citeauthoryear {%
{Beech}%
}{%
{Beech}%
}{%
{\protect \APACyear {1987}}%
}]{%
Beech1987}
\APACinsertmetastar {%
Beech1987}%
\begin{APACrefauthors}%
{Beech}, M.%
\end{APACrefauthors}%
\unskip\
\newblock
\APACrefYearMonthDay{1987}{}{}.
\newblock
{\BBOQ}\APACrefatitle {{On the trail of meteor trains}} {{On the trail of meteor trains}}.{\BBCQ}
\newblock
\APACjournalVolNumPages{Quarterly Journal of the Royal Astronomical Society}{28}{}{445}.
\PrintBackRefs{\CurrentBib}

\bibitem [\protect \citeauthoryear {%
Bettonvil%
}{%
Bettonvil%
}{%
{\protect \APACyear {2005}}%
}]{%
Bettonvil2005}
\APACinsertmetastar {%
Bettonvil2005}%
\begin{APACrefauthors}%
Bettonvil, F.%
\end{APACrefauthors}%
\unskip\
\newblock
\APACrefYearMonthDay{2005}{}{}.
\newblock
{\BBOQ}\APACrefatitle {{Fisheye lenses}} {{Fisheye lenses}}.{\BBCQ}
\newblock
\APACjournalVolNumPages{WGN, Journal of the International Meteor Organization}{33}{1}{9-14}.
\PrintBackRefs{\CurrentBib}

\bibitem [\protect \citeauthoryear {%
{Borovi{\v{c}}ka}%
}{%
{Borovi{\v{c}}ka}%
}{%
{\protect \APACyear {2006}}%
}]{%
Borovicka2006}
\APACinsertmetastar {%
Borovicka2006}%
\begin{APACrefauthors}%
{Borovi{\v{c}}ka}, J.%
\end{APACrefauthors}%
\unskip\
\newblock
\APACrefYearMonthDay{2006}{}{}.
\newblock
{\BBOQ}\APACrefatitle {{Meteor trains - Terminology and physical interpretation}} {{Meteor trains - Terminology and physical interpretation}}.{\BBCQ}
\newblock
\APACjournalVolNumPages{Journal of the Royal Astronomical Society of Canada}{100}{}{194}.
\PrintBackRefs{\CurrentBib}

\bibitem [\protect \citeauthoryear {%
Borovička%
\ \BBA {} Jenniskens%
}{%
Borovička%
\ \BBA {} Jenniskens%
}{%
{\protect \APACyear {2000}}%
}]{%
Borovicka2000}
\APACinsertmetastar {%
Borovicka2000}%
\begin{APACrefauthors}%
Borovička, J.%
\BCBT {}\ \BBA {} Jenniskens, P.%
\end{APACrefauthors}%
\unskip\
\newblock
\APACrefYearMonthDay{2000}{}{}.
\newblock
{\BBOQ}\APACrefatitle {{Time resolved spectroscopy of a Leonid fireball afterglow}} {{Time resolved spectroscopy of a Leonid fireball afterglow}}.{\BBCQ}
\newblock
\APACjournalVolNumPages{Earth, Moon, and Planets}{82-83}{}{399-428}.
\newblock
\begin{APACrefDOI} \doi{10.1023/A:1017071524899} \end{APACrefDOI}
\PrintBackRefs{\CurrentBib}

\bibitem [\protect \citeauthoryear {%
Borovička%
\ \BBA {} Koten%
}{%
Borovička%
\ \BBA {} Koten%
}{%
{\protect \APACyear {2003}}%
}]{%
Borovicka2003}
\APACinsertmetastar {%
Borovicka2003}%
\begin{APACrefauthors}%
Borovička, J.%
\BCBT {}\ \BBA {} Koten, P.%
\end{APACrefauthors}%
\unskip\
\newblock
\APACrefYearMonthDay{2003}{}{}.
\newblock
{\BBOQ}\APACrefatitle {{Three phases in the evolution of Leonid meteor trains}} {{Three phases in the evolution of Leonid meteor trains}}.{\BBCQ}
\newblock
\BIn{} H.~Yano, S.~Abe\BCBL {}\ \BBA {} M.~Yoshikawa\ (\BEDS), \APACrefbtitle {{ISAS report SP: Proceedings of the 2002 International Science Symposium on the Leonid meteor storms}} {{ISAS report SP: Proceedings of the 2002 International Science Symposium on the Leonid meteor storms}}\ (\BVOL~15, \BPG~165-173).
\newblock
\APACaddressPublisher{}{{Institute of Space and Astronautical Science}}.
\PrintBackRefs{\CurrentBib}

\bibitem [\protect \citeauthoryear {%
Buccongello%
, Brown%
, Vida%
\BCBL {}\ \BBA {} Pinhas%
}{%
Buccongello%
\ \protect \BOthers {.}}{%
{\protect \APACyear {2024}}%
}]{%
Buccongello2024}
\APACinsertmetastar {%
Buccongello2024}%
\begin{APACrefauthors}%
Buccongello, N.%
, Brown, P.%
, Vida, D.%
\BCBL {}\ \BBA {} Pinhas, A.%
\end{APACrefauthors}%
\unskip\
\newblock
\APACrefYearMonthDay{2024}{}{}.
\newblock
{\BBOQ}\APACrefatitle {{A physical survey of meteoroid streams: Comparing cometary reservoirs}} {{A physical survey of meteoroid streams: Comparing cometary reservoirs}}.{\BBCQ}
\newblock
\APACjournalVolNumPages{Icarus}{410}{}{115907}.
\newblock
\begin{APACrefDOI} \doi{10.1016/j.icarus.2023.115907} \end{APACrefDOI}
\PrintBackRefs{\CurrentBib}

\bibitem [\protect \citeauthoryear {%
{Ceplecha}%
}{%
{Ceplecha}%
}{%
{\protect \APACyear {1958}}%
}]{%
Ceplecha1958}
\APACinsertmetastar {%
Ceplecha1958}%
\begin{APACrefauthors}%
{Ceplecha}, Z.%
\end{APACrefauthors}%
\unskip\
\newblock
\APACrefYearMonthDay{1958}{}{}.
\newblock
{\BBOQ}\APACrefatitle {{On the composition of meteors}} {{On the composition of meteors}}.{\BBCQ}
\newblock
\APACjournalVolNumPages{Bulletin of the Astronomical Institutes of Czechoslovakia}{9}{}{154}.
\PrintBackRefs{\CurrentBib}

\bibitem [\protect \citeauthoryear {%
{Ceplecha}%
}{%
{Ceplecha}%
}{%
{\protect \APACyear {1967}}%
}]{%
Ceplecha1967}
\APACinsertmetastar {%
Ceplecha1967}%
\begin{APACrefauthors}%
{Ceplecha}, Z.%
\end{APACrefauthors}%
\unskip\
\newblock
\APACrefYearMonthDay{1967}{}{}.
\newblock
{\BBOQ}\APACrefatitle {{Classification of meteor orbits}} {{Classification of meteor orbits}}.{\BBCQ}
\newblock
\APACjournalVolNumPages{Smithsonian Contributions to Astrophysics}{11}{}{35}.
\PrintBackRefs{\CurrentBib}

\bibitem [\protect \citeauthoryear {%
{Ceplecha}%
}{%
{Ceplecha}%
}{%
{\protect \APACyear {1988}}%
}]{%
Ceplecha1988}
\APACinsertmetastar {%
Ceplecha1988}%
\begin{APACrefauthors}%
{Ceplecha}, Z.%
\end{APACrefauthors}%
\unskip\
\newblock
\APACrefYearMonthDay{1988}{}{}.
\newblock
{\BBOQ}\APACrefatitle {{Earth's influx of different populations of sporadic meteoroids from photographic and television data}} {{Earth's influx of different populations of sporadic meteoroids from photographic and television data}}.{\BBCQ}
\newblock
\APACjournalVolNumPages{Bulletin of the Astronomical Institutes of Czechoslovakia}{39}{}{221}.
\PrintBackRefs{\CurrentBib}

\bibitem [\protect \citeauthoryear {%
Chapman%
}{%
Chapman%
}{%
{\protect \APACyear {1939}}%
}]{%
Chapman1939}
\APACinsertmetastar {%
Chapman1939}%
\begin{APACrefauthors}%
Chapman, S.%
\end{APACrefauthors}%
\unskip\
\newblock
\APACrefYearMonthDay{1939}{}{}.
\newblock
{\BBOQ}\APACrefatitle {Notes on atmospheric sodium} {Notes on atmospheric sodium}.{\BBCQ}
\newblock
\APACjournalVolNumPages{The Astrophysical Journal}{90}{}{309-316}.
\newblock
\begin{APACrefDOI} \doi{10.1086/144109} \end{APACrefDOI}
\PrintBackRefs{\CurrentBib}

\bibitem [\protect \citeauthoryear {%
Chapman%
}{%
Chapman%
}{%
{\protect \APACyear {1955}}%
}]{%
Chapman1955}
\APACinsertmetastar {%
Chapman1955}%
\begin{APACrefauthors}%
Chapman, S.%
\end{APACrefauthors}%
\unskip\
\newblock
\APACrefYearMonthDay{1955}{}{}.
\newblock
{\BBOQ}\APACrefatitle {Note on persistent meteor trains} {Note on persistent meteor trains}.{\BBCQ}
\newblock
\BIn{} E\BPBI B.~Armstrong\ \BBA {} A.~Dalgarno\ (\BEDS), \APACrefbtitle {The airglow and aurora} {The airglow and aurora}\ (\BPG~204-205).
\newblock
\APACaddressPublisher{New York, NY}{Pergamon Press}.
\PrintBackRefs{\CurrentBib}

\bibitem [\protect \citeauthoryear {%
Clemesha%
\ \protect \BOthers {.}}{%
Clemesha%
\ \protect \BOthers {.}}{%
{\protect \APACyear {2001}}%
}]{%
Clemesha2001}
\APACinsertmetastar {%
Clemesha2001}%
\begin{APACrefauthors}%
Clemesha, B\BPBI R.%
, de Medeiros, A\BPBI F.%
, Gobbi, D.%
, Takahashi, H.%
, Batista, P\BPBI P.%
\BCBL {}\ \BBA {} Taylor, M\BPBI J.%
\end{APACrefauthors}%
\unskip\
\newblock
\APACrefYearMonthDay{2001}{}{}.
\newblock
{\BBOQ}\APACrefatitle {Multiple wavelength optical observations of a long-lived meteor trail} {Multiple wavelength optical observations of a long-lived meteor trail}.{\BBCQ}
\newblock
\APACjournalVolNumPages{Geophysical Research Letters}{28}{14}{2779-2782}.
\newblock
\begin{APACrefDOI} \doi{10.1029/2000GL012605} \end{APACrefDOI}
\PrintBackRefs{\CurrentBib}

\bibitem [\protect \citeauthoryear {%
Cordonnier%
, Obenberger%
, Holmes%
\BCBL {}\ \BBA {} Taylor%
}{%
Cordonnier%
\ \protect \BOthers {.}}{%
{\protect \APACyear {2023}}%
{\protect \APACexlab {{\protect \BCnt {1}}}}}]{%
CordonnierTables}
\APACinsertmetastar {%
CordonnierTables}%
\begin{APACrefauthors}%
Cordonnier, L\BPBI E.%
, Obenberger, K\BPBI S.%
, Holmes, J\BPBI M.%
\BCBL {}\ \BBA {} Taylor, G\BPBI B.%
\end{APACrefauthors}%
\unskip\
\newblock
\APACrefYearMonthDay{2023{\protect \BCnt {1}}}{}{}.
\newblock
\APACrefbtitle {{Persistent train and associated meteor data}} {{Persistent train and associated meteor data}}\ [dataset].
\newblock
\APACaddressPublisher{}{LWA Data Archive}.
\newblock
\begin{APACrefURL} \url{https://lda10g.alliance.unm.edu/~pasi/PTs/PT_tables/} \end{APACrefURL}
\PrintBackRefs{\CurrentBib}

\bibitem [\protect \citeauthoryear {%
Cordonnier%
, Obenberger%
, Holmes%
\BCBL {}\ \BBA {} Taylor%
}{%
Cordonnier%
\ \protect \BOthers {.}}{%
{\protect \APACyear {2023}}%
{\protect \APACexlab {{\protect \BCnt {2}}}}}]{%
CordonnierImages}
\APACinsertmetastar {%
CordonnierImages}%
\begin{APACrefauthors}%
Cordonnier, L\BPBI E.%
, Obenberger, K\BPBI S.%
, Holmes, J\BPBI M.%
\BCBL {}\ \BBA {} Taylor, G\BPBI B.%
\end{APACrefauthors}%
\unskip\
\newblock
\APACrefYearMonthDay{2023{\protect \BCnt {2}}}{}{}.
\newblock
\APACrefbtitle {{Persistent train images from the WiPT2 camera}} {{Persistent train images from the WiPT2 camera}}\ [dataset].
\newblock
\APACaddressPublisher{}{LWA Data Archive}.
\newblock
\begin{APACrefURL} \url{https://lda10g.alliance.unm.edu/~pasi/PTs/PT_images/} \end{APACrefURL}
\PrintBackRefs{\CurrentBib}

\bibitem [\protect \citeauthoryear {%
Dijkema%
\ \protect \BOthers {.}}{%
Dijkema%
\ \protect \BOthers {.}}{%
{\protect \APACyear {2021}}%
}]{%
Dijkema2021}
\APACinsertmetastar {%
Dijkema2021}%
\begin{APACrefauthors}%
Dijkema, T.%
, Bassa, C.%
, Kuiack, M.%
, Jenniskens, P.%
, Johannink, C.%
, Bettonvil, F.%
\BDBL {}Fallows, R.%
\end{APACrefauthors}%
\unskip\
\newblock
\APACrefYearMonthDay{2021}{}{}.
\newblock
{\BBOQ}\APACrefatitle {{Simultaneous broadband radio and optical emission of meteor trains imaged by LOFAR/AARTFAAC and CAMS}} {{Simultaneous broadband radio and optical emission of meteor trains imaged by LOFAR/AARTFAAC and CAMS}}.{\BBCQ}
\newblock
\APACjournalVolNumPages{WGN, Journal of the International Meteor Organization}{49}{5}{137-141}.
\PrintBackRefs{\CurrentBib}

\bibitem [\protect \citeauthoryear {%
Egal%
, Wiegert%
, Brown%
\BCBL {}\ \BBA {} Vida%
}{%
Egal%
\ \protect \BOthers {.}}{%
{\protect \APACyear {2023}}%
}]{%
Egal2023}
\APACinsertmetastar {%
Egal2023}%
\begin{APACrefauthors}%
Egal, A.%
, Wiegert, P\BPBI A.%
, Brown, P\BPBI G.%
\BCBL {}\ \BBA {} Vida, D.%
\end{APACrefauthors}%
\unskip\
\newblock
\APACrefYearMonthDay{2023}{}{}.
\newblock
{\BBOQ}\APACrefatitle {{Modeling the 2022 $\tau$-Herculid outburst}} {{Modeling the 2022 $\tau$-Herculid outburst}}.{\BBCQ}
\newblock
\APACjournalVolNumPages{The Astrophysical Journal}{949}{2}{96}.
\newblock
\begin{APACrefDOI} \doi{10.3847/1538-4357/acb93a} \end{APACrefDOI}
\PrintBackRefs{\CurrentBib}

\bibitem [\protect \citeauthoryear {%
Emmert%
\ \protect \BOthers {.}}{%
Emmert%
\ \protect \BOthers {.}}{%
{\protect \APACyear {2022}}%
}]{%
Emmert2022}
\APACinsertmetastar {%
Emmert2022}%
\begin{APACrefauthors}%
Emmert, J\BPBI T.%
, Jones~Jr, M.%
, Siskind, D\BPBI E.%
, Drob, D\BPBI P.%
, Picone, J\BPBI M.%
, Stevens, M\BPBI H.%
\BDBL {}Pérot, K.%
\end{APACrefauthors}%
\unskip\
\newblock
\APACrefYearMonthDay{2022}{}{}.
\newblock
{\BBOQ}\APACrefatitle {{NRLMSIS 2.1: An empirical model of nitric oxide incorporated into MSIS}} {{NRLMSIS 2.1: An empirical model of nitric oxide incorporated into MSIS}}.{\BBCQ}
\newblock
\APACjournalVolNumPages{Journal of Geophysical Research: Space Physics}{127}{10}{e2022JA030896}.
\newblock
\begin{APACrefDOI} \doi{10.1029/2022JA030896} \end{APACrefDOI}
\PrintBackRefs{\CurrentBib}

\bibitem [\protect \citeauthoryear {%
Friedman%
}{%
Friedman%
}{%
{\protect \APACyear {2001}}%
}]{%
Friedman2001}
\APACinsertmetastar {%
Friedman2001}%
\begin{APACrefauthors}%
Friedman, J\BPBI H.%
\end{APACrefauthors}%
\unskip\
\newblock
\APACrefYearMonthDay{2001}{}{}.
\newblock
{\BBOQ}\APACrefatitle {{Greedy function approximation: A gradient boosting machine.}} {{Greedy function approximation: A gradient boosting machine.}}{\BBCQ}
\newblock
\APACjournalVolNumPages{The Annals of Statistics}{29}{5}{1189-1232}.
\newblock
\begin{APACrefDOI} \doi{10.1214/aos/1013203451} \end{APACrefDOI}
\PrintBackRefs{\CurrentBib}

\bibitem [\protect \citeauthoryear {%
{Gritsevich}%
}{%
{Gritsevich}%
}{%
{\protect \APACyear {2008}}%
}]{%
Gritsevich2008}
\APACinsertmetastar {%
Gritsevich2008}%
\begin{APACrefauthors}%
{Gritsevich}, M\BPBI I.%
\end{APACrefauthors}%
\unskip\
\newblock
\APACrefYearMonthDay{2008}{}{}.
\newblock
{\BBOQ}\APACrefatitle {{Validity of the photometric formula for estimating the mass of a fireball projectile}} {{Validity of the photometric formula for estimating the mass of a fireball projectile}}.{\BBCQ}
\newblock
\APACjournalVolNumPages{Doklady Physics}{53}{2}{97-102}.
\newblock
\begin{APACrefDOI} \doi{10.1134/S1028335808020110} \end{APACrefDOI}
\PrintBackRefs{\CurrentBib}

\bibitem [\protect \citeauthoryear {%
{Hapgood}%
}{%
{Hapgood}%
}{%
{\protect \APACyear {1980}}%
}]{%
Hapgood1980}
\APACinsertmetastar {%
Hapgood1980}%
\begin{APACrefauthors}%
{Hapgood}, M\BPBI A.%
\end{APACrefauthors}%
\unskip\
\newblock
\APACrefYearMonthDay{1980}{}{}.
\newblock
{\BBOQ}\APACrefatitle {{IR observation of a persistent meteor train}} {{IR observation of a persistent meteor train}}.{\BBCQ}
\newblock
\APACjournalVolNumPages{Nature}{286}{5773}{582-583}.
\newblock
\begin{APACrefDOI} \doi{10.1038/286582a0} \end{APACrefDOI}
\PrintBackRefs{\CurrentBib}

\bibitem [\protect \citeauthoryear {%
{Hawkins}%
\ \BBA {} {Howard}%
}{%
{Hawkins}%
\ \BBA {} {Howard}%
}{%
{\protect \APACyear {1959}}%
}]{%
Hawkins1959}
\APACinsertmetastar {%
Hawkins1959}%
\begin{APACrefauthors}%
{Hawkins}, G\BPBI S.%
\BCBT {}\ \BBA {} {Howard}, W\BPBI E.%
\end{APACrefauthors}%
\unskip\
\newblock
\APACrefYearMonthDay{1959}{}{}.
\newblock
{\BBOQ}\APACrefatitle {{Decay of light from a meteor train}} {{Decay of light from a meteor train}}.{\BBCQ}
\newblock
\APACjournalVolNumPages{The Astrophysical Journal}{130}{}{1003-1007}.
\newblock
\begin{APACrefDOI} \doi{10.1086/146789} \end{APACrefDOI}
\PrintBackRefs{\CurrentBib}

\bibitem [\protect \citeauthoryear {%
{Higa}%
, {Yamamoto}%
, {Toda}%
, {Maeda}%
\BCBL {}\ \BBA {} {Watanabe}%
}{%
{Higa}%
\ \protect \BOthers {.}}{%
{\protect \APACyear {2005}}%
}]{%
Higa2005}
\APACinsertmetastar {%
Higa2005}%
\begin{APACrefauthors}%
{Higa}, Y.%
, {Yamamoto}, M\BHBI Y.%
, {Toda}, M.%
, {Maeda}, K.%
\BCBL {}\ \BBA {} {Watanabe}, J\BHBI I.%
\end{APACrefauthors}%
\unskip\
\newblock
\APACrefYearMonthDay{2005}{}{}.
\newblock
{\BBOQ}\APACrefatitle {{Catalogue of persistent trains II: Images of Leonid meteor trains during the METRO campaign 1998-2002}} {{Catalogue of persistent trains II: Images of Leonid meteor trains during the METRO campaign 1998-2002}}.{\BBCQ}
\newblock
\APACjournalVolNumPages{Publications of the National Astronomical Observatory of Japan}{7}{4}{67-131}.
\PrintBackRefs{\CurrentBib}

\bibitem [\protect \citeauthoryear {%
Hirsch%
}{%
Hirsch%
}{%
{\protect \APACyear {2018}}%
}]{%
Hirsch2018}
\APACinsertmetastar {%
Hirsch2018}%
\begin{APACrefauthors}%
Hirsch, M.%
\end{APACrefauthors}%
\unskip\
\newblock
\APACrefYearMonthDay{2018}{}{}.
\newblock
{\BBOQ}\APACrefatitle {{PyMap3D: 3-D coordinate conversions for terrestrial and geospace environments}} {{PyMap3D: 3-D coordinate conversions for terrestrial and geospace environments}}.{\BBCQ}
\newblock
\APACjournalVolNumPages{Journal of Open Source Software}{3}{23}{580}.
\newblock
\begin{APACrefDOI} \doi{10.21105/joss.00580} \end{APACrefDOI}
\PrintBackRefs{\CurrentBib}

\bibitem [\protect \citeauthoryear {%
Hunter%
}{%
Hunter%
}{%
{\protect \APACyear {2007}}%
}]{%
Hunter2007}
\APACinsertmetastar {%
Hunter2007}%
\begin{APACrefauthors}%
Hunter, J\BPBI D.%
\end{APACrefauthors}%
\unskip\
\newblock
\APACrefYearMonthDay{2007}{}{}.
\newblock
{\BBOQ}\APACrefatitle {{Matplotlib: A 2D graphics environment}} {{Matplotlib: A 2D graphics environment}}.{\BBCQ}
\newblock
\APACjournalVolNumPages{Computing in Science \& Engineering}{9}{3}{90--95}.
\newblock
\begin{APACrefDOI} \doi{10.1109/MCSE.2007.55} \end{APACrefDOI}
\PrintBackRefs{\CurrentBib}

\bibitem [\protect \citeauthoryear {%
Jenniskens%
}{%
Jenniskens%
}{%
{\protect \APACyear {2003}}%
}]{%
Jenniskens2003}
\APACinsertmetastar {%
Jenniskens2003}%
\begin{APACrefauthors}%
Jenniskens, P.%
\end{APACrefauthors}%
\unskip\
\newblock
\APACrefYearMonthDay{2003}{}{}.
\newblock
{\BBOQ}\APACrefatitle {{Morphology of persistent trains is due to fragmentation}} {{Morphology of persistent trains is due to fragmentation}}.{\BBCQ}
\newblock
\APACjournalVolNumPages{WGN, Journal of the International Meteor Organization}{31}{}{88-92}.
\PrintBackRefs{\CurrentBib}

\bibitem [\protect \citeauthoryear {%
Jenniskens%
}{%
Jenniskens%
}{%
{\protect \APACyear {2006}}%
}]{%
JenniskensBook}
\APACinsertmetastar {%
JenniskensBook}%
\begin{APACrefauthors}%
Jenniskens, P.%
\end{APACrefauthors}%
\unskip\
\newblock
\APACrefYear{2006}.
\newblock
\APACrefbtitle {{Meteor showers and their parent comets}} {{Meteor showers and their parent comets}}.
\newblock
\APACaddressPublisher{}{Cambridge University Press}.
\newblock
\begin{APACrefDOI} \doi{10.1017/CBO9781316257104} \end{APACrefDOI}
\PrintBackRefs{\CurrentBib}

\bibitem [\protect \citeauthoryear {%
Jenniskens%
\ \BBA {} Butow%
}{%
Jenniskens%
\ \BBA {} Butow%
}{%
{\protect \APACyear {1999}}%
}]{%
Jenniskens1999}
\APACinsertmetastar {%
Jenniskens1999}%
\begin{APACrefauthors}%
Jenniskens, P.%
\BCBT {}\ \BBA {} Butow, S\BPBI J.%
\end{APACrefauthors}%
\unskip\
\newblock
\APACrefYearMonthDay{1999}{}{}.
\newblock
{\BBOQ}\APACrefatitle {{The 1998 Leonid multi-instrument aircraft campaign—an early review}} {{The 1998 Leonid multi-instrument aircraft campaign—an early review}}.{\BBCQ}
\newblock
\APACjournalVolNumPages{Meteoritics \& Planetary Science}{34}{6}{933-943}.
\newblock
\begin{APACrefDOI} \doi{10.1111/j.1945-5100.1999.tb01411.x} \end{APACrefDOI}
\PrintBackRefs{\CurrentBib}

\bibitem [\protect \citeauthoryear {%
Jenniskens%
, Lacey%
, Allan%
, Self%
\BCBL {}\ \BBA {} Plane%
}{%
Jenniskens%
\ \protect \BOthers {.}}{%
{\protect \APACyear {1998}}%
}]{%
Jenniskens1998}
\APACinsertmetastar {%
Jenniskens1998}%
\begin{APACrefauthors}%
Jenniskens, P.%
, Lacey, M.%
, Allan, B.%
, Self, D.%
\BCBL {}\ \BBA {} Plane, J.%
\end{APACrefauthors}%
\unskip\
\newblock
\APACrefYearMonthDay{1998}{}{}.
\newblock
{\BBOQ}\APACrefatitle {{FeO ``orange arc'' emission detected in optical spectrum of Leonid persistent train}} {{FeO ``orange arc'' emission detected in optical spectrum of Leonid persistent train}}.{\BBCQ}
\newblock
\APACjournalVolNumPages{Earth, Moon, and Planets}{82-83}{}{429-438}.
\newblock
\begin{APACrefDOI} \doi{10.1023/A:1017079725808} \end{APACrefDOI}
\PrintBackRefs{\CurrentBib}

\bibitem [\protect \citeauthoryear {%
Jenniskens%
\ \BBA {} Moskovitz%
}{%
Jenniskens%
\ \BBA {} Moskovitz%
}{%
{\protect \APACyear {2022}}%
}]{%
Jenniskens2022}
\APACinsertmetastar {%
Jenniskens2022}%
\begin{APACrefauthors}%
Jenniskens, P.%
\BCBT {}\ \BBA {} Moskovitz, N.%
\end{APACrefauthors}%
\unskip\
\newblock
\APACrefYearMonthDay{2022}{}{}.
\newblock
{\BBOQ}\APACrefatitle {{An outburst of Andromedids on November 28, 2021}} {{An outburst of Andromedids on November 28, 2021}}.{\BBCQ}
\newblock
\APACjournalVolNumPages{eMeteorNews}{7}{1}{3-5}.
\PrintBackRefs{\CurrentBib}

\bibitem [\protect \citeauthoryear {%
{Jewitt}%
, {Hsieh}%
\BCBL {}\ \BBA {} {Agarwal}%
}{%
{Jewitt}%
\ \protect \BOthers {.}}{%
{\protect \APACyear {2015}}%
}]{%
Jewitt2015}
\APACinsertmetastar {%
Jewitt2015}%
\begin{APACrefauthors}%
{Jewitt}, D.%
, {Hsieh}, H.%
\BCBL {}\ \BBA {} {Agarwal}, J.%
\end{APACrefauthors}%
\unskip\
\newblock
\APACrefYearMonthDay{2015}{}{}.
\newblock
{\BBOQ}\APACrefatitle {{The active asteroids}} {{The active asteroids}}.{\BBCQ}
\newblock
\BIn{} P.~Michel, F\BPBI E.~DeMeo\BCBL {}\ \BBA {} W\BPBI F.~Bottke\ (\BEDS), \APACrefbtitle {{Asteroids IV}} {{Asteroids IV}}\ (\BPG~221-241).
\newblock
\APACaddressPublisher{}{University of Arizona Press}.
\newblock
\begin{APACrefDOI} \doi{10.2458/azu_uapress_9780816532131-ch012} \end{APACrefDOI}
\PrintBackRefs{\CurrentBib}

\bibitem [\protect \citeauthoryear {%
Kelley%
\ \protect \BOthers {.}}{%
Kelley%
\ \protect \BOthers {.}}{%
{\protect \APACyear {2000}}%
}]{%
Kelley2000}
\APACinsertmetastar {%
Kelley2000}%
\begin{APACrefauthors}%
Kelley, M\BPBI C.%
, Gardner, C.%
, Drummond, J.%
, Armstrong, T.%
, Liu, A.%
, Chu, X.%
\BDBL {}Engelman, J.%
\end{APACrefauthors}%
\unskip\
\newblock
\APACrefYearMonthDay{2000}{}{}.
\newblock
{\BBOQ}\APACrefatitle {First observations of long-lived meteor trains with resonance lidar and other optical instruments} {First observations of long-lived meteor trains with resonance lidar and other optical instruments}.{\BBCQ}
\newblock
\APACjournalVolNumPages{Geophysical Research Letters}{27}{13}{1811-1814}.
\newblock
\begin{APACrefDOI} \doi{10.1029/1999GL011175} \end{APACrefDOI}
\PrintBackRefs{\CurrentBib}

\bibitem [\protect \citeauthoryear {%
Kelley%
, Williamson%
\BCBL {}\ \BBA {} Vlasov%
}{%
Kelley%
\ \protect \BOthers {.}}{%
{\protect \APACyear {2013}}%
}]{%
Kelley2013}
\APACinsertmetastar {%
Kelley2013}%
\begin{APACrefauthors}%
Kelley, M\BPBI C.%
, Williamson, C\BPBI H\BPBI K.%
\BCBL {}\ \BBA {} Vlasov, M\BPBI N.%
\end{APACrefauthors}%
\unskip\
\newblock
\APACrefYearMonthDay{2013}{}{}.
\newblock
{\BBOQ}\APACrefatitle {Double laminar and turbulent meteor trails observed in space and simulated in the laboratory} {Double laminar and turbulent meteor trails observed in space and simulated in the laboratory}.{\BBCQ}
\newblock
\APACjournalVolNumPages{Journal of Geophysical Research: Space Physics}{118}{6}{3622-3625}.
\newblock
\begin{APACrefDOI} \doi{10.1002/jgra.50339} \end{APACrefDOI}
\PrintBackRefs{\CurrentBib}

\bibitem [\protect \citeauthoryear {%
Kikwaya%
, Campbell-Brown%
\BCBL {}\ \BBA {} Brown%
}{%
Kikwaya%
\ \protect \BOthers {.}}{%
{\protect \APACyear {2011}}%
}]{%
Kikwaya2011}
\APACinsertmetastar {%
Kikwaya2011}%
\begin{APACrefauthors}%
Kikwaya, J\BHBI B.%
, Campbell-Brown, M.%
\BCBL {}\ \BBA {} Brown, P\BPBI G.%
\end{APACrefauthors}%
\unskip\
\newblock
\APACrefYearMonthDay{2011}{}{}.
\newblock
{\BBOQ}\APACrefatitle {Bulk density of small meteoroids} {Bulk density of small meteoroids}.{\BBCQ}
\newblock
\APACjournalVolNumPages{Astronomy \& Astrophysics}{530}{}{A113}.
\newblock
\begin{APACrefDOI} \doi{10.1051/0004-6361/201116431} \end{APACrefDOI}
\PrintBackRefs{\CurrentBib}

\bibitem [\protect \citeauthoryear {%
{Koschny}%
\ \BBA {} {Borovička}%
}{%
{Koschny}%
\ \BBA {} {Borovička}%
}{%
{\protect \APACyear {2017}}%
}]{%
Koschny2017}
\APACinsertmetastar {%
Koschny2017}%
\begin{APACrefauthors}%
{Koschny}, D.%
\BCBT {}\ \BBA {} {Borovička}, J.%
\end{APACrefauthors}%
\unskip\
\newblock
\APACrefYearMonthDay{2017}{}{}.
\newblock
{\BBOQ}\APACrefatitle {{Definitions of terms in meteor astronomy}} {{Definitions of terms in meteor astronomy}}.{\BBCQ}
\newblock
\APACjournalVolNumPages{WGN, Journal of the International Meteor Organization}{45}{5}{91-92}.
\PrintBackRefs{\CurrentBib}

\bibitem [\protect \citeauthoryear {%
{Koten, P.}%
\ \protect \BOthers {.}}{%
{Koten, P.}%
\ \protect \BOthers {.}}{%
{\protect \APACyear {2023}}%
}]{%
Koten2023}
\APACinsertmetastar {%
Koten2023}%
\begin{APACrefauthors}%
{Koten, P.}%
, {Shrben\'y, L.}%
, {Spurn\'y, P.}%
, {Borovička, J.}%
, {Stork, R.}%
, {Henych, T.}%
\BDBL {}{M\'anek, Jan}%
\end{APACrefauthors}%
\unskip\
\newblock
\APACrefYearMonthDay{2023}{}{}.
\newblock
{\BBOQ}\APACrefatitle {{$\tau$ Herculid meteor shower in the night of 30/31 May 2022 and the meteoroid properties}} {{$\tau$ Herculid meteor shower in the night of 30/31 May 2022 and the meteoroid properties}}.{\BBCQ}
\newblock
\APACjournalVolNumPages{Astronomy \& Astrophysics}{675}{}{A70}.
\newblock
\begin{APACrefDOI} \doi{10.1051/0004-6361/202346537} \end{APACrefDOI}
\PrintBackRefs{\CurrentBib}

\bibitem [\protect \citeauthoryear {%
{Kres{\'a}k}%
}{%
{Kres{\'a}k}%
}{%
{\protect \APACyear {1949}}%
}]{%
Kresak1949}
\APACinsertmetastar {%
Kresak1949}%
\begin{APACrefauthors}%
{Kres{\'a}k}, L.%
\end{APACrefauthors}%
\unskip\
\newblock
\APACrefYearMonthDay{1949}{}{}.
\newblock
{\BBOQ}\APACrefatitle {{On the connection between long-enduring meteor trains and changes of solar activity}} {{On the connection between long-enduring meteor trains and changes of solar activity}}.{\BBCQ}
\newblock
\APACjournalVolNumPages{Bulletin of the Astronomical Institutes of Czechoslovakia}{1}{}{87}.
\PrintBackRefs{\CurrentBib}

\bibitem [\protect \citeauthoryear {%
Kronk%
}{%
Kronk%
}{%
{\protect \APACyear {2014}}%
}]{%
Kronk2014}
\APACinsertmetastar {%
Kronk2014}%
\begin{APACrefauthors}%
Kronk, G\BPBI W.%
\end{APACrefauthors}%
\unskip\
\newblock
\APACrefYear{2014}.
\newblock
\APACrefbtitle {{Meteor showers: An annotated catalog}} {{Meteor showers: An annotated catalog}}.
\newblock
\APACaddressPublisher{}{Springer}.
\newblock
\begin{APACrefDOI} \doi{10.1007/978-1-4614-7897-3} \end{APACrefDOI}
\PrintBackRefs{\CurrentBib}

\bibitem [\protect \citeauthoryear {%
Kruschwitz%
\ \protect \BOthers {.}}{%
Kruschwitz%
\ \protect \BOthers {.}}{%
{\protect \APACyear {2001}}%
}]{%
Kruschwitz2001}
\APACinsertmetastar {%
Kruschwitz2001}%
\begin{APACrefauthors}%
Kruschwitz, C\BPBI A.%
, Kelley, M\BPBI C.%
, Gardner, C\BPBI S.%
, Swenson, G.%
, Liu, A\BPBI Z.%
, Chu, X.%
\BDBL {}Jenniskens, P.%
\end{APACrefauthors}%
\unskip\
\newblock
\APACrefYearMonthDay{2001}{}{}.
\newblock
{\BBOQ}\APACrefatitle {{Observations of persistent Leonid meteor trails: 2. Photometry and numerical modeling}} {{Observations of persistent Leonid meteor trails: 2. Photometry and numerical modeling}}.{\BBCQ}
\newblock
\APACjournalVolNumPages{Journal of Geophysical Research: Space Physics}{106}{A10}{21525-21541}.
\newblock
\begin{APACrefDOI} \doi{10.1029/2000JA000174} \end{APACrefDOI}
\PrintBackRefs{\CurrentBib}

\bibitem [\protect \citeauthoryear {%
Lema{{\^i}}tre%
, Nogueira%
\BCBL {}\ \BBA {} Aridas%
}{%
Lema{{\^i}}tre%
\ \protect \BOthers {.}}{%
{\protect \APACyear {2017}}%
}]{%
Guillaume2017}
\APACinsertmetastar {%
Guillaume2017}%
\begin{APACrefauthors}%
Lema{{\^i}}tre, G.%
, Nogueira, F.%
\BCBL {}\ \BBA {} Aridas, C\BPBI K.%
\end{APACrefauthors}%
\unskip\
\newblock
\APACrefYearMonthDay{2017}{}{}.
\newblock
{\BBOQ}\APACrefatitle {{Imbalanced-learn: A Python toolbox to tackle the curse of imbalanced datasets in machine learning}} {{Imbalanced-learn: A Python toolbox to tackle the curse of imbalanced datasets in machine learning}}.{\BBCQ}
\newblock
\APACjournalVolNumPages{Journal of Machine Learning Research}{18}{17}{1-5}.
\PrintBackRefs{\CurrentBib}

\bibitem [\protect \citeauthoryear {%
Lucas%
}{%
Lucas%
}{%
{\protect \APACyear {2023}}%
}]{%
pymsis}
\APACinsertmetastar {%
pymsis}%
\begin{APACrefauthors}%
Lucas, G.%
\end{APACrefauthors}%
\unskip\
\newblock
\APACrefYearMonthDay{2023}{}{}.
\newblock
\APACrefbtitle {{pymsis}} {{pymsis}}\ [software].
\newblock
\APACaddressPublisher{}{Zenodo}.
\newblock
\begin{APACrefDOI} \doi{10.5281/zenodo.8403883} \end{APACrefDOI}
\PrintBackRefs{\CurrentBib}

\bibitem [\protect \citeauthoryear {%
McKinney%
}{%
McKinney%
}{%
{\protect \APACyear {2010}}%
}]{%
pandasBook}
\APACinsertmetastar {%
pandasBook}%
\begin{APACrefauthors}%
McKinney, W.%
\end{APACrefauthors}%
\unskip\
\newblock
\APACrefYearMonthDay{2010}{}{}.
\newblock
{\BBOQ}\APACrefatitle {Data structures for statistical computing in {P}ython} {Data structures for statistical computing in {P}ython}.{\BBCQ}
\newblock
\BIn{} S.~van~der Walt\ \BBA {} J.~Millman\ (\BEDS), \APACrefbtitle {{P}roceedings of the 9th {P}ython in {S}cience {C}onference} {{P}roceedings of the 9th {P}ython in {S}cience {C}onference}\ (\BPG~56-61).
\newblock
\begin{APACrefDOI} \doi{10.25080/Majora-92bf1922-00a} \end{APACrefDOI}
\PrintBackRefs{\CurrentBib}

\bibitem [\protect \citeauthoryear {%
{Moorhead}%
\ \protect \BOthers {.}}{%
{Moorhead}%
\ \protect \BOthers {.}}{%
{\protect \APACyear {2017}}%
}]{%
Moorhead2017}
\APACinsertmetastar {%
Moorhead2017}%
\begin{APACrefauthors}%
{Moorhead}, A\BPBI V.%
, {Blaauw}, R\BPBI C.%
, {Moser}, D\BPBI E.%
, {Campbell-Brown}, M\BPBI D.%
, {Brown}, P\BPBI G.%
\BCBL {}\ \BBA {} {Cooke}, W\BPBI J.%
\end{APACrefauthors}%
\unskip\
\newblock
\APACrefYearMonthDay{2017}{}{}.
\newblock
{\BBOQ}\APACrefatitle {{A two-population sporadic meteoroid bulk density distribution and its implications for environment models}} {{A two-population sporadic meteoroid bulk density distribution and its implications for environment models}}.{\BBCQ}
\newblock
\APACjournalVolNumPages{Monthly Notices of the Royal Astronomical Society}{472}{4}{3833-3841}.
\newblock
\begin{APACrefDOI} \doi{10.1093/mnras/stx2175} \end{APACrefDOI}
\PrintBackRefs{\CurrentBib}

\bibitem [\protect \citeauthoryear {%
{Murad}%
}{%
{Murad}%
}{%
{\protect \APACyear {2001}}%
}]{%
Murad2001}
\APACinsertmetastar {%
Murad2001}%
\begin{APACrefauthors}%
{Murad}, E.%
\end{APACrefauthors}%
\unskip\
\newblock
\APACrefYearMonthDay{2001}{}{}.
\newblock
{\BBOQ}\APACrefatitle {{Heterogeneous chemical processes as a source of persistent meteor trains}} {{Heterogeneous chemical processes as a source of persistent meteor trains}}.{\BBCQ}
\newblock
\APACjournalVolNumPages{Meteoritics \& Planetary Science}{36}{9}{1217-1224}.
\newblock
\begin{APACrefDOI} \doi{10.1111/j.1945-5100.2001.tb01955.x} \end{APACrefDOI}
\PrintBackRefs{\CurrentBib}

\bibitem [\protect \citeauthoryear {%
Obenberger%
\ \protect \BOthers {.}}{%
Obenberger%
\ \protect \BOthers {.}}{%
{\protect \APACyear {2020}}%
}]{%
Obenberger2020}
\APACinsertmetastar {%
Obenberger2020}%
\begin{APACrefauthors}%
Obenberger, K\BPBI S.%
, Holmes, J\BPBI M.%
, Ard, S\BPBI G.%
, Dowell, J.%
, Shuman, N\BPBI S.%
, Taylor, G\BPBI B.%
\BDBL {}Viggiano, A\BPBI A.%
\end{APACrefauthors}%
\unskip\
\newblock
\APACrefYearMonthDay{2020}{}{}.
\newblock
{\BBOQ}\APACrefatitle {Association between meteor radio afterglows and optical persistent trains} {Association between meteor radio afterglows and optical persistent trains}.{\BBCQ}
\newblock
\APACjournalVolNumPages{Journal of Geophysical Research: Space Physics}{125}{9}{e2020JA028053}.
\newblock
\begin{APACrefDOI} \doi{10.1029/2020JA028053} \end{APACrefDOI}
\PrintBackRefs{\CurrentBib}

\bibitem [\protect \citeauthoryear {%
Obenberger%
\ \protect \BOthers {.}}{%
Obenberger%
\ \protect \BOthers {.}}{%
{\protect \APACyear {2016}}%
}]{%
Obenberger2016}
\APACinsertmetastar {%
Obenberger2016}%
\begin{APACrefauthors}%
Obenberger, K\BPBI S.%
, Holmes, J\BPBI M.%
, Dowell, J\BPBI D.%
, Schinzel, F\BPBI K.%
, Stovall, K.%
, Sutton, E\BPBI K.%
\BCBL {}\ \BBA {} Taylor, G\BPBI B.%
\end{APACrefauthors}%
\unskip\
\newblock
\APACrefYearMonthDay{2016}{}{}.
\newblock
{\BBOQ}\APACrefatitle {Altitudinal dependence of meteor radio afterglows measured via optical counterparts} {Altitudinal dependence of meteor radio afterglows measured via optical counterparts}.{\BBCQ}
\newblock
\APACjournalVolNumPages{Geophysical Research Letters}{43}{17}{8885-8892}.
\newblock
\begin{APACrefDOI} \doi{10.1002/2016GL070059} \end{APACrefDOI}
\PrintBackRefs{\CurrentBib}

\bibitem [\protect \citeauthoryear {%
Obenberger%
\ \protect \BOthers {.}}{%
Obenberger%
\ \protect \BOthers {.}}{%
{\protect \APACyear {2014}}%
}]{%
Obenberger2014}
\APACinsertmetastar {%
Obenberger2014}%
\begin{APACrefauthors}%
Obenberger, K\BPBI S.%
, Taylor, G\BPBI B.%
, Hartman, J\BPBI M.%
, Dowell, J.%
, Ellingson, S\BPBI W.%
, Helmboldt, J\BPBI F.%
\BDBL {}Wilson, T\BPBI L.%
\end{APACrefauthors}%
\unskip\
\newblock
\APACrefYearMonthDay{2014}{}{}.
\newblock
{\BBOQ}\APACrefatitle {Detection of radio emission from fireballs} {Detection of radio emission from fireballs}.{\BBCQ}
\newblock
\APACjournalVolNumPages{Astrophysical Journal Letters}{788}{2}{L26}.
\newblock
\begin{APACrefDOI} \doi{10.1088/2041-8205/788/2/L26} \end{APACrefDOI}
\PrintBackRefs{\CurrentBib}

\bibitem [\protect \citeauthoryear {%
Obenberger%
\ \protect \BOthers {.}}{%
Obenberger%
\ \protect \BOthers {.}}{%
{\protect \APACyear {2015}}%
}]{%
Obenberger2015}
\APACinsertmetastar {%
Obenberger2015}%
\begin{APACrefauthors}%
Obenberger, K\BPBI S.%
, Taylor, G\BPBI B.%
, Lin, C\BPBI S.%
, Dowell, J.%
, Schinzel, F\BPBI K.%
\BCBL {}\ \BBA {} Stovall, K.%
\end{APACrefauthors}%
\unskip\
\newblock
\APACrefYearMonthDay{2015}{}{}.
\newblock
{\BBOQ}\APACrefatitle {Dynamic radio spectra from two fireballs} {Dynamic radio spectra from two fireballs}.{\BBCQ}
\newblock
\APACjournalVolNumPages{Journal of Geophysical Research: Space Physics}{120}{11}{9916-9928}.
\newblock
\begin{APACrefDOI} \doi{10.1002/2015JA021229} \end{APACrefDOI}
\PrintBackRefs{\CurrentBib}

\bibitem [\protect \citeauthoryear {%
Olivier%
}{%
Olivier%
}{%
{\protect \APACyear {1942}}%
}]{%
Olivier1942}
\APACinsertmetastar {%
Olivier1942}%
\begin{APACrefauthors}%
Olivier, C\BPBI P.%
\end{APACrefauthors}%
\unskip\
\newblock
\APACrefYearMonthDay{1942}{}{}.
\newblock
{\BBOQ}\APACrefatitle {Long enduring meteor trains} {Long enduring meteor trains}.{\BBCQ}
\newblock
\APACjournalVolNumPages{Proceedings of the American Philosophical Society}{85}{2}{93-135}.
\PrintBackRefs{\CurrentBib}

\bibitem [\protect \citeauthoryear {%
Olivier%
}{%
Olivier%
}{%
{\protect \APACyear {1947}}%
}]{%
Olivier1947}
\APACinsertmetastar {%
Olivier1947}%
\begin{APACrefauthors}%
Olivier, C\BPBI P.%
\end{APACrefauthors}%
\unskip\
\newblock
\APACrefYearMonthDay{1947}{}{}.
\newblock
{\BBOQ}\APACrefatitle {{Long enduring meteor trains: Second paper}} {{Long enduring meteor trains: Second paper}}.{\BBCQ}
\newblock
\APACjournalVolNumPages{Proceedings of the American Philosophical Society}{91}{4}{315-327}.
\PrintBackRefs{\CurrentBib}

\bibitem [\protect \citeauthoryear {%
Olivier%
}{%
Olivier%
}{%
{\protect \APACyear {1957}}%
}]{%
Olivier1957}
\APACinsertmetastar {%
Olivier1957}%
\begin{APACrefauthors}%
Olivier, C\BPBI P.%
\end{APACrefauthors}%
\unskip\
\newblock
\APACrefYearMonthDay{1957}{}{}.
\newblock
{\BBOQ}\APACrefatitle {{Long enduring meteor trains and fireball orbits: Third paper}} {{Long enduring meteor trains and fireball orbits: Third paper}}.{\BBCQ}
\newblock
\APACjournalVolNumPages{Proceedings of the American Philosophical Society}{101}{3}{296-315}.
\PrintBackRefs{\CurrentBib}

\bibitem [\protect \citeauthoryear {%
Pedregosa%
\ \protect \BOthers {.}}{%
Pedregosa%
\ \protect \BOthers {.}}{%
{\protect \APACyear {2011}}%
}]{%
Pedregosa2011}
\APACinsertmetastar {%
Pedregosa2011}%
\begin{APACrefauthors}%
Pedregosa, F.%
, Varoquaux, G.%
, Gramfort, A.%
, Michel, V.%
, Thirion, B.%
, Grisel, O.%
\BDBL {}Duchesnay, E.%
\end{APACrefauthors}%
\unskip\
\newblock
\APACrefYearMonthDay{2011}{}{}.
\newblock
{\BBOQ}\APACrefatitle {{Scikit-learn: Machine learning in Python}} {{Scikit-learn: Machine learning in Python}}.{\BBCQ}
\newblock
\APACjournalVolNumPages{Journal of Machine Learning Research}{12}{}{2825--2830}.
\PrintBackRefs{\CurrentBib}

\bibitem [\protect \citeauthoryear {%
Popov%
\ \protect \BOthers {.}}{%
Popov%
\ \protect \BOthers {.}}{%
{\protect \APACyear {2020}}%
}]{%
Popov2020}
\APACinsertmetastar {%
Popov2020}%
\begin{APACrefauthors}%
Popov, A\BPBI M.%
, Berezhnoy, A\BPBI A.%
, Borovička, J.%
, Labutin, T\BPBI A.%
, Zaytsev, S\BPBI M.%
\BCBL {}\ \BBA {} Stolyarov, A\BPBI V.%
\end{APACrefauthors}%
\unskip\
\newblock
\APACrefYearMonthDay{2020}{}{}.
\newblock
{\BBOQ}\APACrefatitle {{Tackling the FeO orange band puzzle in meteor and airglow spectra through combined astronomical and laboratory studies}} {{Tackling the FeO orange band puzzle in meteor and airglow spectra through combined astronomical and laboratory studies}}.{\BBCQ}
\newblock
\APACjournalVolNumPages{Monthly Notices of the Royal Astronomical Society}{500}{4}{4296-4306}.
\newblock
\begin{APACrefDOI} \doi{10.1093/mnras/staa3487} \end{APACrefDOI}
\PrintBackRefs{\CurrentBib}

\bibitem [\protect \citeauthoryear {%
{Russell}%
\ \protect \BOthers {.}}{%
{Russell}%
\ \protect \BOthers {.}}{%
{\protect \APACyear {1998}}%
}]{%
Russell1998}
\APACinsertmetastar {%
Russell1998}%
\begin{APACrefauthors}%
{Russell}, R\BPBI W.%
, {Rossano}, G\BPBI S.%
, {Chatelain}, M\BPBI A.%
, {Lynch}, D\BPBI K.%
, {Tessensohn}, T\BPBI K.%
, {Abendroth}, E.%
\BDBL {}{Jenniskens}, P.%
\end{APACrefauthors}%
\unskip\
\newblock
\APACrefYearMonthDay{1998}{}{}.
\newblock
{\BBOQ}\APACrefatitle {{Mid-infrared spectroscopy of persistent Leonid trains}} {{Mid-infrared spectroscopy of persistent Leonid trains}}.{\BBCQ}
\newblock
\APACjournalVolNumPages{Earth, Moon, and Planets}{82}{}{439-456}.
\newblock
\begin{APACrefDOI} \doi{10.1023/A:1017083811695} \end{APACrefDOI}
\PrintBackRefs{\CurrentBib}

\bibitem [\protect \citeauthoryear {%
Scott%
}{%
Scott%
}{%
{\protect \APACyear {1992}}%
}]{%
Scott1992}
\APACinsertmetastar {%
Scott1992}%
\begin{APACrefauthors}%
Scott, D\BPBI W.%
\end{APACrefauthors}%
\unskip\
\newblock
\APACrefYear{1992}.
\newblock
\APACrefbtitle {Multivariate density estimation theory, practice, and visualization} {Multivariate density estimation theory, practice, and visualization}.
\newblock
\APACaddressPublisher{}{John Wiley \& Sons, Inc.}
\newblock
\begin{APACrefDOI} \doi{10.1002/9780470316849} \end{APACrefDOI}
\PrintBackRefs{\CurrentBib}

\bibitem [\protect \citeauthoryear {%
Smieton%
}{%
Smieton%
}{%
{\protect \APACyear {1885}}%
}]{%
Smieton1885}
\APACinsertmetastar {%
Smieton1885}%
\begin{APACrefauthors}%
Smieton, J.%
\end{APACrefauthors}%
\unskip\
\newblock
\APACrefYearMonthDay{1885}{}{}.
\newblock
{\BBOQ}\APACrefatitle {{The November meteors}} {{The November meteors}}.{\BBCQ}
\newblock
\APACjournalVolNumPages{Nature}{33}{840}{104}.
\newblock
\begin{APACrefDOI} \doi{10.1038/033101b0} \end{APACrefDOI}
\PrintBackRefs{\CurrentBib}

\bibitem [\protect \citeauthoryear {%
Smith%
\ \protect \BOthers {.}}{%
Smith%
\ \protect \BOthers {.}}{%
{\protect \APACyear {2013}}%
}]{%
Smith2013}
\APACinsertmetastar {%
Smith2013}%
\begin{APACrefauthors}%
Smith, A\BPBI K.%
, Harvey, V\BPBI L.%
, Mlynczak, M\BPBI G.%
, Funke, B.%
, García-Comas, M.%
, Hervig, M.%
\BDBL {}Walker, K\BPBI A.%
\end{APACrefauthors}%
\unskip\
\newblock
\APACrefYearMonthDay{2013}{}{}.
\newblock
{\BBOQ}\APACrefatitle {Satellite observations of ozone in the upper mesosphere} {Satellite observations of ozone in the upper mesosphere}.{\BBCQ}
\newblock
\APACjournalVolNumPages{Journal of Geophysical Research: Atmospheres}{118}{11}{5803-5821}.
\newblock
\begin{APACrefDOI} \doi{10.1002/jgrd.50445} \end{APACrefDOI}
\PrintBackRefs{\CurrentBib}

\bibitem [\protect \citeauthoryear {%
{Spurn{\'y}}%
\ \protect \BOthers {.}}{%
{Spurn{\'y}}%
\ \protect \BOthers {.}}{%
{\protect \APACyear {2014}}%
}]{%
Spurny2014}
\APACinsertmetastar {%
Spurny2014}%
\begin{APACrefauthors}%
{Spurn{\'y}}, P.%
, {Shrben{\'y}}, L.%
, {Borovi{\v{c}}ka}, J.%
, {Koten}, P.%
, {Voj{\'a}{\v{c}}ek}, V.%
\BCBL {}\ \BBA {} {{\v{S}}tork}, R.%
\end{APACrefauthors}%
\unskip\
\newblock
\APACrefYearMonthDay{2014}{}{}.
\newblock
{\BBOQ}\APACrefatitle {{Bright Perseid fireball with exceptional beginning height of 170 km observed by different techniques}} {{Bright Perseid fireball with exceptional beginning height of 170 km observed by different techniques}}.{\BBCQ}
\newblock
\APACjournalVolNumPages{Astronomy \& Astrophysics}{563}{}{A64}.
\newblock
\begin{APACrefDOI} \doi{10.1051/0004-6361/201323261} \end{APACrefDOI}
\PrintBackRefs{\CurrentBib}

\bibitem [\protect \citeauthoryear {%
Subasinghe%
, Campbell-Brown%
\BCBL {}\ \BBA {} Stokan%
}{%
Subasinghe%
\ \protect \BOthers {.}}{%
{\protect \APACyear {2016}}%
}]{%
Subasinghe2016}
\APACinsertmetastar {%
Subasinghe2016}%
\begin{APACrefauthors}%
Subasinghe, D.%
, Campbell-Brown, M\BPBI D.%
\BCBL {}\ \BBA {} Stokan, E.%
\end{APACrefauthors}%
\unskip\
\newblock
\APACrefYearMonthDay{2016}{}{}.
\newblock
{\BBOQ}\APACrefatitle {{Physical characteristics of faint meteors by light curve and high-resolution observations, and the implications for parent bodies}} {{Physical characteristics of faint meteors by light curve and high-resolution observations, and the implications for parent bodies}}.{\BBCQ}
\newblock
\APACjournalVolNumPages{Monthly Notices of the Royal Astronomical Society}{457}{2}{1289-1298}.
\newblock
\begin{APACrefDOI} \doi{10.1093/mnras/stw019} \end{APACrefDOI}
\PrintBackRefs{\CurrentBib}

\bibitem [\protect \citeauthoryear {%
{The pandas Development Team}%
}{%
{The pandas Development Team}%
}{%
{\protect \APACyear {2023}}%
}]{%
pandas2023}
\APACinsertmetastar {%
pandas2023}%
\begin{APACrefauthors}%
{The pandas Development Team}.%
\end{APACrefauthors}%
\unskip\
\newblock
\APACrefYearMonthDay{2023}{}{}.
\newblock
\APACrefbtitle {{pandas-dev/pandas: Pandas}} {{pandas-dev/pandas: Pandas}}\ [software].
\newblock
\APACaddressPublisher{}{Zenodo}.
\newblock
\begin{APACrefDOI} \doi{10.5281/zenodo.8092754} \end{APACrefDOI}
\PrintBackRefs{\CurrentBib}

\bibitem [\protect \citeauthoryear {%
Trowbridge%
}{%
Trowbridge%
}{%
{\protect \APACyear {1907}}%
}]{%
Trowbridge1907}
\APACinsertmetastar {%
Trowbridge1907}%
\begin{APACrefauthors}%
Trowbridge, C\BPBI C.%
\end{APACrefauthors}%
\unskip\
\newblock
\APACrefYearMonthDay{1907}{}{}.
\newblock
{\BBOQ}\APACrefatitle {Physical nature of meteor trains} {Physical nature of meteor trains}.{\BBCQ}
\newblock
\APACjournalVolNumPages{The Astrophysical Journal}{26}{2}{191-205}.
\newblock
\begin{APACrefDOI} \doi{10.1086/141478} \end{APACrefDOI}
\PrintBackRefs{\CurrentBib}

\bibitem [\protect \citeauthoryear {%
Trowbridge%
}{%
Trowbridge%
}{%
{\protect \APACyear {1911}}%
}]{%
Trowbridge1911}
\APACinsertmetastar {%
Trowbridge1911}%
\begin{APACrefauthors}%
Trowbridge, C\BPBI C.%
\end{APACrefauthors}%
\unskip\
\newblock
\APACrefYearMonthDay{1911}{}{}.
\newblock
{\BBOQ}\APACrefatitle {The origin of luminous meteor trains} {The origin of luminous meteor trains}.{\BBCQ}
\newblock
\APACjournalVolNumPages{Popular Science Monthly}{79}{}{191-203}.
\PrintBackRefs{\CurrentBib}

\bibitem [\protect \citeauthoryear {%
van~der Walt%
\ \protect \BOthers {.}}{%
van~der Walt%
\ \protect \BOthers {.}}{%
{\protect \APACyear {2014}}%
}]{%
skimage2014}
\APACinsertmetastar {%
skimage2014}%
\begin{APACrefauthors}%
van~der Walt, S.%
, Sch\"onberger, J\BPBI L.%
, Nunez-Iglesias, J.%
, Boulogne, F.%
, Warner, J\BPBI D.%
, Yager, N.%
\BDBL {}the scikit-image Contributors%
\end{APACrefauthors}%
\unskip\
\newblock
\APACrefYearMonthDay{2014}{}{}.
\newblock
{\BBOQ}\APACrefatitle {{scikit-image: Image processing in Python}} {{scikit-image: Image processing in Python}}.{\BBCQ}
\newblock
\APACjournalVolNumPages{PeerJ}{2}{}{e453}.
\newblock
\begin{APACrefDOI} \doi{10.7717/peerj.453} \end{APACrefDOI}
\PrintBackRefs{\CurrentBib}

\bibitem [\protect \citeauthoryear {%
Varghese%
, Dowell%
, Obenberger%
, Taylor%
\BCBL {}\ \BBA {} Malins%
}{%
Varghese%
\ \protect \BOthers {.}}{%
{\protect \APACyear {2021}}%
}]{%
Varghese2021}
\APACinsertmetastar {%
Varghese2021}%
\begin{APACrefauthors}%
Varghese, S\BPBI S.%
, Dowell, J.%
, Obenberger, K\BPBI S.%
, Taylor, G\BPBI B.%
\BCBL {}\ \BBA {} Malins, J.%
\end{APACrefauthors}%
\unskip\
\newblock
\APACrefYearMonthDay{2021}{}{}.
\newblock
{\BBOQ}\APACrefatitle {Broadband imaging to study the spectral distribution of meteor radio afterglows} {Broadband imaging to study the spectral distribution of meteor radio afterglows}.{\BBCQ}
\newblock
\APACjournalVolNumPages{Journal of Geophysical Research: Space Physics}{126}{10}{e2021JA029296}.
\newblock
\begin{APACrefDOI} \doi{10.1029/2021JA029296} \end{APACrefDOI}
\PrintBackRefs{\CurrentBib}

\bibitem [\protect \citeauthoryear {%
Varghese%
, Obenberger%
, Taylor%
\BCBL {}\ \BBA {} Dowell%
}{%
Varghese%
\ \protect \BOthers {.}}{%
{\protect \APACyear {2019}}%
}]{%
Varghese2019}
\APACinsertmetastar {%
Varghese2019}%
\begin{APACrefauthors}%
Varghese, S\BPBI S.%
, Obenberger, K\BPBI S.%
, Taylor, G\BPBI B.%
\BCBL {}\ \BBA {} Dowell, J.%
\end{APACrefauthors}%
\unskip\
\newblock
\APACrefYearMonthDay{2019}{}{}.
\newblock
{\BBOQ}\APACrefatitle {Testing the radiation pattern of meteor radio afterglow} {Testing the radiation pattern of meteor radio afterglow}.{\BBCQ}
\newblock
\APACjournalVolNumPages{Journal of Geophysical Research: Space Physics}{124}{12}{10749-10759}.
\newblock
\begin{APACrefDOI} \doi{10.1029/2019JA026922} \end{APACrefDOI}
\PrintBackRefs{\CurrentBib}

\bibitem [\protect \citeauthoryear {%
Vasilyev%
\ \protect \BOthers {.}}{%
Vasilyev%
\ \protect \BOthers {.}}{%
{\protect \APACyear {2021}}%
}]{%
Vasilyev2021}
\APACinsertmetastar {%
Vasilyev2021}%
\begin{APACrefauthors}%
Vasilyev, R\BPBI V.%
, Syrenova, T\BPBI E.%
, Beletsky, A\BPBI B.%
, Artamonov, M\BPBI F.%
, Merzlyakov, E\BPBI G.%
, Podlesny, A\BPBI V.%
\BCBL {}\ \BBA {} Cedric, M\BPBI V.%
\end{APACrefauthors}%
\unskip\
\newblock
\APACrefYearMonthDay{2021}{}{}.
\newblock
{\BBOQ}\APACrefatitle {Studying a long-lasting meteor trail from stereo images and radar data} {Studying a long-lasting meteor trail from stereo images and radar data}.{\BBCQ}
\newblock
\APACjournalVolNumPages{Atmosphere}{12}{7}{}.
\newblock
\begin{APACrefDOI} \doi{10.3390/atmos12070841} \end{APACrefDOI}
\PrintBackRefs{\CurrentBib}

\bibitem [\protect \citeauthoryear {%
{Vida}%
, {Campbell-Brown}%
, {Brown}%
, {Egal}%
\BCBL {}\ \BBA {} {Mazur}%
}{%
{Vida}%
\ \protect \BOthers {.}}{%
{\protect \APACyear {2020}}%
}]{%
Vida2020}
\APACinsertmetastar {%
Vida2020}%
\begin{APACrefauthors}%
{Vida}, D.%
, {Campbell-Brown}, M.%
, {Brown}, P\BPBI G.%
, {Egal}, A.%
\BCBL {}\ \BBA {} {Mazur}, M\BPBI J.%
\end{APACrefauthors}%
\unskip\
\newblock
\APACrefYearMonthDay{2020}{}{}.
\newblock
{\BBOQ}\APACrefatitle {{A new method for measuring the meteor mass index: Application to the 2018 Draconid meteor shower outburst}} {{A new method for measuring the meteor mass index: Application to the 2018 Draconid meteor shower outburst}}.{\BBCQ}
\newblock
\APACjournalVolNumPages{Astronomy \& Astrophysics}{635}{}{A153}.
\newblock
\begin{APACrefDOI} \doi{10.1051/0004-6361/201937296} \end{APACrefDOI}
\PrintBackRefs{\CurrentBib}

\bibitem [\protect \citeauthoryear {%
Vida%
, Gural%
, Brown%
, Campbell-Brown%
\BCBL {}\ \BBA {} Wiegert%
}{%
Vida%
\ \protect \BOthers {.}}{%
{\protect \APACyear {2019}}%
}]{%
Vida2019}
\APACinsertmetastar {%
Vida2019}%
\begin{APACrefauthors}%
Vida, D.%
, Gural, P\BPBI S.%
, Brown, P\BPBI G.%
, Campbell-Brown, M.%
\BCBL {}\ \BBA {} Wiegert, P.%
\end{APACrefauthors}%
\unskip\
\newblock
\APACrefYearMonthDay{2019}{}{}.
\newblock
{\BBOQ}\APACrefatitle {{Estimating trajectories of meteors: an observational Monte Carlo approach – I. Theory}} {{Estimating trajectories of meteors: an observational Monte Carlo approach – I. Theory}}.{\BBCQ}
\newblock
\APACjournalVolNumPages{Monthly Notices of the Royal Astronomical Society}{491}{2}{2688-2705}.
\newblock
\begin{APACrefDOI} \doi{10.1093/mnras/stz3160} \end{APACrefDOI}
\PrintBackRefs{\CurrentBib}

\bibitem [\protect \citeauthoryear {%
Vida%
\ \protect \BOthers {.}}{%
Vida%
\ \protect \BOthers {.}}{%
{\protect \APACyear {2021}}%
}]{%
Vida2021}
\APACinsertmetastar {%
Vida2021}%
\begin{APACrefauthors}%
Vida, D.%
, Šegon, D.%
, Gural, P\BPBI S.%
, Brown, P\BPBI G.%
, McIntyre, M\BPBI J\BPBI M.%
, Dijkema, T\BPBI J.%
\BDBL {}Zubović, D.%
\end{APACrefauthors}%
\unskip\
\newblock
\APACrefYearMonthDay{2021}{}{}.
\newblock
{\BBOQ}\APACrefatitle {{The Global Meteor Network – Methodology and first results}} {{The Global Meteor Network – Methodology and first results}}.{\BBCQ}
\newblock
\APACjournalVolNumPages{Monthly Notices of the Royal Astronomical Society}{506}{4}{5046-5074}.
\newblock
\begin{APACrefDOI} \doi{10.1093/mnras/stab2008} \end{APACrefDOI}
\PrintBackRefs{\CurrentBib}

\bibitem [\protect \citeauthoryear {%
Voj\'acek%
, Borovi{\v{c}}ka%
, Koten%
, Spurn\'y%
\BCBL {}\ \BBA {} Stork%
}{%
Voj\'acek%
\ \protect \BOthers {.}}{%
{\protect \APACyear {2019}}%
}]{%
Vojacek2019}
\APACinsertmetastar {%
Vojacek2019}%
\begin{APACrefauthors}%
Voj\'acek, V.%
, Borovi{\v{c}}ka, J.%
, Koten, P.%
, Spurn\'y, P.%
\BCBL {}\ \BBA {} Stork, R.%
\end{APACrefauthors}%
\unskip\
\newblock
\APACrefYearMonthDay{2019}{}{}.
\newblock
{\BBOQ}\APACrefatitle {Properties of small meteoroids studied by meteor video observations} {Properties of small meteoroids studied by meteor video observations}.{\BBCQ}
\newblock
\APACjournalVolNumPages{Astronomy \& Astrophysics}{621}{}{A68}.
\newblock
\begin{APACrefDOI} \doi{10.1051/0004-6361/201833289} \end{APACrefDOI}
\PrintBackRefs{\CurrentBib}

\bibitem [\protect \citeauthoryear {%
{West}%
\ \BBA {} {Broida}%
}{%
{West}%
\ \BBA {} {Broida}%
}{%
{\protect \APACyear {1975}}%
}]{%
West1975}
\APACinsertmetastar {%
West1975}%
\begin{APACrefauthors}%
{West}, J\BPBI B.%
\BCBT {}\ \BBA {} {Broida}, H\BPBI P.%
\end{APACrefauthors}%
\unskip\
\newblock
\APACrefYearMonthDay{1975}{}{}.
\newblock
{\BBOQ}\APACrefatitle {{Chemiluminescence and photoluminescence of diatomic iron oxide}} {{Chemiluminescence and photoluminescence of diatomic iron oxide}}.{\BBCQ}
\newblock
\APACjournalVolNumPages{The Journal of Chemical Physics}{62}{7}{2566-2574}.
\newblock
\begin{APACrefDOI} \doi{10.1063/1.430837} \end{APACrefDOI}
\PrintBackRefs{\CurrentBib}

\bibitem [\protect \citeauthoryear {%
Wiegert%
, Brown%
, Weryk%
\BCBL {}\ \BBA {} Wong%
}{%
Wiegert%
\ \protect \BOthers {.}}{%
{\protect \APACyear {2013}}%
}]{%
Wiegert2013}
\APACinsertmetastar {%
Wiegert2013}%
\begin{APACrefauthors}%
Wiegert, P\BPBI A.%
, Brown, P\BPBI G.%
, Weryk, R\BPBI J.%
\BCBL {}\ \BBA {} Wong, D\BPBI K.%
\end{APACrefauthors}%
\unskip\
\newblock
\APACrefYearMonthDay{2013}{}{}.
\newblock
{\BBOQ}\APACrefatitle {{The return of the Andromedids meteor shower}} {{The return of the Andromedids meteor shower}}.{\BBCQ}
\newblock
\APACjournalVolNumPages{The Astronomical Journal}{145}{3}{70}.
\newblock
\begin{APACrefDOI} \doi{10.1088/0004-6256/145/3/70} \end{APACrefDOI}
\PrintBackRefs{\CurrentBib}

\bibitem [\protect \citeauthoryear {%
{Yamamoto}%
, {Toda}%
, {Higa}%
, {Maeda}%
\BCBL {}\ \BBA {} {Watanabe}%
}{%
{Yamamoto}%
\ \protect \BOthers {.}}{%
{\protect \APACyear {2004}}%
}]{%
Yamamoto2004}
\APACinsertmetastar {%
Yamamoto2004}%
\begin{APACrefauthors}%
{Yamamoto}, M\BHBI Y.%
, {Toda}, M.%
, {Higa}, Y.%
, {Maeda}, K.%
\BCBL {}\ \BBA {} {Watanabe}, J\BHBI I.%
\end{APACrefauthors}%
\unskip\
\newblock
\APACrefYearMonthDay{2004}{}{}.
\newblock
{\BBOQ}\APACrefatitle {{Altitudinal distribution of 20 persistent meteor trains: Estimates derived from Metro campaign archives}} {{Altitudinal distribution of 20 persistent meteor trains: Estimates derived from Metro campaign archives}}.{\BBCQ}
\newblock
\APACjournalVolNumPages{Earth, Moon, and Planets}{95}{1-4}{279-287}.
\newblock
\begin{APACrefDOI} \doi{10.1007/s11038-005-9048-4} \end{APACrefDOI}
\PrintBackRefs{\CurrentBib}

\bibitem [\protect \citeauthoryear {%
Yuan%
\ \protect \BOthers {.}}{%
Yuan%
\ \protect \BOthers {.}}{%
{\protect \APACyear {2008}}%
}]{%
Yuan2008}
\APACinsertmetastar {%
Yuan2008}%
\begin{APACrefauthors}%
Yuan, T.%
, She, C\BHBI Y.%
, Krueger, D\BPBI A.%
, Sassi, F.%
, Garcia, R.%
, Roble, R\BPBI G.%
\BDBL {}Schmidt, H.%
\end{APACrefauthors}%
\unskip\
\newblock
\APACrefYearMonthDay{2008}{}{}.
\newblock
{\BBOQ}\APACrefatitle {{Climatology of mesopause region temperature, zonal wind, and meridional wind over Fort Collins, Colorado (41°N, 105°W), and comparison with model simulations}} {{Climatology of mesopause region temperature, zonal wind, and meridional wind over Fort Collins, Colorado (41°N, 105°W), and comparison with model simulations}}.{\BBCQ}
\newblock
\APACjournalVolNumPages{Journal of Geophysical Research: Atmospheres}{113}{D3}{D03105}.
\newblock
\begin{APACrefDOI} \doi{10.1029/2007JD008697} \end{APACrefDOI}
\PrintBackRefs{\CurrentBib}

\bibitem [\protect \citeauthoryear {%
Zhu%
, Byrd%
, Lu%
\BCBL {}\ \BBA {} Nocedal%
}{%
Zhu%
\ \protect \BOthers {.}}{%
{\protect \APACyear {1997}}%
}]{%
LBFGSB}
\APACinsertmetastar {%
LBFGSB}%
\begin{APACrefauthors}%
Zhu, C.%
, Byrd, R\BPBI H.%
, Lu, P.%
\BCBL {}\ \BBA {} Nocedal, J.%
\end{APACrefauthors}%
\unskip\
\newblock
\APACrefYearMonthDay{1997}{}{}.
\newblock
{\BBOQ}\APACrefatitle {{Algorithm 778: L-BFGS-B: Fortran subroutines for large-scale bound-constrained optimization}} {{Algorithm 778: L-BFGS-B: Fortran subroutines for large-scale bound-constrained optimization}}.{\BBCQ}
\newblock
\APACjournalVolNumPages{ACM Transactions on Mathematical Software}{23}{4}{550–560}.
\newblock
\begin{APACrefDOI} \doi{10.1145/279232.279236} \end{APACrefDOI}
\PrintBackRefs{\CurrentBib}

\bibitem [\protect \citeauthoryear {%
Zinn%
\ \BBA {} Drummond%
}{%
Zinn%
\ \BBA {} Drummond%
}{%
{\protect \APACyear {2005}}%
}]{%
Zinn2005}
\APACinsertmetastar {%
Zinn2005}%
\begin{APACrefauthors}%
Zinn, J.%
\BCBT {}\ \BBA {} Drummond, J.%
\end{APACrefauthors}%
\unskip\
\newblock
\APACrefYearMonthDay{2005}{}{}.
\newblock
{\BBOQ}\APACrefatitle {{Observations of persistent Leonid meteor trails: 4. Buoyant rise/vortex formation as mechanism for creation of parallel meteor train pairs}} {{Observations of persistent Leonid meteor trails: 4. Buoyant rise/vortex formation as mechanism for creation of parallel meteor train pairs}}.{\BBCQ}
\newblock
\APACjournalVolNumPages{Journal of Geophysical Research: Space Physics}{110}{A4}{}.
\newblock
\begin{APACrefDOI} \doi{10.1029/2004JA010575} \end{APACrefDOI}
\PrintBackRefs{\CurrentBib}

\end{thebibliography}
\end{document}

%% ------------------------------------------------------------------------ %%
%
%  TITLE
%
%% ------------------------------------------------------------------------ %%

%\includegraphics{agu_pubart-white_reduced.eps}

\title{Supporting Information for ``Not So Fast: A New Catalog of Meteor Persistent Trains"}
%
%
%DOI: 10.1002/%insert paper number here%

%% ------------------------------------------------------------------------ %%
%
%  AUTHORS AND AFFILIATIONS
%
%% ------------------------------------------------------------------------ %%

\authors{L. E. Cordonnier\affil{1,2}, K. S. Obenberger\affil{1}, J. M. Holmes\affil{1}, and G. B. Taylor\affil{2}, and D. Vida\affil{3}}

\affiliation{1}{Space Vehicles Directorate, Air Force Research Laboratory, Kirtland AFB, NM, USA}
\affiliation{2}{Department of Physics and Astronomy, University of New Mexico, Albuquerque, NM, USA}
\affiliation{3}{Department of Physics and Astronomy, University of Western Ontario, London, Ontario, Canada}

\begin{article}

%% ------------------------------------------------------------------------ %%
%
%  TEXT
%
%% ------------------------------------------------------------------------ %%

\noindent\textbf{Contents of this file}
%%%Remove or add items as needed%%%
\begin{enumerate}
\item Captions for Movies S1 to S6
\end{enumerate}

\noindent\textbf{Introduction}
%Type or paste your text here. The introduction gives a brief overview of the supporting information. You should include information %about as many of the following as possible (when appropriate):
% 1. a general overview of the kind of data files;
% 2. information about when and how the data were collected or created;
% 3. a general description of processing steps used;
% 4. any known imperfections or anomalies in the data.

%\clearpage

In order to support our paper, we have included six movie files of persistent train (PT) evolution. These movies are meant to represent the diverse structural morphologies exhibited by PTs, as well as to showcase some of the visually interesting trains that we observed. The movies were created using images taken by the Widefield Persistent Train Camera 2nd Edition (WiPT2 Cam), which was deployed at the Sevilleta National Wildlife Refuge in New Mexico. The WiPT2 uses a ZWO ASI6200MM camera with an IMX455 CMOS sensor. It is fairly sensitive to NIR emission, having a quantum efficiency (QE) of about 45\% at 700 nm and 15\% at 900 nm. A Canon EF 15mm f/2.8 fisheye lens is mounted atop the camera. The camera is configured to take long exposure (5 second) images and has an inherent $\sim$1.5 second readout time between images. These images are written to HDF5 files and saved locally to a System 76 Meerkat computer with 4 TBs of storage. The camera only collects data while the moon is not in the sky (to prevent the images from becoming overexposed) and requires the sun to be at least 15$^{\circ}$ below the horizon. Typical conditions in a clear, dark patch of sky provide a limiting apparent stellar magnitude around +8 mag which we take to also represent the dimmest train visible under the best of conditions. All movies included here are in shown in the camera's full, non-archival resolution. Airplanes and satellites are commonly seen throughout the movies; these appear as streaks which travel across multiple consecutive frames. 

\noindent\textbf{Movie S1.}
%upload your movie(s) to AGU's journal submission site and select, "Supporting Information %(SI)" as the file type. Following naming convention: ms01.

Meteor occurred around 2021-11-02 04:45:44 UTC. Movie shows about 11 minutes (100 frames) worth of PT evolution. A portion of the meteor occurred during the readout gap between frames causing the trailing region to appear dimmer---though the PT seems to be unassociated with the (apparent) brightest part of the meteor, without the complete light curve this is a meaningless claim.  

\noindent\textbf{Movie S2.}

Meteor occurred around 2021-12-06 11:28:52 UTC. Movie shows about 11 minutes (100 frames) worth of PT evolution. This PT becomes stretched and appears to double back on itself.

\noindent\textbf{Movie S3.}

Meteor occurred around 2023-01-22 11:54:38 UTC. Movie shows about 22 minutes (200 frames) worth of PT evolution and corresponds to the PT seen in Figure 1 of the paper. A bright knot appears in the train during its early stages. Unfortunately, the bottom boundary of the movie represents the edge of the WiPT2's field of view, so we were unable to determine the true duration of this PT. 

\noindent\textbf{Movie S4.}

Meteor occurred around 2023-01-27 08:59:34 UTC. Movie shows about 17 minutes (155 frames) worth of PT evolution. The train in this example becomes diffuse relatively quickly, but continues its emission through the end of the movie. 

\noindent\textbf{Movie S5.}

Meteor occurred around 2023-04-01 11:12:59 UTC. Movie shows about 30 minutes (275 frames) worth of PT evolution. The meteor occurs at the far right of the movie on top of the galactic plane; the subsequent PT gradually drifts from right to left.

\noindent\textbf{Movie S6.}

Meteor occurred around 2023-04-23 09:34:28 UTC. Movie shows about 11 minutes (100 frames) worth of PT evolution. The example shows a meteor with a relatively small apparent angular extent producing a PT that expands nearly perpendicularly to the direction of the meteor.

\end{article}
\clearpage